# Main Belt Binary Asteroidal Systems With Circular Mutual Orbits[*]

F. Marchis[a,b,c], P. Descamps[b], M. Baek[c], A. W. Harris[d], M. Kaasalainen[e], J. Berthier[b], D. Hestroffer[b], F. Vachier[b]

[a] University of California at Berkeley, Department of Astronomy, 601 Campbell Hall, Berkeley, CA 94720, USA
[b] Institut de Mécanique Céleste et de Calcul des Éphémérides, Observatoire de Paris, 75014 Paris, France
c. SETI Institute, Carl Sagan Center, 515 N. Whismann Road, Mountain View CA 94043, USA
d. DLR Institute of Planetary Research, Rutherfordstrasse 2, 12489 Berlin, Germany
e. Department of Mathematics and Statistics, Gustaf Hallstromin katu 2b, P.O. Box 68, FIN-00014 University of Helsinki, Finland

Pages: 72
Tables: 9
Figures: 4

**Proposed running head:** circular mutual orbits of binary asteroidal systems

**Editorial correspondence to:**
Franck Marchis
Department of Astronomy
601 Campbell Hall
Berkeley CA 94720
USA
Phone: +1 510 642 3928
Fax: +1 510 642 3411
Email: fmarchis@berkeley.edu

**ABSTRACT**

In 2003, we initiated a long-term adaptive optics campaign to study the orbit of various main-belt asteroidal systems. Here we present a consistent solution for the mutual orbits of four binary systems: 22 Kalliope, 45 Eugenia, 107 Camilla and 762 Pulcova. With the exception of 45 Eugenia, we did not detect any additional satellites around these systems although we have the capability of detecting a loosely-bound fragment (located at 1/4 × $R_{Hill}$) that is ~40 times smaller in diameter than the primary. The common characteristic of these mutual orbits is that they are roughly circular. Three of these binary systems belong to a C-"group" taxonomic class. Our estimates of their bulk densities are consistently lower (~1 g/cm$^3$) than their associated meteorite analogs, suggesting an interior porosity of 30-50% (taking CI-CO meteorites as analogs). 22 Kalliope, a W-type asteroid, has a significantly higher bulk density of ~3 g/cm$^3$, derived based on IRAS radiometric size measurement. We compare the characteristics of these orbits in the light of tidal-effect evolution.

## 1. Introduction

The frequent discovery in recent years of companions orbiting large asteroids is largely due to the increased image quality provided by Adaptive Optics (AO) on large-aperture ground-based telescopes and by direct imaging cameras on the Hubble Space Telescope. At the time of writing, 70 binary or multiple systems have been imaged. Five years after the discovery by the Galileo spacecraft of Dactyl, a companion of (243) Ida (Belton *et al.* 1996), the Canada-France-Hawaii Telescope (CFHT) AO system provided the first ground–based observations revealing the presence of a binary system: Petit-Prince, a small companion of 45 Eugenia (Merline *et al.* 1999). To date, companions around 17 NEAs have been resolved by means of radar observations, 15 multiple main-belt asteroids have been imaged with AO and HST, 2 Jupiter-Trojans with AO and ~45 multiple TNOs with HST and by means of classical observations.

The total number of suspected binary asteroid systems is in fact significantly higher (~155) since many of them display mutual event signatures (Behrend *et al.* 2006, Descamps *et al.* 2007a) and/or multi-period components (Pravec *et al.* 2006) in their lightcurves.

In 2003, as part of our quest to better understand the formation of these binary systems, we initiated an observational program using high-resolution capabilities of various 8m-class telescopes equipped with AO systems (Yepun-UT4 of the Very Large Telescope, W.M. Keck-II and Gemini-North). The goal of our program is to gain insight into the physical characteristics these binary systems, such as the orbital parameters of the satellites, the size and shape of the components of the systems, the composition of their surfaces, the bulk densities and distribution of materials in the interiors of the primaries. We have supplemented our observational data with archive data from various other observations (LAOSA, Large Adaptive Optics Survey of Asteroids, see Marchis *et al.* 2006c). Descamps *et al.* (2007) have published a complete analysis of 90 Antiope, a double system, and Marchis *et al.* (2008) have published the orbits of 4 binary asteroid moonlets having significantly eccentric orbits.

In this work, we focus on four binary asteroids with small companions, or moonlets, revolving around their primaries in almost circular orbits. In Section 2 we present the AO observations of the four resolved binary systems, 22 Kalliope, 45 Eugenia, 107 Camilla, and 762 Pulcova. Section 3 describes how we derive the orbits of these systems. In Section 4 we estimate the bulk density of the primaries using reprocessed IRAS radiometric data and direct resolved measurements. Finally, in Section 5 we discuss the origin of these systems based on their measured characteristics.

## 2. Adaptive optics observations

### 2.1 Collected data and basic data reduction

It is only more than thirty years after Babcock (1953) proposed his concept of real time atmospheric correction that this idea was technologically applicable and developed in the 1980's on 3-4m size ground-based telescopes. These types of instruments, named Adaptive Optics (AO) procure a correction that is only partial and slightly variable in time in the near-infrared (1-5μm). Several AO systems are now available on 8m-class telescopes, such as Keck-10m II and Gemini-8m North both at Mauna Kea (Hawaii, USA) and the Yepun-8m of the Very Large Telescope observatory at Paranal (Chile).

These systems provide a stable correction in K-band (2.2 μm), with an angular resolution close to the diffraction limit of the telescope which is 60 milli-arcsec (*mas*) for the Gemini and the VLT, and 50 mas for the Keck under good exterior seeing conditions (<0.8") on targets brighter than the 13-14[th] magnitude in visible.

In 2003, we initiated a large campaign of observations using the UT4 of the VLT of the European Southern Observatory and its AO system called NAOS (Nasmyth Adaptive Optics System). The observations were recorded in direct imaging using the CONICA near-infrared camera equipped with an ALADDIN2 1024×1024 pixel InSb array of 27 μm pixels. Most of the data were recorded with the S13 camera (13.27 *mas*/pixel scale) in Ks band (central wavelength 2.18 μm and bandwidth of 0.35 μm). NACO, which stands for NAOS-CONICA, provides the best angular correction in this wavelength range (Lenzen *et al.* 2003, Rousset *et al.* 2003). Approximately 70 hours of observations were allocated to this program in service observing. In 2005 and 2006, we continued this program using the Gemini North telescope, which is equipped with the ALTAIR AO system (Herriot *et al.* 2000) and NIRI, the Gemini Near-infrared Imager (Hodapp *et al.* 2003). NIRI contains a 1024×1024 pixel ALADDIN InSb array, which is sensitive from 1 to 5 microns. It was used in imaging mode along with the f/32 cameras providing a pixel scale of 22 mas. Twelve hours of observations were recorded in queue scheduling under median seeing conditions of ~1.0" with this instrument. On a few occasions during this campaign, complementary Ks band observations taken with the Keck-II AO and its Near-InfraRed Camera (NIRC2) were added to our analysis. We also included in the LAOSA database all observations retrieved from Gemini-North and VLT archive centers totaling ~1100 AO observations of various minor planets.

The basic data processing (sky subtraction, bad-pixel removal, and flat-field correction) applied on all these raw data was performed using the *eclipse* data reduction package (Devillard, 1997). Successive frames taken over a time span of less than 6 min, were combined into one single average image after applying an accurate shift-and-add process through the *Jitter* pipeline offered in the same package. Data processing with this software on such high S/N data (>1000) is relatively straightforward. Since these data respect the Shannon's theorem, it is possible to retrieve completely the continuous signal from the knowledge of enough samples. After re-sampling each image to $1/10^{th}$ of the pixel level, the centroid position on each frame can be accurately measured by a Gaussian fit. The final image is obtained by stacking the set of frames with individual shifts determined from cross-correlation.

**2.2 Targets**

This work describes the analysis of four main-belt minor planets already known to have a satellite: 22 Kalliope, 45 Eugenia, 107 Camilla, and 762 Pulcova. The discovery of a satellite around 22 Kalliope, nammed "Kalliope Linus I" was announced simultaneously by Merline *et al.* (2001) and Margot and Brown (2001) in September 2001 from Keck AO observations. At the time of the discovery, this was the first and only supposedly M-type asteroid (Tholen and Barucci, 1989) with a moonlet companion. The average size of the primary estimated by radiometric measurement from IRAS is 181 km (Tedesco *et al.* 2002). "Petit-Prince" moonlet was discovered around 45 Eugenia by Merline *et al.* (1999) using PUEO at CFHT. Its primary is a C-type (Bus and Binzel, 2002) asteroid with an estimated diameter of 215 km (Tedesco *et al.* 2002). 107 Camilla companion was revealed through HST observations and announced in March 2001 by Storrs et al (2001). The taxonomic classification of this 223 km diameter asteroid is 'X' by Bus and Binzel (2002). In June 2000, the binary nature of 762 Pulcova was shown by Merline *et al.* (2000) based on CFHT AO (PUEO) observations. The IRAS radiometric diameter of this Cb-type asteroid (Lazzaro *et al.* 2004) is estimated to be 137 km (Tedesco *et al.* 2002). Table 1 summarizes the known characteristics of these minor planets extracted from various published sources.

For 107 Camilla and 762 Pulcova, the orbital parameters of the companion orbit were previously unknown or poorly defined. The orbital elements of 22 Kalliope's companion was first published in Marchis *et al.* (2002). Two symmetrical solutions were later proposed in Marchis et al (2003) and Margot and Brown (2003) based on Lick-3m and Palomar-5m observations. The orbit of Petit-Prince was partially published in Merline *et al.* (1999). The main motivation of this work is to obtain an accurate knowledge of their orbit that allow us to calculate directly the mass of the system from the Kepler's third law, the size of the moonlet and the primary, and eventually the bulk density and porosity of the primary. New observations were acquired using 8-10m class telescopes. Tables 2a-2d contain the observing log of all reduced observations of these binary systems extracted from the LAOSA database (Marchis *et al.* 2006 and available in SSODNET[1]).

**Material:**

**Table 1: Characteristics of the studied minor planets**

**Table 2a,b,c,d: Observing conditions of AO observations**

---

[1] http://www.imcce.fr/page.php?nav=fr/observateur/support/ssodnet/

**2.3 Search for moonlet companions**

We describe in Marchis *et al.* (2006b) and Marchis *et al.* (2008) the challenge of searching a faint point source around a bright asteroid. The Point Spread Function (PSF) of an AO system is composed of a coherent peak surrounded by a halo in which speckle patterns are also present. Because these speckle artifacts are variable in time and have a size corresponding to the diffraction-limit of the telescope, as well as a faint intensity ($\Delta m > 7$), they could be easily mistaken for moonlet satellites. Additionally the presence of a continuous halo around the primary limits the signal-to-noise ratio of the detected moonlet and thus the accuracy of its position and its photometry.

We applied the algorithm described by Marchis et al (2006b) to all observations of the four binary systems. Tables 3a -3d summarize the characteristics of their synthesized 2-$\sigma$ detection profiles. As previously shown in Marchis *et al.* (2006b), the three parameters ($\alpha$, $\Delta m_{lim}$, $r_{lim}$) that characterize the synthesized detection profile are quite variable. They depend on parameters such as the seeing conditions, the airmass, the brightness of the object, and the total integration time during the observations, but also the telescope and the design of its AO system. In the case of 45 Eugenia, the average $\Delta m_{lim}$ is -7.6 with a 1$\sigma$ variation of 0.8 for $r_{lim}= 0.9 \pm 0.2$. This variation in the detection profile can be directly translated into the minimum diameter size (12±8 km or 7±3 km) for a moonlet to be detected if located at $2/100 \times R_{Hill}$ or $1/4 \times R_{Hill}$, respectively.

In Fig. 1a-1d we detail each step of the detection curve profile analysis for one observation of each asteroid. Subtracting the azimuthally averaged function improves the detection of the moonlet. The characteristics of the synthesized detection profile are also displayed. The two regimes separated by $r_{lim}$ are obvious on these detection profiles. We detect the companion unambiguously in 23 out of 25 observations of 22 Kalliope, 32 out of 49 for 45 Eugenia, 23 out of 42 for 107 Camilla, and 10 out of 13 for 762 Pulcova. The almost 100% detection rate for Kalliope companion is mostly due to the brightness of the primary (mv=11-12) and the low contrast between the primary and the secondary. We could not see the companion under extremely poor seeing and weather conditions. As the lowest detection rate for 45 Eugenia, it is chiefly due to poor seeing conditions combined with an edge-on appearance of the orbit. The moonlet was sometime located too close to the primary and its flux was lost in the halo due to the uncorrected residual phase of the AO.

Our accurate analysis of each frame in the database revealed the presence of a second and closer moonlet around 45 Eugenia (Marchis *et al.* 2007a). This second moonlet is very faint (1.6-2.3 x σ above the background, i.e. 6 km in diameter) and close to the primary (0.4", corresponding to a projected distance of ~600 km), hence at the limit of detection of our analysis. We were able to detect the second satellite in three observations (detection rate of 7%). Further observations are necessary to better characterize its orbital elements and the system will be described in Marchis et al. (in preparation, 2008). Including 87 Sylvia (Marchis *et al.* 2005a), two triple asteroid systems are now known in the main-belt.

**Material:**

**Table 3a and 3b**

**Fig 1a,1b,1c,1d**

## 2.4 Size and Shape of Primary from direct AO images

### 22 Kalliope

The average angular size of 22 Kalliope's primary measured on 14 AO data is 116 mas (Table 4a) varying from 90 to 150 mas. To improve the sharpness of these 14 resolved images of Kalliope's primary, we applied AIDA deconvolution algorithm, which is thoroughly described in Hom *et al.* (2007) and extensively tested in Marchis *et al.* (2006b) on asteroid-type images. Figure 2a shows the resulting deconvolved images. The various sizes and shapes of the observed primary were approximated by fitting the reduced image with an ellipse, of which major-axes and orientation are listed in Table 4a. With this technique, using the error function described in Marchis *et al.* (2006b), we estimate our size measurement accurate to 5% on average, corresponding to 10 km. This error remains low only if the asteroid projected shape is close to an ellipse.

We compared the apparent shape of Kalliope's primary with the model developed by lightcurve inversion (Kaasalainen *et al.* 2002), as well as a refined version developed specifically for a new pole solution, ($\lambda = 197\pm2º$, $\beta=-3\pm2º$ in ECJ2000) described in Descamps *et al.* (2007c). We adjusted the timing of the 3D model using a lightcurve observation recorded by our group in November 2006. This solution reproduces quite well the general orientation of the observations. In Fig. 2a, we display the calculated appearance of Kalliope at the date of the observations considering the shading of the shape model and the illumination geometry. A quantitative analysis indicates that the image orientations are shifted by 10º on average. The global shape which was measured by comparing a/b between the model and the observations is significantly offseted. Two observations taken on June 5, 2004 at 13:23 UT and July 26, 2004 at 4:51 UT have been recorded while the asteroid had almost the same geometry (SEP[2] longitude and latitude of 159º and -38º). The a/b ratios measured on the observed images are consistant (a/b= 1.24 & 1.12) but significantly different to the expected one (a/b~1.60) suggesting that Kalliope's primary is less elongated than the shape model.

The average diameter estimated from our AO observations is 196±20 km, which is 9% larger than IRAS radiometric diameter from Tedesco *et al.* (2002). The tendency of IRAS radiometric measurements to underestimate the size of large and elongated asteroids has already been noted for various main-belt asteroids (Marchis *et al.* 2008). However, in the case of 22 Kalliope, the shape of the primary is quite irregular. We noticed that because of its pear shape (see March 4, 2004 or July 26, 2004 images in Fig. 2a), the asteroid is often well resolved in one direction and overestimated in the smaller one, leading to a significant error in the size measurement by the technique of ellipse fitting that we used. Finally between 2003-2004, the asteroid was almost always pole-on showing is larger apparent size. In the rest of this manuscript, we will consider the IRAS radiometric measurements to discuss the bulk density of 22 Kalliope (Section 4).

**Material:**

**Table 4a,4b**

**Figure 2a & 2b**

### 45 Eugenia

45 Eugenia's primary is resolved in 35 observations collected with Keck, VLT and

---

[2] Sub-Earth Point

Gemini AO. Its shape and size were measured using the same technique detailed in Section 2.4 for 22 Kalliope. Table 4b detailed the orientation and size ratio after fitting by an ellipsoid. The angular size of this asteroid is ~140 mas (more than twice the angular resolution of the AO system) so the error on the measurement should be lower than for Kalliope (estimated to 3%). The average diameter extracted from our observations is 193 ± 15 km, with an average a/b = 1.2. This measurement is 9% smaller than the IRAS diameter (215 km) reported in Tedesco *et al.* (2002).

Figure 2b displays a close-up images of the primary obtained after deconvolution and a comparison with the shape predicted by Kaasalainen et al (2002) based on a reconstructed shape model from lightcurve with a pole solution ($\lambda = 124°$, $\beta=-30°$ in ECJ2000) and considering the illumination geometry. The projected model shape of Eugenia is in rough agreement with the observed ones. This model is based on lightcurve published before 1994, we adjusted the timing using a more recent lightcurve published on the CdR+CdL website[3] recorded in 2005. We notice, however, a remaining shift in the orientation by ~20°. By combining AO images with lightcurve observations including newer observations, we should be able to improve the shape model of this asteroid (Kaasalainen *et al.* 2007).

**107 Camilla**

Camilla's primary is resolved on 26 frames from data taken mostly with VLT-UT4 in 2004, and more recently in 2006 with Gemini North. The angular size of the primary along the major axis is twice the angular resolution of the telescope. The error on the size estimate, therefore, is significantly low (~5%). The average diameter is estimated to 246 ± 13 km, which is 10% larger than the IRAS STM measurement reported in Tedesco *et al.* (2002). We show in Fig. 2c a close-up of the primary and a comparison with Torppa *et al.* (2003) shape model ($\lambda$ = 72º, $\beta$=51º ECJ2000). Camilla is located in the outer part of the main belt with a low inclination (~10º), therefore the sub-Earth point latitude of the primary does not vary significantly (less than 27º) over the 2004 and 2006 period of observations. Table 4c contains the fitting parameter for this asteroid and a comparison with the model. Our AO images reveals the elongated shape of the primary with a maximum a/b up to 1.6-1.9 on September 14 2004 with an inclination of ~15º. The projected shape model displays in Fig. 2c is in agreement with the observations. The average shift in the orientation of a fitted ellipse on both images per run is only 8º.

---

[3] CdR+CdL by R. Behrend et al., http://obswww.unige.ch/~behrend/page1cou.html#000045

**(762) Pulcova**

Pulcova's primary is marginally resolved on only four images collected with Gemini AO only. This AO system was commissioned in 2004 a few months after the data were recorded. It provided an unstable correction on 13$^{th}$ magnitude target like Pulcova (FWHM=100 mas). The average angular size measured on the data is 140 mas and the average diameter is only accurate to ~20%, (216 ± 40 km). We did not consider this size measurement due to its poor quality in the rest of this work. Instead, we will use IRAS radiometric diameter of 137 km diameter (Tedesco *et al.* 2002). There is no shape model derived from lightcurve for this asteroid.

> **Material:**
>
> Table 4c,4d
>
> Figure 2c

**2.5 Astrometric positions and photometric measurements on the satellite**

We described in Marchis et al (2005b) and Marchis *et al.* (2008) how we measure the position of the satellite with respect to its primary. Details about the method can be found in these articles. Our algorithm is based on a Moffat-Gauss profile in two dimensions which is adjusted to fit the position of the satellite and the primary. An estimate of the background due to residual of the wavefront correction is also added.

The astrometric positions relative to the primary in arcsec are labeled X and Y in Table 5a-5d. They correspond to the projected separation on the celestial sphere between the primary and the satellite: $X = \delta RA \times \cos(DEC)$ and $Y = \delta DEC$ with X positive when the satellite is located on the astronomical East of the primary and Y positive when it is locate North.

From the Moffat-Gauss profile we also estimate the relative integrated flux between the moonlet and the primary. Since the primary is resolved in most of the images, we derived the size of the secondary ($D_{sat}$) using the relation

$$D_{sat} = D_{av} \times (\Phi_{sat}/\Phi_{primary})^{1/2} \quad (1)$$

with $\Phi_{primary} = \int F_{primary}$, or the integrated flux of the Moffat-Gauss profile on the primary and $\Phi_{sat} = \int F_{sat}$, or the integrated flux on the secondary, and $D_{av}$ the average size measured on the primary. We assumed the same albedo for the satellite and the primary.

The size of each moonlet is given in Table 1. The uncertainties in the size measurements of the moonlets are large (up to 50% in the case of Kaliope and Pulcova) which is due to the difficulty in extracting the weak flux of the moonlet orbiting close to the primary asteroid.

**Material:**

**Table 5a,b,c,d:**

**Figure 3**

**3. Orbit determination**

**3.1 Method**

Descamps (2005) developed and published a dedicated algorithm, named BOF for Binary Orbit Fit, to estimate the true orbit of a binary asteroid. The application of this algorithm is described in Marchis *et al.* (2008). It is a Keplerian algorithm including precession effect due to the oblateness of the gravitational field that is introduced by the ellipsoidal shape of the primary. To deliver a unique orbit solution, a sufficient number of relative positions as regards to the required eight orbital parameters should be collected. Additionnaly, these astrometric positions must be well distributed along the orbit to carefully determine the degree of eccentricity. In the case of misalignment of the orbital and primary's equatorial planes, a long-term follow-up can help to assess the level of precession through the $J_2$ parameter[4] of the central body. Figures 3a-3d display the measured and predicted satellite positions of the four studied binary systems. Their orbital parameters are summarized in Table 6. BOF has already been used to estimate the orbits of various other binary asteroids (Marchis *et al.* 2005a, 2005b; Marchis *et al.* 2006a; Marchis *et al.*, 2008). We confirmed the orbital elements derived for all these mutual orbits using independently StatOrbit another algorithm described in Hestroffer *et al.* (2005).

---

[4] $J_2$ or ($C_{20}$) is the second parameter of the decomposition of the gravitational field of the primary which is directly related to the precession of an almost equatorial orbit

**3.2 Discussion on the moonlet orbits**

**22 Kalliope**

The orbit of Kalliope's moonlet is the very well constrained in this study. Table 5a and Figure 3a shows that the positions of the moonlet are well distributed around the primary and in time. In this orbital analysis we also included the astrometric positions from the previous observations published by Marchis *et al.* (2003) and Margot and Brown (2003) recorded in 2001-2002 with Lick-3m and Palomar-5m telescopes. Our model fits the observations with a average root mean square (rms) error of ~30 mas. Linus, a 20-km diameter companion orbits around Kalliope (~180 km) well inside its Hill sphere (a ~ 1100 km = 1/40 × $R_{Hill}$ = 6 × $R_p$) in ~3.6 days describing an almost circular orbit (e<0.006). The average root mean square residual remains significant (35 mas) as shown in Fig. 3a. We did not detect a precession of the orbit suggesting that the pole orientation of Linus and the spin pole orientation of the primary are the same. Based on the shape of Kalliope developed by Kaasalainen *et al.* (2002) assuming an homogeneous distribution of material in its interior, the nodal precession effect introduced by Kalliope oblateness should be ~0.15° per day (on the basis of a theoretical $J_2$=0.19 and a zero inclination over the equatorial plane of the primary) so a small inclination of the moonlet's orbit, larger than a few degrees, would have been detected over 6 years of observations. Our solution, which mirrors the one adopted by Marchis *et al.* (2003), is very close to the orbital solution proposed by Margot and Brown (2003).

In November 2006, a group of astronomers reported the observations of a secondary stellar occultation event by Linus and Kalliope's primary. The event was detected 70 km away from the position predicted by our model (Soma *et al.* 2006), providing independent validation of our orbit solution. The duration of the event was 3.6 s corresponding to a moonlet with a minimum size of 22 km in agreement with our AO measurements.

Based on these orbital parameters, Descamps *et al.* (2007c) predicted a series of mutual events which were observed successfully by a large team of observers (Descamps *et al.* 2007d). It provided detailed and numerous insights about this well-known binary system. The knowledge of the orbital parameters of Linus allow us to conduct complex and challenging observation programs, which in return will help understanding how this system formed. We will expand this kind of study to other binary asteroidal systems.

**45 Eugenia**

Eugenia system received a fair amount of attention recently when Marchis *et al.* (2007a) announced the discovery of the second, smaller and closer, moonlet in this multiple asteroidal system. In this work, we emphasize the orbit of Petit-Prince only, the outermost and largest satellite discovered in 1998. We processed and included in the orbit analysis positions of Petit-Prince recorded at the time of its discovery in 1998 using the CFHT and its PUEO AO system. These data from the programs with ID numbers 98IIH07a (PI: F. Roddier) and 98IIH07b (PI: L. Close) were found and downloaded from the CFHT Science Archive[5]. As shown in Figure 3b, the astrometric positions span the entire orbit. The 7-km satellite orbits at 1170 km (3/100 × $R_{Hill}$ or 5.5 × $R_p$) away from the primary in 4.766 days describing a circular orbit (e<0.002). The satellites of Eugenia and Kalliope have orbits fundamentally different since Eugenia Petit-Prince I has an significantly inclined orbit (12±1°) therefore the appearance of the orbit changed significantly over the 9 years of observation due to precession effects. From our orbit model we estimate the dynamical flattening $J_2$=0.15, very close to the theoretical $J_2$=0.19 calculated for Eugenia's primary based on Kaasalainen *et al.* (2002) 3D shape model, assuming an uniform distribution of material in its interior.

Our orbital model is accurate enough (average rms error ~48 mas) to allow predictions of the positions of Petit-Prince. Using our orbital analysis, Berthier *et al.* (2007) announced the predicted positions of the satellite at the time of three occultations predicted in 2007 & 2008. Successful future stellar occultation observations by Petit-Prince are very important to refine our orbital model and estimate its size and shape.

**107 Camilla**

The orbit of Camilla's moonlet is well defined thanks to data recorded over 3 years (Table 5c). The 16-km satellite orbits in a prograde manner at 1250 km from the primary in 3.722 days. No eccentricity of the orbit has been detected (e<0.002). The moonlet orbits very close to the equatorial plane of the primary, so no precession of the orbit was detected. Our orbital model fits the data with an average rms error of 22 mas. Figure 3c

---

[5] CADC: URL http://www2.cadc-ccda.hia-iha.nrc-cnrc.gc.ca/cfht/overview.html

shows that its orbit is roughly seen in its edge-on configuration from Earth.

**762 Pulcova**

The orbit of Pulcova's moonlet is difficult to constrain despite 3 years of observation with Gemini and VLT (Table 5d). The astrometric positions (see Fig. 3d) are not well distributed along the orbit. We could however find a reliable solution with an average rms error ~15 mas. The error on the semi-major axis measurement is 2%. More observations will help to refine this orbit improving the accuracy on the orbital elements and possibly detect a precession of the orbital plane. We can say with confidence that the ~20-km diameter satellite of Pulcova revolves around the primary at ~700 km in 4.44 days describing a roughly circular orbit (e=0.03±0.01).

Our fitted orbital elements for the orbits of satellites are shown in Table 6. Using Kepler's third law (Kepler, 1609), it is possible to compute the mass of the system (Table 7). The 1-σ error on the mass (~2-7%) is dominated by the accuracy on the semi-major axis measurement (0.5-2%).

**Material**

**Table 6 & 7**

All four binary systems are quite similar. The period of revolution is between 3.6 and 4.4 days with a semi-major axis ~0.3-2% × $R_{Hill}$ or ~10 × $R_p$. The Hill sphere size is calculated using the mass of the system derived from the orbit analysis. Previous orbits of binary main-belt systems, such as 87 Sylvia in Marchis *et al.* 2005a; 121 Hermione in Marchis *et al.* (2005b)) display the same characteristics: a km size satellite orbiting well inside the Hill sphere of the primary and describing a circular orbit. Marchis *et al.* (2008) shows that four other binary asteroidal systems (130 Elektra, 283 Emma, 379 Huenna, 3749 Balam) located in the main belt are quite different with a significant detected eccentricity ($e$>0.1). Their formation must be different to those studied here.

**4. Internal structure: bulk density and porosity**

The determination of the mass and size of a binary asteroid system allows us to directly estimate its bulk density. Very few measurements on bulk densities on small solar system bodies are yet available in the literature. The density of a few NEAs are available thanks to space mission visits. Two S-type asteroids (25143) Itokawa and (433) Eros have a bulk density estimated to 1.9 and 2.7 g/cm$^3$ respectively (Fijiwara et al., 2006; Wilkinson et al., 2002). A binary NEA named (66391) 1999 KW4 and studied by radar observations combined with lightcurves (Ostro et al., 2006) seems also to have a bulk density of ~2 g/cm$^3$. Recent studies of binary main-belt revealed a low bulk-density for these asteroids. We can cite for instance the bulk density of (90) Antiope, a C-type asteroid, estimated to 1.25+/-0.05 g/cm3 combining adaptive optics and lightcurve observations (Descamps et al., 2007b). Study of the binary Jupiter Trojan (617) Patroclus-Menoetius by adaptive optics combined with lightcurve observations during mutual events (Marchis et al., 2006a; Berthier et al. 2007b) showed that the bulk density of this asteroid is less than water (0.9 +/-0.2 g/cm2). Lightcurve observations of possible binary Trojan asteroids confirmed the low density (0.6 & 0.8 g/cm3 for 17635 and 29314) of this population. Bulk density measurements of TNOs show a large diversity varying from 0.6 g/cm3 for 2001QG298 (D ~240 km) (Lacerda et al., 2007) up to 2.6 g/cm3 for 2003 EL61, a large D~1000 km TNO (Rabinowitz et al., 2006). The density of comet nuclei is not well constrained since their measurements involved non-gravitational forces. A value of 0.4 g/cm3 was derived after the Deep impact encounter with the comet 9/P Tempel 1 (Richardson et al., 2007) based on the ballistic analysis of the ejecta plume after the impact.

Using the mass from the analysis of the orbit ($M_{system}$) together with the average radius estimated from radiometric IRAS measurements, we can derive the bulk density of these binary systems. The SIMPS catalog (Tedesco *et al.* 2002) contains the diameter (called $D_{STM}$) of ~2200 small solar system bodies detected by IRAS extracted using the Standard Thermal Model (STM) designed for large asteroids with low thermal inertia and/or slow rotation. Using a modified approach with a model, called NEATM (Harris, 1998), which was developed specifically for near-Earth asteroids, which can spin rapidly and have significant thermal inertia, we reanalyzed the IRAS measurements and derived a new average diameter ($D_{NEATM}$) shown in Table 1. NEATM is a more complex model which to first order includes the effects of thermal inertia, surface roughness, and spin vector via the parameter $\eta$. In contrast, in the STM the value of $\eta$ is kept constant (0.756) (see Harris, 2006 and references therein). In Table 1 we see that the NEATM and the STM methods give average radii with significant differences between one another, leading to differences in bulk density of up to 30% (e.g. 107 Camilla). The larger size of 107 Camilla's primary is also measured on AO observations. Table 7 summarizes the bulk densities derived from the masses and the average diameters for both the NEATM and the AO size estimate (only for 45 Eugenia and 107 Camilla).

**4.1 Bulk Density of 22 Kalliope, a W-type asteroid?**

The bulk density derived using IRAS radiometric diameter for this asteroid is 2.7±0.2 g/cm$^3$. This density is almost three times larger than the ones reported for C-type asteroids (45 Eugenia, Merline *et al.* 1999 or 121 Hermione Marchis *et al.* 2005b). 22 Kalliope is a large asteroid classified by Tholen and Barucci (1989) as an M-type. Because many of these M-type asteroids (Lupishko and Belskaya, 1989) show the same polarimetric properties as iron meteorites, it was thought they could be the best meteorite analogs. On the basis of this assumption and an approximated mutual orbit, Marchis *et al.* (2003) shows that assuming the grain density of iron meteorite (~7.4 g.cm-3), the porosity of this asteroid should be 60-70%.

In a more recent work by Rivkin *et al.* (2000), a 3-micron absorption feature interpreted as the presence of hydrated material on the surface was detected by spectrophotometry. 22 Kalliope, and various other large M-type asteroids, was placed in a new W taxonomic class. No other W-type asteroid listed in their work is a known binary for which we have a mass estimate, therefore this could be considered the first measurement of a W-type bulk density.

**4.2 bulk density of 45 Eugenia, a typical C-type asteroid**

Bus and Binzel (2002) classified 45 Eugenia as a C-type asteroid. On the basis of IRAS radiometric diameters, we derived a low bulk density ~1.1 g/cm$^3$. Several binary asteroids belonging to this C class or any of its sub-class (X, Cb, G) are known and have well constrained orbit, thus bulk density measurement available. In Table 8, we can see how consistently low (0.9-1.4 g/cm$^3$) are their bulk density measurements. As described in Marchis *et al.* (2005a,), considering CI or CM carbonaceous chondrites as meteorites analogues (with 9% micro-porosity and an average bulk density ~2.1 g/cm$^3$, see Britt and Consolmagno, 2003), the macroporosity of Eugenia could be ~40%. This value is significantly higher than the porosity of meteorites (Britt and Consolmagno, 2004), and S-type asteroids (433 Eros, see Wilkinson et al. 2002). It is, however, similar to the porosity inferred for other asteroids studied by space missions: 25143 Itokawa (Fujiwara et al. 2006) and 253 Mathilde (Yeomans et al. 1997). Like these asteroids, 45 Eugenia's interior could be made of loosely packed fragments or, most likely, evolved into a "rubble-pile" of material. If we consider that this low bulk density is the result of a mix of carbonaceous material and water ice, we obtain an unrealistic large portion of water ice (90%) for an asteroid.

**4.3 slightly higher bulk density of 107 Camilla, an X-type**

107 Camilla, X-type asteroid, has a bulk density estimated to 1.4±0.3 g/cm$^3$. This measurement departs by 3-$\sigma$ from the averaged C and X-type bulk density measurements reported in Table 8. The orbit of 107 Camilla is well constrain and leads to 3% error on the mass estimate (less than 1% on the bulk density error). The radiometric diameters estimated using 9 individual IRAS insights are quite different between results from the STM and the NEATM models. With the large NEATM diameter (249 km), the bulk density is reduced by 30%. The NEATM analysis indicates that the measurement has an error of 18 km. Consequently the bulk density could be as low as 1.1 g/cm$^3$, so very close to other C- and X-types bulk densities. Without knowing exactly the meteorite analog of an X-type asteroid, we must consider a large sample of carbonaceous chondrites bulk measurements from 2.2 g/cm$^3$ for CI to 3.5 g/cm$^3$ for CV/CO (Britt and Consolmagno, 2003) to derive a large range of macroporosity (30-55%). This binary asteroid has a significant macroporosity which is in agreement with a rubble-pile interior (see Section 4.3).

**4.4 Bulk density of a Cb-type asteroid: 762 Pulcova**

762 Pulcova is part of the large C-group as well. Based on recent spectroscopic study by Lazzaro *et al.* (2004) it is known to be a Cb-class asteroid. Its low density (0.9±0.1 g/cm$^3$) is similar with other C-group binary asteroid bulk density (Table 8). Assuming a composition similar to CI or CO carbonaceous chondrite (density 2.11, 2.95 g/cm$^3$ and average microporosity 11, 20% respectively; see Britt and Consolmagno, 2004), Pulcova could be made of a very porous material with a macroporosity up to ~50%, suggesting a rubble-pile interior (see discussion in Section 4.3).

**Material**: include here Table 8

**5. Tidal effect dissipation**

Dissipation by tides between the satellite and the primary of a binary asteroidal system can force the orbital elements of the satellite to evolve. Weidenschilling *et al.* (1989) (and references therein) described how the theory of tides could apply to binary asteroids. This precursor work was published before the existence of binary asteroid was proven. Marchis *et al.* (2008) described and re-discussed this work on the basis of known binary asteroid systems. We will refer to this work and the equations therein to discuss the analysis of the tidal effect dissipation specifically for these 4 binary main-belt asteroid systems (22 Kalliope, 45 Eugenia, 107 Camilla, (672) Pulcova). We will include also other binary asteroid system with circular mutual orbits (121 Hermione (Marchis *et al.* 2005b) and 87 Sylvia (Marchis *et al.* 2005a)). 130 Elektra, 283 Emma, 379 Huenna, and 3749 Balam, 4 main-belt binary systems with eccentric mutual orbits studied by Marchis *et al.* (2008), and 90 Antiope, a double binary asteroid system (Descamps *et al.* 2007b) are also included for sake of completeness.

**Material**: include here Figure 4 & Table 9

## 5.1 Orbital stability

The condition of stability for which the mutual orbit of a binary asteroid remains stable despite the effect of tides is defined by Weidenschilling *et al.* (1989):

$$\left(\frac{a}{R_p}\right)^2 < \frac{6}{5}\frac{(\lambda_p + q)(\lambda_s + q^{5/3})}{q} \quad (2)$$

where $q = M_s/M_p$ and $R_p$ is the radius of the primary and $\lambda_p$ and $\lambda_s$ are the non-sphericity parameters[6]. Assuming the same bulk density for the primary and satellite, we can approximate $q$ using the radius estimate ($q \sim (R_s/R_p)^3$ where $R_s$ is the radius of the satellite. The first column in Table 9 indicates if the condition of stability is reached for these binary systems. None of the listed binary systems, besides 90 Antiope, 379 Huenna, and 3749 Balam, satisfy the stability condition; we can therefore expect the tides to drive the orbital parameters of their mutual orbits. Figure 4 shows the domains of relative separation a/R against mass ratio $q$. All, but 90 Antiope, fall well short of synchronous stability, indicating that the orbits of their moonlets will evolve due to tidal dissipation.

## 5.2 Time scale for semi-major axes

---

[6] $\lambda = I/(2/5 \times m \times R^2)$ where I is the body's moment of inertia. $\lambda = 1$ for a homogeneous sphere is assumed here.

For a satellite that was formed outside the synchronous orbit ($a_{syn}$), the tides raised by the satellite on the primary will increase the semi-major axis ($a$) and decrease the spin rate of the primary ($\Omega$). Weidenschilling *et al.* (1989) estimated the tidal evolution timescale $\tau$ from initial, $a_i$, to final semi-major axis, $a_f$, as

$$\left(\frac{a_f}{R_p}\right)^{13/2} - \left(\frac{a_i}{R_p}\right)^{13/2} = K\tau \frac{\rho^{5/2} q\sqrt{1+q} R_p^2}{\mu Q} \qquad (3)$$

where $K = 10\pi^{3/2} G^{3/2}$, $\rho$ is the bulk density, and $\mu Q$ is the tidal parameter, consisting of the product of rigidity ($\mu$) by the specific dissipation parameter ($Q$). $\mu \sim 10^8$ N m$^{-2}$ is a typical value for a moderately fractured asteroid and $\mu > 10^{10}$ N m$^{-2}$ for consolidated rocky material. $Q \sim 100$ was measured for Phobos by Yoder (1982). $\mu Q$ can be in a large range between $10^{10}$ to $10^{12}$ N m$^{-2}$. Using the approximation described in Marchis *et al.* (2008), we derive an approximate age for all binary asteroid systems (Table 9). Considering the age of our solar system as an upper limit (4.5 Byrs), it is obvious that a tidal parameter $\mu Q$ of $10^{10}$ N m$^{-2}$ is a realistic value for these binary systems (see Fig. 4). If we could constrain independently the age of these asteroids by studying their collisional family (Zappala *et al.* 1995) and the aging by space weathering, we will be able to determine the tidal parameter $\mu Q$ for a rubble-pile asteroid. However, none of these binary asteroids, but 283 Emma, are listed as member of a collisional family.

**5.3 Evolution of eccentricity**

Tidal evolution also modifies the satellite's eccentricity; the tidal forces on the satellite vary along the orbit and will tend to circularize the orbit, whereas the tide on the planet will increase the eccentricity. From Harris and Ward (1982), assuming that the physical properties (such as density, rigidity and $Q$) of the primary and secondary are similar, we

derive:

$$\frac{\dot{e}}{e} = \left(\frac{19}{8}\text{sgn}(2\Omega - 3n) - \frac{7Rs}{2Rp}\right)\frac{\dot{a}}{a} \quad (4)$$

where *sgn* is the *sign* function. $\Omega$, the spin rate of the primary, is derived from the rotation period ($P=2\pi\Omega$) that is measured accurately by lightcurve observations. Harris and Warner (Minor Planet Lightcurve Parameters[7]) report consistent measurements for 10 of these binary systems (with exception of 3749 Balam).

In Table 9, we calculate the coefficient $\alpha_{evol}$ (*de/e* = $\alpha_{evol}$ × *da/a*) from Eq. 4. *da/a* >0 since the moonlet is moving away from the primary (see Section 5.2). All the calculated $\alpha_{evol}$ suggests that the eccentricity of the mutual orbit should increase due to the tide. This result is in contradiction with the low eccentricity reported in this work for four binary asteroidal systems.

We displayed in Fig. 4 the eccentricity excitation limit assuming $Q_s=Q_p$ and $k_s=k_p$ (Eq. 8 in Weidenschilling *et al.* 1989). As mentioned by Marchis et al (2007), two systems (130 Elektra and 283 Emma) located beneath the limit of *e* excitation do have a significant eccentricity. 22 Kalliope, 87 Sylvia, and 762 Pulcova are located in the area for which the eccentricity is damped by tidal effects, thus their circular orbits reported in this work and in Marchis *et al.* (2005a). However, the locations of binary systems like 45 Eugenia, 107 Camilla, and 121 Hermione suggest that their orbit should be significantly eccentric in contradiction with the orbit estimate presented in this work and by Marchis *et al.* (2005b). This obvious discrepancy between the tidal evolution proposed by Weidenschilling *et al.* (1989) and our orbit determination suggests that the assumptions taken in this simplified formalism could be erroneous. For instance, we assumed that the primary and the satellite have the same love number $k$, however, in reality, if they have the same composition, $k$ is almost proportional to $R^2$. We also supposed that the specific dissipation parameter ($Q$) is similar between the primary and its satellite. For most of these binary systems the satellite is 1/10-1/30 times smaller that the primary (5-20 km in radius). Since the satellites could be more monolithic, the primary dissipation parameter could be different. Harris and Ward (1982) mentioned that the tides from giant planets (with $Q>1000$) compared with their satellites ($Q=100-500$) circularize their mutual orbits. Using the relation between the size and love number, and varying $Q_s$ significantly, it may be possible to retrieve a tidal parameter $\mu Q$ which agrees with the eccentricity found for each binary system.

---

[7] http://cfa-www.harvard.edu/iau/lists/LightcurveDat.html

**Conclusion**

On the basis of observations taken mostly with 8-10m class telescopes (VLT-UT4, Keck II, and Gemini North) equipped with adaptive optic systems, we have derived the mutual orbits of four binary asteroids located in the main belt. Our monitoring of the satellites over more than 4 years allows us to determine their orbits with high precision. The common characteristic of these binary systems is that each moonlet describes an almost circular orbit. Using the best-fitting orbital parameters and the IRAS radiometric measurements, we have estimated their masses and bulk densities:

- The 22 Kalliope binary system is well characterized, thanks to observations taken since 2001. Kalliope's Linus I satellite, 26 km in diameter, orbits well inside the Hill sphere of the primary (1/40 × $R_{Hill}$). The system does not show precessions, suggesting that Linus orbits very close to the equatorial plane of the primary. The mutual orbit eccentricity has presumably been damped by tidal effects. The bulk density is estimated to be $2.8 \pm 0.2$ g/cm$^3$, based on $D_p$=177 km from the IRAS radiometric measurements. This is the first bulk density estimate of a W-type asteroid for which we do not know the meteorite analog.

- 45 Eugenia was the first binary asteroid discovered from the ground in 1998. We were able to construct an orbital model of its companion which fits all data taken between 1998 and 2006. Eugenia Petit-Prince I is a 7-km satellite, orbiting at $3/100 \times R_{Hill}$ from the 217-km primary. Since Petit-Prince's orbit is inclined by ~12°, the model includes a precession effect to fit the data assuming $J_2$=0.15 (very close to the theoretical $J_2$=0.19, with the assumption of a homogeneous primary). The low bulk density derived for this C-type asteroid ($1.1\pm0.1$ g/cm$^3$) is in agreement with other C-type bulk densities already published, and suggests a rubble-pile internal structure. We cannot explain why the orbit of Petit-Prince is circular based on our tidal evolution model. Marchis *et al.* (2007a) announced the discovery of a second moonlet (temporary named S/2004(45)1) around this asteroid.

- The 16-km satellite (S/2002(107)1), orbiting around 107 Camilla (a 250-km X-type asteroid) at $1/40 \times R_{Hill}$ away from the primary, leads once again to a low bulk density estimate ($1.4 \pm 0.3$ g/cm$^3$), suggesting a rubble-pile interior. Since no orbital precession is needed in our model to fit three years of data, we can assume that the moonlet revolves very close to the equatorial plane of the primary. It is surprising that the mutual orbit is circular since tidal evolution should have excited it significantly.

- S/2001(762)1, the 20-km diameter companion of 762 Pulcova (143 km Cb-type) orbits at $1/40 \times R_{Hill}$ and describes a circular orbit in 4.438 days. The bulk density of Pulcova is at lower end of other C-group asteroids (0.9±0.1 g/cm$^3$). The companion orbits almost in the equatorial plane of the primary. Tidal effects between the system components have presumably circularized the mutual orbit.

The low bulk densities (~1.1 g/cm$^3$) inferred for the asteroids 45 Eugenia, 107 Camilla, and 762 Pulcova which belong to the C-"group" indicate that these asteroids have a significant macro-porosity (30-50%). Spacecraft measurements of 253 Mathilde (Yeomans et al. 1997) and more recently of 25143 Itokawa (Fujiwara et al. 2006) also suggest that these asteroids have an important macro-porosity. By the analysis of surface features and/or impact crater size and quantity, these authors concluded that these two asteroids have a rubble-pile interior. In the case of these binary asteroids, we do not have access to close-up images of the surface, however the same macro-porosity measurements imply that they could be as well rubble-pile bodies.

Knowledge of the orbital parameters of these binary systems is crucial for a complete understanding of how they formed and evolved. For instance, Descamps and Marchis (2007) noticed that this type of binary asteroid (large primary $R_p$~100 km with a much smaller moonlet of $R_s$<20 km) has a total angular momentum of ~0.27, which is very close to the Maclaurin-Jacobi bifurcation point and the stability limit defined by Holsapple (2001), implying a small friction angle (10-14°). These companions could be the outcome of mass shedding of a parent body presumably subjected to an external torque.

The ability to predict the position of a moonlet around its primary can also help to provide more information about the system itself. A recent stellar occultation (Soma *et al.* 2006) and mutual event observing campaign (Descamps *et al.* 2007d) involving the binary system 22 Kalliope have brought unprecedented information about the system that is not attainable with current AO systems. Since instrumentation sensitivity and angular resolution available from the ground or in space are improving continuously, it should soon become possible to consider more complex observation programs on these binary systems, such as comparative spectroscopy, or a search for fainter and closer moonlets around them. The number of binary asteroidal systems will continue to increase over the coming years with the advent of larger telescopes (TMT-30m, ESO-ELT) and large-scale photometric survey (Pan-Starrs).

**Acknowledgements**

This work was supported by the National Science Foundation Science and Technology Center for Adaptive Optics, and managed by the University of California at Santa Cruz under cooperative agreement No. AST-9876783. Part of these data was obtained at the W.M. Keck observatory, which is operated as a scientific partnership between the California Institute of Technology, the University of California and the National Aeronautics and Space Administration. The observatory and its AO system were made possible by the generous financial support of the W. M. Keck Foundation. Other observations were obtained at the Gemini Observatory and the Gemini Science Archive, which is operated by the Association of Universities for Research in Astronomy, Inc., under a cooperative agreement with the NSF on behalf of the Gemini partnership. We thank two anonymous reviewers for numerous constructive suggestions on how to improve this manuscript.**References**

Babcock, H.W., 1953. The possibility of compensating seeing, PASP 65, 229-239

Belton, M.J.S. *et al.* 1996. The Discovery and orbit of 1993(243)1 Dactyl, Icarus 120, 185-199.

Behrend, R. , Bernasconi, L., Roy, R., and 46 collaborators, 2006. Four new binary minor planets: (854) Frostia, (1089) Tama, (1313) Berna, (4492) Debussy, A&A, 446, 1177

Berthier, J, Descamps, P., Marchis, F., Baek, M., *I. de pater, H. Hammel, M. Showalter.* 2007a. Occutations by (45) Eugenia, CBET 1073, Edited by Green, D. W. E

Berthier, J., Marchis, F., Descamps, P., Assafin, M., Bouley, S., Colas, F., Dubos, G., Emery, J.P., De Caat, P., Farell, J.A., Leroy, A., Pauwels, T., Pollock, J.T., Reddy, V., Sada, P.V., Vingerhoets, P., Vachier, F., Viera-Martins, R., Wong, M.H., Reichart, D.E., Ivarsen, K.M., Crain, J.A, La Cluye, A.P., Nysewander, M.C. 2007b. An Observing Campaign of the Mutual Events Within (617) Patroclus-Menoetius Binary Trojan System, American Astronomical Society, DPS #39, #35.05

Britt, D.T. & Consolmagno, G.J., 2003. Stony meteorite porosities and densities: A review of the data through 2001, MP&S, 38, 8, 1161-1180

Britt, D.T. & Consolmagno, G.J., 2004. Meteorite Porosities and Densities: A Review of Trends in the Data, 35th Lunar and Planetary Science Conference, March 15-19, 2004, League City, Texas, abstract no.2108

Bus, S. J. and Binzel, R. P.,2002. Phase II of the Small Main-Belt Asteroid Spectroscopic Survey: A Feature-Based Taxonomy", Icarus 158, 146-177

Descamps, P. 2005. Orbit of an Astrometric Binary System. Celestial Mechanics and Dynamical Astronomy, 92, 381-402

Descamps, P., Marchis, F., Michalowski, T., *et al.* 2007a. Nature of the small main belt asteroid 3169 Ostro, Icarus 189, 2, 363-369.

Descamps, P., Marchis, F., Michalowski, T., *et al.* 2007b. Figure of the double asteroid 90 Antiope from AO and lightcurves observations. Icarus, Icarus, 187, 2, 484-499.

Descamps, P., Marchis, F., Pollock, J., Berthier, J., Birlan, M., Vachier, F., Colas, F., 2007c. 2007 Mutual Events within the Binary System of (22) Kalliope, PSS, in press.

Descamps, P., Marchis, F., Pollock, J., *et al.* 2007d. New determination of the density of the binary asteroid 22 Kalliope from mutual event observations, Icarus, submitted.

Devillard, N., 1997. The eclipse software, The Messenger 87 (1997), pp. 19–20.

Fujiwara,A., J. Kawaguchi, D. K. Yeomans, M. Abe, T. Mukai, T. Okada, J. Saito, H. Yano, M. Yoshikawa, D. J. Scheeres, O. Barnouin-Jha, A. F. Cheng, H. Demura, R. W. Gaskell, N. Hirata, H. Ikeda, T. Kominato, H. Miyamoto, A. M. Nakamura, R. Nakamura, S. Sasaki, K. Uesugi, , 2006. The Rubble-Pile Asteroid Itokawa as Observed by Hayabusa, Science, 312, 5578, 1330-1334.

Harris, A.W & Ward, W.R. 1982. Dynamical constraints on the formation and evolution of planetary bodies, In: Annual review of earth and planetary sciences, 10., Palo Alto, CA, Annual Reviews, Inc., 1982, 61-108.

Harris, A.W. 1998. A Thermal Model for Near-Earth Asteroids, Icarus 131, 291.

Harris, A.W. 2006. The surface properties of small asteroids from thermal-infrared observations, Asteroids, Comets, Meteors, Proceedings of the 229th Symposium of the International Astronomical Union held in Búzios, Rio de Janeiro, Brasil August 7-12, 2005, Edited by Daniela, L.; Sylvio Ferraz, M.; Angel, F. Julio Cambridge: Cambridge University Press, 449-463

Herriot, G., Morris, S., Anthony, A., Derdall, D., Duncan, D. Dunn, J., Ebbers, A.W., Fletcher, J.M., Hardy, T., Leckie, B., Mirza, A., Morbey, C.L., Pfleger, M., Roberts, S., Shott,, P., Smith, M., Saddlemyer, L.K., Sebesta, J., Szeto, K., Wooff, R., Windels, W., Veran, J.-P, 2000. Progress on ALTAIR: the Gemini North adaptive optics system, Proc. SPIE vol. 4007, 115-125, Adaptive Optical Systems Technology, Peter L. Wizinowich; Ed.

Hestroffer, D., Vachier, F., Balat, B. 2005. Orbit Determination of Binary Asteroids, EM&P. 97, 3-4, 245-260.

Hodapp , K.W., Jensen , J.B., Irwin , E.M., Yamada , H., Chung , R., Fletcher , K., Robertson , L., Hora , J.L., Simons , D.A., Mays , W., Nolan , R., Bec , M., Merrill , M., Fowler , A.M. 2003. The Gemini Near-Infrared Imager (NIRI), PASP 115, 1388-1406

Hom, E.F.Y, Marchis, F., Lee, T.K., Haase, S., Agard, D.A., Sedat, J.W. 2007. AIDA: an adaptive image deconvolution algorithm with application to multi-frame and three-dimensional data, Journal of the Optical Society of America A, vol. 24, Issue 6, pp.1580-1600

Kaasalainen, M, Torppa, J., Piironen, J., 2002. Models of Twenty Asteroids from Photometric Data, Icarus, Volume 159, Issue 2, p. 369-395.

Kaasalainen, M., Marchis, F., Carry, B. 2007. Asteroid Maps From Photometry and Adaptive Optics, AAS/DPS meeting, 30.12, Orlando, Florida, USA

Kepler, J., 1609. Astronomia nova, Pragae.

Lacerda, P., & Jewitt, D. C. 2007, Densities of Solar system objects from their rotational light curves. Astron. J., 133, 1393

Lazzaro, D., Angeli, C.A., Carvano, J.M., Mothe-Diniz, T., Duffard, R., Florczak, M. 2004. $S^3OS^2$: the visible spectroscopic survey of 820 asteroids, Icarus, Volume 172, Issue 1, p. 179-220

Lenzen, R., Hartung, M., Brandner, W., Finger, G., Hubin, N.N., Lacombe, F., Lagrange, A.-M., Lehnert, M.D., Moorwood, A.F.M., Mouillet, D., 2003. instrument Design and Performance for Optical/Infrared Ground-based Telescopes. Edited by Iye M, Moorwood, A. F. M. Proceedings of the SPIE, Volume 4841, pp. 944-952

Marchis, F., P. Descamps, D. Hestroffer, J. Berthier, I. de Pater and D. Gavel 2002. Adaptive Optics observations of the binary system (22) Kalliope: A three dimensional orbit solution, ESA Special Publications series, Asteroid Comet and Meteor, Berlin, pp. 725-728

Lupishko, D.F. and Belskaya, I.N., 1989. On the surface composition of the M-type asteroids, Icarus, 78, 295-401

Marchis, F., Descamps, P., Hestroffer, D., Berthier, J., Vachier, F., Boccaletti, A., de Pater, I., Gavel, D., 2003, A three-dimensional solution for the orbit of the asteroidal satellite of 22 Kalliope. Icarus, 165, 112-120.

Marchis, F., Descamps, P., Hestroffer, D., Berthier, J., 2005a. Discovery of the triple asteroidal system 87 Sylvia. Nature, 436, 822-824.

Marchis, F., Hestroffer, Descamps, P., D., Berthier, J., Laver, C., de Pater, I. 2005b. Mass and density of Asteroid 121 Hermione from an analysis of its companion orbit, Icarus, 178, 2, 450-464

Marchis, F., Hestroffer, D., Descamps, P., Berthier, J., Bouchez, A. H., Campbell, R. D., Chin, J. C. Y., van Dam, M. A., Hartman, S. K., Johansson, E. M., Lafon, R. E., Le Mignant, D., de Pater, Imke, Stomski, P. J., Summers, D. M., Vachier, F., Wizinovich, P. L., Wong, M. H., 2006a. A low density of 0.8 g.cm$^{-3}$ for the Trojan binary asteroid 617 Patroclus. Nature, 439, 565-567.

Marchis, F., Kaasalainen, M., Hom, E.F.Y., Berthier, J., Enriquez, J., Hestroffer, D., Le Mignant, D., and de Pater, I. 2006b. Shape, Size, and multiplicity of main-belt asteroids. I Keck Adaptive Optics Survey, Icarus, 185, 39-63.

Marchis, F., Baek, M., Berthier, J., Descamps, P., Hestroffer, D., Kaasalainen, M., Vachier, F., 2006c. Large adaptive optics of asteroids (LAOSA): Size, shape, and occasionally density via multiplicity. Workshop on Spacecraft Reconnaissance of Asteroid and Comet Interiors, Abstract #3042.

Marchis, F., P. Descamps, J. Berthier, D. Hestroffer, F. Vachier, M. Baek, A. Harris, D. Nesvorny, 2008. Main Belt Binary Asteroidal Systems with eccentric orbits, Icarus, in press.

Marchis, F., Baek, M., Descamps, P., Berthier, J., Hestroffer, D., Vachier, F., 2007a. S/2004(45)1, IAU Circ., 8817, 1, Edited by Green, D. W. E.

Margot, J.-L. and Brown, M.E. 2001. S/2001(22)1, IAU Circ., 7703, 3, Edited by Green, D. W. E.

Margot, J.-L. and Brown, M.E. 2003. A Low-Density M-type Asteroid in the Main Belt, Science, 300, 5627, 1939-1942

Merline, W.J., Close, L.M., Dumas, C., Chapman, C.R., Roddier, F., Menard, F., Slater, D.C., Duvert, G., Shelton, C., Morgan, T. 1999. Discovery of a moon orbiting the asteroid 45 Eugenia, Nature, 401, 565.

Merline, W.J., Close, L.M., Dumas, C., Shelton, J.C., Menard, F., Chapman, C.R., Slater, D.C., Discovery of Companions to asteroids 762 Pulcova and 90 Antiope by Dircet Imaging, American Astronomical Society, DPS Meeting #32, #13.06; Bulletin of the American Astronomical Society, 32, p.1017

Merline, W.J., Menard, F., Close, L., Dumas, C, Chapman, C.R., and Slater, D.C . 2001. S/2001(22)1, IAU Circ., 7703, 2.

Nesvorny, D.; Vokrouhlicky, D., 2006. New Candidates for Recent Asteroid Breakups, Astron. J., 132, 5, 1950-1958.

Nesvorny, D., Bottke, W. F., Vokrouhlicky, D., Morbidelli, A,, Jedicke, R., 2006. Asteroids, Comets, Meteors, Proceedings of the 229th Symposium of the International Astronomical Union held in Búzios, Rio de Janeiro, Brasil August 7-12, 2005, Edited by Daniela, L.; Sylvio Ferraz, M.; Angel, F. Julio Cambridge: Cambridge University Press, 2006., pp.289-299

Pravec, P. *et al.* 2006. Photometric survey of binary near-Eat asteroids, Icarus 181, 1, 63-93

Rabinowitz, D.L., K. Barkume, M.E. Brown, H. Roe, M. Schwartz, S. Tourtellotte and C. Trujillo, Photometric observations constraining the size, shape, and albedo of 2003 $EL_{61}$, a rapidly rotating, Pluto-sized object in the Kuiper belt, 2006. *Astrophys. J.* **639**, pp. 1238–1251.

Richardson, J.E., Melosh, H.J., Lisse, C.<., Carcich, B. 2007. A ballistics analysis of the Deep Impact ejecta plume: Determining Comet Tempel 1's gravity, mass, and density, Icarus 191, 2, 176-209.

Rivkin, A.S., Howell, E.S., Lebofsky, L.A., Clark, B.E., Britt, D.T. 2000. The nature of M-class asteroids from 3-micron observations. Icarus, 145, 351-368

Rousset, G., Lacombe, F., Puget, P., Hubin, N.N., Gendron, E., Fusco, T., Arsenault, R., Charton, J., Feautrier, P., Gigan, P., Kern, P.Y., Lagrange, A.-M., Madec, P.-Y., Mouillet, D., Rabaud, D., Rabou, P., Stadler, E., Zins, G. 2003. Adaptive Optical System Technologies II. Edited by Wizinowich, Peter L.; Bonaccini, Domenico. Proceedings of the SPIE, Volume 4839, pp. 140-149

Soma, M., Hayamizu, T., Berthier, J., Lecacheux, J. 2006. (22) Kalliope and (22) Kalliope I, Central Bureau Electronic Telegrams, 732, 1. Edited by Green, D. W. E.

Storrs, A, F. Vilas, R. Landis, E. Wells, C. Woods, B., Zellner, and M. Gaffey, 2001. Iau circular, Ed. D. W.E Green, 7599,1.

Tedesco, E.F., Noah, P.V., Noah, M. and Price, S.D. 2002. The supplemental IRAS Minor Planet Survey, Astron. J. 123, 1056-1085.

Tholen, D.J and Barucci, M.A., 1989. Asteroid taxonomy, In Asteroids II (R.P. Binzel, *et al.* eds), pp. 806-825. Univ. of Arizona, Tucson.

Torppa, J., Kaasalainen, M., Michalowski, T., Kwiatkowski, T., Kryszczynska, A., Denchev, P., Kowalski, R., 2003. Shapes and rotational properties of thirty asteroids from photometric data, Icarus, Volume 164, Issue 2, p. 346-383

Yeomans, D.K., J.P., Barriot, D.W. Dunham, R. W. Farquhar, J. D. Giorgini, C. E. Helfrich, A. S. Konopliv, J. V. McAdams, J. K. Miller, W. M. Owen Jr., D. J. Scheeres, S. P. Synnott, B. G. Williams, 1997. Estimating the Mass of Asteroid 253 Mathilde from Tracking Data During the NEAR Flyby, Science, 278, 5346, 2106-2109.

Yoder, C.F., 1982. Tidal rigidity of Phobos, Icarus, 49, 327-346

Weidenschilling, S.J., Paolicchi, P.Zappala, V. 1989. Do asteroids have satellites? In Asteroids II; Proceedings of the Conference, Tucson, AZ, Mar. 8-11, 1988 (A90-27001 10-91). Tucson, AZ, University of Arizona Press, 1989, p. 643-658.

Wilkison, S.L., Robinson, M.S., Thomas, P.C., Veverka, J., McCoy, T.J., Murchie, S.L.,Prokter, L.M., Yeomans, D.K. 2002. An Estimate of Eros's Porosity and Implications for Internal Structure, Icarus, Volume 155, 94-103

Zappalà, V., Bendjoya, Ph., Cellino A., Farinella P., Froeschlé C., 1995. Asteroid families: Search of a 12,487-asteroid sample using two different clustering techniques, Icarus 116, 291-314.

**Table 1**

Characteristics of the studied binary minor planets. IRAS radiometric diameters estimated by means of the STM or NEATM models from IRAS data are indicated.

| Asteroid | Primary Diameter (km) | | | Rotational Period / max(a/b)[4] in hours | Sp. type | Secondary | |
|---|---|---|---|---|---|---|---|
| | IRAS STM | IRAS NEATM | AO | | | Name | $D_{satellite}$ |
| 22 Kalliope | 181±5 | 177±4 | 196±20[6] | 4.148/1.30 | W[5]/M[1] | S/2001 (22)1 "Linus" | 26±11 |
| 45 Eugenia | 215±4 | 217±8 | 193±15 | 5.699/1.41 | C[2] | S/1998 (45)1 "Petit-Prince" | 7±2 |
| 107 Camilla | 223±17 | 249±18 | 246±13 | 4.843/1.52 | X[2] | S/2001 (107)1 | 16±6 |
| 762 Pulcova | 137±3 | 143±2 | 216±40[6] | 5.839/1.30 | Cb[3] | S/2000 (762)1 | 19±7 |

1. Tholen and Barucci, (1989)
2. Bus and Binzel, (2002)
3. S3OS2 BB Lazarro *et al.* (2004)
4. Minor Planet Lightcurve Parameters, A.W. Harris and B. D. Warner, http://cfa-www.harvard.edu/iau/lists/LightcurveDat.html
5. Rivkin *et al.* (2000)
6. This size measurement is overestimated and will not be used in this work (see text)

**Table 2a**
Summary of our AO Observations of 22 Kalliope collected with the VLT and Gemini North telescopes. The predicted magnitude in visible ($m_v$), celestial coordinates (RA, DEC), and distance from Earth are extracted from the IMCCE ephemeris web site (http://www.imcce.fr).

| ID | Name | Date | UT | Telescope | Filter | Mv predicted | Airmass | RA J2000 | DEC J2000 | Distance from Earth (AU) |
|---|---|---|---|---|---|---|---|---|---|---|
| 22 | Kalliope | 14-Jan-03 | 08:53:59 | VLT | Ks | 11.8 | 1.30 | 12 57 54.04 | 10 35 53.7 | 2.56729 |
| 22 | Kalliope | 14-Jan-03 | 09:00:19 | VLT | H | 11.8 | 1.28 | 12 57 54.17 | 10 35 54.0 | 2.56723 |
| 22 | Kalliope | 14-Jan-03 | 09:07:39 | VLT | J | 11.8 | 1.27 | 12 57 54.31 | 10 35 54.4 | 2.56717 |
| 22 | Kalliope | 14-Feb-03 | 08:21:57 | VLT | H | 11.4 | 1.25 | 13 02 34.74 | 12 19 18.3 | 2.23262 |
| 22 | Kalliope | 15-Jul-03 | 00:58:10 | VLT | H | 12.3 | 1.72 | 12 31 27.71 | 06 28 13.5 | 3.26785 |
| 22 | Kalliope | 16-Jul-03 | 00:01:32 | VLT | H | 12.3 | 1.42 | 12 32 14.85 | 06 19 05.7 | 3.28098 |
| 22 | Kalliope | 17-Jul-03 | 00:24:36 | VLT | H | 12.3 | 1.54 | 12 33 05.25 | 06 09 24.8 | 3.29483 |
| 22 | Kalliope | 03-Mar-04 | 10:08:26 | VLT | H | 12.4 | 1.03 | 17 11 35.85 | -21 37 31.8 | 3.14872 |
| 22 | Kalliope | 04-Mar-04 | 10:13:54 | VLT | H | 12.4 | 1.02 | 17 12 23.24 | -21 40 26.5 | 3.13418 |
| 22 | Kalliope | 04-Jun-04 | 09:46:02 | Gemini | Kp | 10.9 | 1.47 | 16 58 44.72 | -26 01 07.0 | 2.18463 |
| 22 | Kalliope | 04-Jun-04 | 09:48:13 | Gemini | Kp | 10.9 | 1.46 | 16 58 44.63 | -26 01 07.2 | 2.18463 |
| 22 | Kalliope | 05-Jun-04 | 08:21:22 | Gemini | Kp | 10.9 | 1.73 | 16 57 50.45 | -26 03 02.7 | 2.18393 |
| 22 | Kalliope | 05-Jun-04 | 11:07:04 | Gemini | Kp | 10.9 | 1.46 | 16 57 43.60 | -26 03 16.9 | 2.18386 |
| 22 | Kalliope | 05-Jun-04 | 12:07:29 | Gemini | Kp | 10.9 | 1.63 | 16 57 41.11 | -26 03 21.9 | 2.18384 |
| 22 | Kalliope | 05-Jun-04 | 13:22:58 | Gemini | Kp | 10.9 | 2.22 | 16 57 38.00 | -26 03 27.9 | 2.18382 |
| 22 | Kalliope | 28-Jun-04 | 02:20:44 | VLT | Ks | 11.2 | 1.01 | 16 37 17.16 | -26 37 19.3 | 2.24340 |
| 22 | Kalliope | 24-Jul-04 | 04:06:17 | VLT | H | 11.7 | 1.34 | 16 24 22.40 | -27 02 30.6 | 2.46552 |
| 22 | Kalliope | 25-Jul-04 | 05:40:48 | VLT | H | 11.7 | 2.24 | 16 24 11.16 | -27 03 37.4 | 2.47724 |
| 22 | Kalliope | 25-Jul-04 | 05:47:42 | VLT | H | 11.7 | 2.35 | 16 24 11.11 | -27 03 37.7 | 2.47729 |
| 22 | Kalliope | 26-Jul-04 | 01:26:47 | VLT | H | 11.7 | 1.01 | 16 24 03.86 | -27 04 31.4 | 2.48636 |
| 22 | Kalliope | 26-Jul-04 | 03:23:10 | VLT | H | 11.7 | 1.21 | 16 24 03.07 | -27 04 36.4 | 2.48727 |
| 22 | Kalliope | 26-Jul-04 | 04:51:14 | VLT | H | 11.7 | 1.68 | 16 24 02.50 | -27 04 39.9 | 2.48797 |
| 22 | Kalliope | 07-Oct-04 | 00:31:15 | VLT | Ks | 12.4 | 1.51 | 17 09 25.70 | -29 13 51.3 | 3.43409 |
| 22 | Kalliope | 07-Oct-04 | 00:45:10 | VLT | Ks | 12.4 | 1.61 | 17 09 26.39 | -29 13 52.4 | 3.43421 |
| 22 | Kalliope | 12-Dec-06 | 13:41:59 | Keck-NIRC2 | Kp | 10.0 | 1.28 | 05 54 43.84 | 30 48 12.1 | 1.65845 |

**Table 2b:**
Summary of our AO Observations of 45 Eugenia collected with Keck, VLT, and Gemini AO systems.

| ID | Name | Date | UT | Telescope | Filter | Mv predicted | Airmass | RA J2000 | DEC J2000 | Distance from Earth (AU) |
|---|---|---|---|---|---|---|---|---|---|---|
| 45 | Eugenia | 06-Dec-03 | 12:42:32 | Keck-NIRC2 | Kp | 12.4 | 1.07 | 08 38 24.49 | 13 09 09.9 | 2.1401 |
| 45 | Eugenia | 06-Dec-03 | 14:20:10 | Keck-NIRC2 | Kp | 12.4 | 1.01 | 08 38 24.16 | 13 09 10.2 | 2.1393 |
| 45 | Eugenia | 04-Jan-04 | 07:03:54 | VLT | Ks | 11.9 | 1.31 | 08 26 33.26 | 14 02 51.7 | 1.8712 |
| 45 | Eugenia | 05-Jan-04 | 05:56:03 | VLT | Ks | 11.9 | 1.29 | 08 25 50.75 | 14 06 21.3 | 1.8655 |
| 45 | Eugenia | 06-Jan-04 | 05:10:22 | VLT | Ks | 11.6 | 1.33 | 08 25 06.56 | 14 09 60.0 | 1.8600 |
| 45 | Eugenia | 07-Jan-04 | 05:43:15 | VLT | Ks | 11.6 | 1.29 | 08 24 18.77 | 14 13 57.1 | 1.8544 |
| 45 | Eugenia | 07-Jan-04 | 06:41:16 | VLT | Ks | 11.6 | 1.30 | 08 24 16.79 | 14 14 06.6 | 1.8542 |
| 45 | Eugenia | 07-Jan-04 | 07:25:36 | VLT | Ks | 11.6 | 1.38 | 08 24 15.28 | 14 14 13.7 | 1.8540 |
| 45 | Eugenia | 11-Feb-04 | 03:11:20 | VLT | Ks | 11.5 | 1.34 | 07 54 06.86 | 17 00 21.0 | 1.8455 |
| 45 | Eugenia | 11-Feb-04 | 03:20:05 | VLT | H | 11.5 | 1.34 | 07 54 06.58 | 17 00 22.8 | 1.8456 |
| 45 | Eugenia | 11-Feb-04 | 03:28:16 | VLT | J | 11.5 | 1.34 | 07 54 06.32 | 17 00 24.4 | 1.8456 |
| 45 | Eugenia | 11-Feb-04 | 03:43:53 | VLT | J | 11.5 | 1.35 | 07 54 05.82 | 17 00 27.5 | 1.8456 |
| 45 | Eugenia | 12-Feb-04 | 02:52:35 | VLT | Ks | 11.5 | 1.34 | 07 53 24.01 | 17 05 01.5 | 1.8503 |
| 45 | Eugenia | 12-Feb-04 | 03:00:54 | VLT | H | 11.5 | 1.34 | 07 53 23.75 | 17 05 03.2 | 1.8503 |
| 45 | Eugenia | 12-Feb-04 | 03:08:06 | VLT | J | 11.5 | 1.34 | 07 53 23.53 | 17 05 04.6 | 1.8503 |
| 45 | Eugenia | 12-Feb-04 | 03:33:25 | VLT | J | 11.5 | 1.35 | 07 53 22.74 | 17 05 09.6 | 1.8504 |
| 45 | Eugenia | 12-Feb-04 | 04:09:55 | VLT | Ks | 11.5 | 1.39 | 07 53 21.61 | 17 05 16.7 | 1.8506 |
| 45 | Eugenia | 14-Feb-04 | 03:41:49 | VLT | Ks | 11.5 | 1.36 | 07 51 59.21 | 17 14 30.8 | 1.8609 |
| 45 | Eugenia | 15-Feb-04 | 03:30:32 | VLT | Ks | 11.7 | 1.36 | 07 51 19.75 | 17 19 03.8 | 1.8664 |
| 45 | Eugenia | 16-Feb-04 | 03:42:16 | VLT | Ks | 11.7 | 1.38 | 07 50 40.93 | 17 23 37.9 | 1.8723 |
| 45 | Eugenia | 01-Mar-04 | 03:55:54 | VLT | H | 12.0 | 1.64 | 07 44 16.16 | 18 20 11.0 | 1.9771 |
| 45 | Eugenia | 01-Mar-04 | 04:03:50 | VLT | H | 12.0 | 1.69 | 07 44 16.06 | 18 20 12.2 | 1.9772 |
| 45 | Eugenia | 03-Mar-04 | 04:10:27 | VLT | H | 12.0 | 1.80 | 07 43 46.58 | 18 27 05.7 | 1.9953 |
| 45 | Eugenia | 04-Mar-04 | 03:32:56 | VLT | Ks | 12.0 | 1.59 | 07 43 34.70 | 18 30 19.7 | 2.0043 |
| 45 | Eugenia | 04-Mar-04 | 03:49:08 | VLT | H | 12.0 | 1.66 | 07 43 34.55 | 18 30 21.9 | 2.0044 |
| 45 | Eugenia | 04-Mar-04 | 04:00:55 | VLT | J | 12.0 | 1.74 | 07 43 34.45 | 18 30 23.5 | 2.0045 |
| 45 | Eugenia | 07-Mar-04 | 03:17:19 | VLT | Ks | 12.2 | 1.58 | 07 43 07.95 | 18 39 45.9 | 2.0330 |
| 45 | Eugenia | 07-Mar-04 | 03:30:42 | VLT | H | 12.2 | 1.63 | 07 43 07.88 | 18 39 47.6 | 2.0331 |
| 45 | Eugenia | 07-Mar-04 | 03:41:49 | VLT | J | 12.2 | 1.70 | 07 43 07.81 | 18 39 49.0 | 2.0332 |
| 45 | Eugenia | 08-Mar-04 | 00:56:15 | VLT | Ks | 12.2 | 1.39 | 07 43 02.94 | 18 42 28.3 | 2.0419 |
| 45 | Eugenia | 08-Mar-04 | 01:06:29 | VLT | H | 12.2 | 1.38 | 07 43 02.90 | 18 42 29.5 | 2.0420 |
| 45 | Eugenia | 08-Mar-04 | 01:17:31 | VLT | J | 12.2 | 1.37 | 07 43 02.85 | 18 42 30.9 | 2.0421 |
| 45 | Eugenia | 09-Mar-04 | 02:09:22 | VLT | Ks | 12.2 | 1.41 | 07 42 58.53 | 18 45 32.0 | 2.0525 |
| 45 | Eugenia | 10-Mar-04 | 01:26:36 | VLT | Ks | 12.2 | 1.38 | 07 42 56.20 | 18 48 16.9 | 2.0623 |
| 45 | Eugenia | 11-Mar-04 | 00:31:09 | VLT | Ks | 12.2 | 1.40 | 07 42 55.44 | 18 50 55.6 | 2.0722 |
| 45 | Eugenia | 11-Mar-04 | 00:45:46 | VLT | Ks | 12.2 | 1.39 | 07 42 55.42 | 18 50 57.3 | 2.0723 |
| 45 | Eugenia | 12-Mar-04 | 01:36:44 | VLT | Ks | 12.2 | 1.39 | 07 42 56.18 | 18 53 42.9 | 2.0831 |
| 45 | Eugenia | 12-Mar-04 | 01:44:32 | VLT | Ks | 12.2 | 1.40 | 07 42 56.18 | 18 53 43.7 | 2.0832 |
| 45 | Eugenia | 12-Mar-04 | 01:53:32 | VLT | Ks | 12.2 | 1.41 | 07 42 56.18 | 18 53 44.7 | 2.0833 |
| 45 | Eugenia | 04-Jun-04 | 06:32:55 | Gemini | H | 13.1 | 1.87 | 08 57 09.91 | 17 30 19.4 | 3.0734 |
| 45 | Eugenia | 04-Jun-04 | 06:34:54 | Gemini | H | 13.1 | 1.90 | 08 57 10.03 | 17 30 19.1 | 3.0734 |
| 45 | Eugenia | 04-Jun-04 | 06:44:05 | Gemini | Kp | 13.1 | 2.01 | 08 57 10.57 | 17 30 17.3 | 3.0735 |
| 45 | Eugenia | 04-Jun-04 | 06:45:55 | Gemini | Kp | 13.1 | 2.05 | 08 57 10.68 | 17 30 16.9 | 3.0735 |
| 45 | Eugenia | 15-Jan-05 | 15:02:35 | Keck-NIRC2 | Kp | 12.8 | 1.61 | 15 01 44.95 | -11 45 56.7 | 2.6943 |
| 45 | Eugenia | 17-Jul-05 | 07:25:44 | Keck-NIRC2 | Kp | 11.8 | 1.23 | 15 23 06.07 | -09 46 46.8 | 1.8909 |
| 45 | Eugenia | 03-Aug-06 | 14:07:40 | Keck-NIRC2 | Kp | 12.1 | 1.10 | 23 55 35.28 | -03 00 04.5 | 1.9824 |
| 45 | Eugenia | 09-Sep-07 | 14:31:52 | Keck-NIRC2 | Kp | 13.2 | 1.17 | 05 20 56.24 | 16 16 18.7 | 2.8369 |
| 45 | Eugenia | 09-Sep-07 | 15:20:50 | Keck-NIRC2 | Kp | 13.2 | 1.05 | 05 20 57.94 | 16 16 16.5 | 2.8364 |

**Table 2c:**
Summary of our AO Observations of 107 Camilla collected mostly with VLT and Gemini AO systems.

| ID | Name | Date | UT | Telescope | Filter | Mv predicted | Airmass | RA J2000 | DEC J2000 | Distance from Earth (AU) |
|---|---|---|---|---|---|---|---|---|---|---|
| 107 | Camilla | 15-Jul-03 | 07:32:49 | VLT | H | 12.6 | 1.32 | 19 40 25.92 | -09 29 28.7 | 2.7504 |
| 107 | Camilla | 15-Jul-03 | 07:37:26 | VLT | H | 12.6 | 1.35 | 19 40 25.78 | -09 29 29.2 | 2.7504 |
| 107 | Camilla | 1-Sep-04 | 05:07:39 | VLT | Ks | 12.6 | 1.09 | 23 30 44.98 | -02 00 49.9 | 2.6694 |
| 107 | Camilla | 1-Sep-04 | 05:17:23 | VLT | H | 12.6 | 1.09 | 23 30 44.74 | -02 00 52.3 | 2.6693 |
| 107 | Camilla | 1-Sep-04 | 08:06:44 | VLT | Ks | 12.6 | 1.38 | 23 30 40.49 | -02 01 33.1 | 2.6689 |
| 107 | Camilla | 3-Sep-04 | 06:51:57 | VLT | Ks | 12.4 | 1.17 | 23 29 31.67 | -02 12 57.0 | 2.6623 |
| 107 | Camilla | 4-Sep-04 | 09:02:35 | VLT | Ks | 12.4 | 1.90 | 23 28 52.44 | -02 19 24.7 | 2.6591 |
| 107 | Camilla | 5-Sep-04 | 04:28:20 | VLT | Ks | 12.4 | 1.10 | 23 28 23.35 | -02 24 14.4 | 2.6569 |
| 107 | Camilla | 5-Sep-04 | 08:34:56 | VLT | Ks | 12.4 | 1.66 | 23 28 16.93 | -02 25 15.9 | 2.6564 |
| 107 | Camilla | 6-Sep-04 | 03:35:31 | VLT | Ks | 12.4 | 1.18 | 23 27 48.29 | -02 30 01.2 | 2.6545 |
| 107 | Camilla | 6-Sep-04 | 07:49:11 | VLT | Ks | 12.4 | 1.39 | 23 27 41.64 | -02 31 04.8 | 2.6541 |
| 107 | Camilla | 7-Sep-04 | 02:20:10 | VLT | Ks | 12.4 | 1.43 | 23 27 13.59 | -02 35 44.2 | 2.6525 |
| 107 | Camilla | 8-Sep-04 | 06:41:20 | VLT | Ks | 12.4 | 1.19 | 23 26 29.77 | -02 42 54.2 | 2.6503 |
| 107 | Camilla | 11-Sep-04 | 04:34:25 | VLT | Ks | 12.4 | 1.08 | 23 24 41.47 | -03 00 42.2 | 2.6468 |
| 107 | Camilla | 13-Sep-04 | 03:42:52 | VLT | Ks | 12.1 | 1.11 | 23 23 27.99 | -03 12 46.6 | 2.6458 |
| 107 | Camilla | 13-Sep-04 | 05:47:29 | VLT | Ks | 12.1 | 1.12 | 23 23 24.63 | -03 13 18.5 | 2.6458 |
| 107 | Camilla | 14-Sep-04 | 04:09:30 | VLT | Ks | 12.1 | 1.08 | 23 22 49.81 | -03 19 02.7 | 2.6457 |
| 107 | Camilla | 14-Sep-04 | 07:06:44 | VLT | Ks | 12.1 | 1.35 | 23 22 45.06 | -03 19 48.1 | 2.6457 |
| 107 | Camilla | 14-Sep-04 | 07:14:32 | VLT | H | 12.1 | 1.39 | 23 22 44.85 | -03 19 50.1 | 2.6457 |
| 107 | Camilla | 15-Sep-04 | 04:18:35 | VLT | Ks | 12.1 | 1.07 | 23 22 12.15 | -03 25 14.1 | 2.6460 |
| 107 | Camilla | 15-Sep-04 | 04:26:56 | VLT | H | 12.1 | 1.07 | 23 22 11.92 | -03 25 16.3 | 2.6460 |
| 107 | Camilla | 16-Sep-04 | 04:48:18 | VLT | Ks | 12.1 | 1.08 | 23 21 34.00 | -03 31 30.5 | 2.6465 |
| 107 | Camilla | 7-Oct-04 | 01:47:18 | VLT | Ks | 12.5 | 1.10 | 23 09 51.59 | -05 30 36.8 | 2.7243 |
| 107 | Camilla | 7-Oct-04 | 02:02:04 | VLT | Ks | 12.5 | 1.08 | 23 09 51.29 | -05 30 39.9 | 2.7244 |
| 107 | Camilla | 7-Oct-04 | 02:12:52 | VLT | H | 12.5 | 1.07 | 23 09 51.07 | -05 30 42.2 | 2.7244 |
| 107 | Camilla | 7-Oct-04 | 02:22:53 | VLT | J | 12.5 | 1.07 | 23 09 50.87 | -05 30 44.2 | 2.7245 |
| 107 | Camilla | 20-Oct-04 | 00:39:22 | VLT | Ks | 12.7 | 1.11 | 23 05 02.47 | -06 26 40.1 | 2.8301 |
| 107 | Camilla | 25-Oct-04 | 06:52:46 | Keck-NIRC2 | Kp | 12.9 | 1.12 | 23 03 50.61 | -06 43 57.0 | 2.8834 |
| 107 | Camilla | 2-Nov-04 | 07:38:36 | Gemini | Kp | 13.1 | 1.16 | 23 02 56.13 | -07 03 38.6 | 2.9744 |
| 107 | Camilla | 2-Nov-04 | 15:39:39 | Gemini | Kp | 13.1 | 1.35 | 23 02 55.25 | -07 04 16.6 | 2.9784 |
| 107 | Camilla | 20-Dec-05 | 10:28:08 | Gemini | Ks | 12.4 | 1.35 | 03 17 18.98 | 05 35 56.7 | 2.5377 |
| 107 | Camilla | 20-Dec-05 | 10:33:57 | Gemini | Ks | 12.4 | 1.37 | 03 17 18.89 | 05 35 56.7 | 2.5378 |
| 107 | Camilla | 21-Dec-05 | 09:01:03 | Gemini | Ks | 12.4 | 1.10 | 03 16 57.62 | 05 36 12.9 | 2.5460 |
| 107 | Camilla | 21-Dec-05 | 09:05:52 | Gemini | Ks | 12.4 | 1.10 | 03 16 57.54 | 05 36 13.0 | 2.5460 |
| 107 | Camilla | 1-Jan-06 | 10:12:21 | Gemini | Ks | 12.6 | 1.56 | 03 14 01.74 | 05 47 39.6 | 2.6555 |
| 107 | Camilla | 1-Jan-06 | 10:17:13 | Gemini | Ks | 12.6 | 1.58 | 03 14 01.70 | 05 47 39.9 | 2.6555 |
| 107 | Camilla | 6-Jan-06 | 09:32:44 | Gemini | Ks | 12.7 | 1.42 | 03 13 30.75 | 05 57 30.7 | 2.7114 |
| 107 | Camilla | 6-Jan-06 | 09:38:56 | Gemini | Ks | 12.7 | 1.45 | 03 13 30.73 | 05 57 31.3 | 2.7114 |
| 107 | Camilla | 8-Jan-06 | 05:03:57 | Gemini | Ks | 12.7 | 1.10 | 03 13 27.25 | 06 01 46.2 | 2.7326 |
| 107 | Camilla | 9-Jan-06 | 05:14:24 | Gemini | Ks | 12.7 | 1.07 | 03 13 26.93 | 06 04 16.8 | 2.7445 |
| 107 | Camilla | 9-Jan-06 | 05:20:13 | Gemini | Ks | 12.7 | 1.07 | 03 13 26.92 | 06 04 17.4 | 2.7446 |
| 107 | Camilla | 16-Jan-06 | 05:16:50 | Gemini | Ks | 12.9 | 1.04 | 03 13 58.79 | 06 24 25.2 | 2.8308 |

**Table 2d:**
Summary of our AO Observations of 762 Pulcova mainly collected with the Gemini AO system.

| ID | Name | Date | UT | Telescope | Filter | Mv predicted | Airmass | RA J2000 | DEC J2000 | Distance from Earth (AU) |
|---|---|---|---|---|---|---|---|---|---|---|
| 762 | Pulcova | 15-Jul-03 | 08:23:30 | VLT | Ks | 14.3 | 1.47 | 00 40 49.63 | 16 39 40.5 | 3.0930 |
| 762 | Pulcova | 30-Oct-04 | 11:45:25 | Gemini | Kp | 13.6 | 1.30 | 06 48 16.93 | 34 19 39.7 | 2.3742 |
| 762 | Pulcova | 30-Oct-04 | 11:55:18 | Gemini | Kp | 13.6 | 1.29 | 06 48 17.01 | 34 19 40.1 | 2.3741 |
| 762 | Pulcova | 30-Oct-04 | 12:06:42 | Gemini | Kp | 13.6 | 1.23 | 06 48 17.09 | 34 19 40.6 | 2.3740 |
| 762 | Pulcova | 30-Oct-04 | 13:45:57 | Gemini | Kp | 13.6 | 1.05 | 06 48 17.81 | 34 19 44.5 | 2.3731 |
| 762 | Pulcova | 30-Oct-04 | 15:29:56 | Gemini | Kp | 13.6 | 1.06 | 06 48 18.55 | 34 19 48.2 | 2.3722 |
| 762 | Pulcova | 2-Nov-04 | 12:16:54 | Gemini | Kp | 13.4 | 1.20 | 06 48 45.99 | 34 22 15.3 | 2.3361 |
| 762 | Pulcova | 2-Nov-04 | 12:22:15 | Gemini | Kp | 13.4 | 1.19 | 06 48 46.02 | 34 22 15.5 | 2.3361 |
| 762 | Pulcova | 2-Nov-04 | 15:13:27 | Gemini | Kp | 13.4 | 1.05 | 06 48 46.63 | 34 22 21.9 | 2.3346 |
| 762 | Pulcova | 18-Mar-06 | 12:43:54 | Gemini | Kp | 12.7 | 1.49 | 12 12 29.32 | -19 50 06.8 | 2.0083 |
| 762 | Pulcova | 28-Apr-06 | 08:32:21 | Gemini | Kp | 13.0 | 1.29 | 11 44 13.61 | -17 05 40.4 | 2.1314 |
| 762 | Pulcova | 16-May-06 | 06:12:35 | Gemini | Kp | 13.3 | 1.24 | 11 41 29.42 | -15 48 36.9 | 2.3045 |
| 762 | Pulcova | 18-May-06 | 07:17:47 | Gemini | Kp | 13.3 | 1.27 | 11 41 38.45 | -15 41 40.1 | 2.3275 |
| 762 | Pulcova | 19-May-06 | 08:26:58 | Gemini | Kp | 13.3 | 1.48 | 11 41 45.18 | -15 38 17.0 | 2.3395 |
| 762 | Pulcova | 23-Jun-06 | 06:18:55 | Gemini | Kp | 13.8 | 1.43 | 11 57 47.74 | -15 01 34.3 | 2.7944 |

**Table 3a**: Search for moonlet companions around 22 Kalliope. The characteristics of the 2-σ detection curve for each asteroid are calculated. α is the slope of the function, and $r_{lim}$ separation between both noise regimes dominated by the Poisson noise close to the primary at $r < r_{lim}$ and by the [detector+sky] noises at $r > r_{lim}$. At $r > r_{lim}$ the detection function can be approximated by a flat function with a value of $\Delta m_{lim}$. The radius of the Hill sphere is calculate based on consideration about the size and density of the asteroid (see Table 7 and details in Marchis *et al.* 2006b). The minimum size for a moonlet to be detected at 1/4 and 2/100 × $R_{Hill}$ is also indicated.

| ID | Name | Date | UT Time | α | Dm Limit | r Limit arcsec | Int Time s | Airmass | FWHM arcsec | Δm at 2/100xRHill | Diameter (km) at 2/100xRHill | Dm at 1/4xRHill | Diameter (km) at 2/100xRHill |
|---|---|---|---|---|---|---|---|---|---|---|---|---|---|
| 22 | Kalliope | 14-Jan-03 | 08:53:59 | -4.51 | -8.53 | 0.78 | 240 | 1.30 | 0.1 | -4.6 | 21.4 | -8.5 | 3.7 |
| 22 | Kalliope | 14-Jan-03 | 09:00:19 | -2.96 | -8.19 | 0.94 | 120 | 1.28 | 0.1 | -4.6 | 22.1 | -8.2 | 4.1 |
| 22 | Kalliope | 14-Jan-03 | 09:07:39 | -2.87 | -8.12 | 1.11 | 240 | 1.27 | 0.1 | -4.6 | 22.3 | -8.1 | 4.3 |
| 22 | Kalliope | 14-Feb-03 | 08:21:57 | -2.46 | -8.17 | 1.92 | 24.1 | 1.25 | 0.16 | -4.5 | 22.9 | -8.2 | 4.1 |
| 22 | Kalliope | 15-Jul-03 | 00:58:10 | -3.78 | -9.65 | 1.98 | 357.8 | 1.72 | 0.2 | -3.7 | 32.5 | -9.7 | 2.1 |
| 22 | Kalliope | 16-Jul-03 | 00:01:32 | -5.41 | -8.86 | 1.03 | 55.2 | 1.42 | 0.1 | -4.7 | 21.3 | -8.9 | 3.0 |
| 22 | Kalliope | 17-Jul-03 | 00:24:36 | -2.81 | -8.58 | 2.36 | 171 | 1.54 | 0.42 | -3.4 | 38.2 | -8.6 | 3.5 |
| 22 | Kalliope | 03-Mar-04 | 10:08:26 | -4.73 | -7.59 | 0.97 | 24 | 1.03 | 0.11 | -5.8 | 12.3 | -7.7 | 5.3 |
| 22 | Kalliope | 04-Jun-04 | 09:46:02 | -5.7 | -9.33 | 1.12 | 9 | 1.47 | 0.12 | -7.9 | 4.8 | -9.3 | 2.5 |
| 22 | Kalliope | 05-Jun-04 | 08:21:22 | -3.24 | -9.11 | 1.07 | 6 | 1.73 | 0.11 | -4.4 | 23.9 | -9.1 | 2.8 |
| 22 | Kalliope | 05-Jun-04 | 11:07:04 | -3.92 | -8.29 | 1.05 | 3 | 1.46 | 0.12 | -5.1 | 17.1 | -8.3 | 4.0 |
| 22 | Kalliope | 05-Jun-04 | 12:07:29 | -4.78 | -9.36 | 1.14 | 10 | 1.63 | 0.11 | -4.4 | 23.7 | -9.4 | 2.4 |
| 22 | Kalliope | 05-Jun-04 | 13:22:58 | -3.68 | -8.43 | 1.36 | 6 | 2.21 | 0.13 | -5.1 | 17.4 | -8.4 | 3.7 |
| 22 | Kalliope | 28-Jun-04 | 02:20:44 | -3.62 | -8.82 | 0.82 | 600 | 1.01 | 0.11 | -7.6 | 5.5 | -8.9 | 3.0 |
| 22 | Kalliope | 24-Jul-04 | 04:06:17 | -3.62 | -8.3 | 1.06 | 120 | 1.34 | 0.11 | -6.1 | 11.0 | -8.5 | 3.6 |
| 22 | Kalliope | 25-Jul-04 | 05:40:48 | -1.81 | -6.25 | 1.29 | 60 | 2.24 | 0.15 | -4.6 | 21.9 | -6.2 | 10.2 |
| 22 | Kalliope | 25-Jul-04 | 05:47:42 | -2.4 | -6.31 | 1.25 | 40 | 2.35 | 0.18 | -4.6 | 21.9 | -6.3 | 9.9 |
| 22 | Kalliope | 26-Jul-04 | 01:26:47 | -4.87 | -8.08 | 0.95 | 60 | 1.01 | 0.1 | -6.6 | 8.8 | -8.1 | 4.4 |
| 22 | Kalliope | 26-Jul-04 | 03:23:10 | -3.62 | -7.54 | 1.33 | 60 | 1.21 | 0.14 | -5.8 | 12.4 | -7.6 | 5.6 |
| 22 | Kalliope | 26-Jul-04 | 04:51:14 | -4.23 | -8 | 1.19 | 45 | 1.68 | 0.12 | -6.2 | 10.5 | -7.7 | 5.2 |
| 22 | Kalliope | 07-Oct-04 | 00:31:15 | -2.91 | -6.64 | 1.31 | 180 | 1.51 | 0.15 | -4.2 | 26.5 | -6.7 | 8.5 |
| 22 | Kalliope | 07-Oct-04 | 00:45:10 | -4.12 | -8.06 | 1.02 | 360 | 1.61 | 0.11 | -4.8 | 19.5 | -8.1 | 4.4 |
| 22 | Kalliope | 12-Dec-06 | 13:41:59 | -3.63 | -8.33 | 0.99 | 180 | 1.28 | 0.14 | -3.9 | 29.6 | -8.4 | 3.9 |

**Table 3b**: Search for moonlet companions around 45 Eugenia.

| ID | Name | Date | UT Time | α | Dm Limit | r Limit arcsec | Int Time s | Airmass | FWHM arcsec | Δm at 2/100xRHill | Diameter (km) at 2/100xRHill | Dm at 1/4xRHill | Diameter (km) at 2/100xRHill |
|---|---|---|---|---|---|---|---|---|---|---|---|---|---|
| 45 | Eugenia | 06-Dec-03 | 12:42:32 | -8.5 | -6.9 | 0.51 | 45 | 1.07 | 0.13 | -6.8 | 9.3 | -6.9 | 9.0 |
| 45 | Eugenia | 06-Dec-03 | 14:20:10 | -5.1 | -7.9 | 0.84 | 180 | 1.01 | 0.14 | -7.0 | 8.4 | -7.9 | 5.6 |
| 45 | Eugenia | 04-Jan-04 | 07:03:54 | -4.6 | -8.4 | 0.90 | 240 | 1.31 | 0.16 | -7.7 | 6.2 | -8.5 | 4.4 |
| 45 | Eugenia | 05-Jan-04 | 05:56:03 | -3.9 | -8.2 | 0.86 | 240 | 1.28 | 0.16 | -7.5 | 7.0 | -8.3 | 4.8 |
| 45 | Eugenia | 06-Jan-04 | 05:10:22 | -2.5 | -7.8 | 1.09 | 240 | 1.33 | 0.16 | -7.1 | 8.1 | -7.8 | 6.0 |
| 45 | Eugenia | 07-Jan-04 | 05:43:15 | -3.2 | -8.2 | 1.17 | 240 | 1.29 | 0.16 | -7.2 | 7.8 | -8.2 | 5.0 |
| 45 | Eugenia | 07-Jan-04 | 06:41:16 | -2.0 | -6.3 | 0.97 | 240 | 1.30 | 0.19 | -5.6 | 16.3 | -6.3 | 11.8 |
| 45 | Eugenia | 07-Jan-04 | 07:25:36 | -2.9 | -6.1 | 0.84 | 240 | 1.38 | 0.20 | -5.5 | 16.8 | -6.2 | 12.6 |
| 45 | Eugenia | 11-Feb-04 | 03:11:20 | -3.7 | -8.1 | 0.90 | 300 | 1.34 | 0.16 | -7.4 | 7.2 | -8.2 | 5.0 |
| 45 | Eugenia | 11-Feb-04 | 03:20:05 | -3.4 | -8.1 | 1.15 | 300 | 1.34 | 0.17 | -6.7 | 10.0 | -8.1 | 5.2 |
| 45 | Eugenia | 11-Feb-04 | 03:28:16 | -3.2 | -7.4 | 1.10 | 300 | 1.34 | 0.19 | -6.0 | 13.6 | -7.4 | 7.0 |
| 45 | Eugenia | 11-Feb-04 | 03:43:53 | -3.4 | -7.1 | 1.09 | 300 | 1.35 | 0.18 | -5.8 | 14.7 | -7.2 | 7.9 |
| 45 | Eugenia | 12-Feb-04 | 02:52:35 | -4.2 | -8.0 | 0.84 | 300 | 1.34 | 0.16 | -7.3 | 7.4 | -8.0 | 5.3 |
| 45 | Eugenia | 12-Feb-04 | 03:00:54 | -4.2 | -7.9 | 0.85 | 300 | 1.34 | 0.16 | -6.7 | 9.9 | -8.0 | 5.4 |
| 45 | Eugenia | 12-Feb-04 | 03:33:25 | -4.0 | -8.0 | 0.86 | 300 | 1.35 | 0.16 | -7.0 | 8.6 | -8.1 | 5.3 |
| 45 | Eugenia | 12-Feb-04 | 04:09:55 | -4.0 | -7.1 | 0.77 | 300 | 1.39 | 0.16 | -6.5 | 10.6 | -7.1 | 8.1 |
| 45 | Eugenia | 14-Feb-04 | 03:41:49 | -3.5 | -8.5 | 0.86 | 300 | 1.36 | 0.16 | -7.2 | 7.7 | -8.5 | 4.2 |
| 45 | Eugenia | 15-Feb-04 | 03:30:32 | -3.7 | -8.4 | 0.91 | 300 | 1.36 | 0.15 | -7.9 | 5.7 | -8.4 | 4.5 |
| 45 | Eugenia | 16-Feb-04 | 03:42:16 | -3.3 | -8.4 | 0.88 | 300 | 1.38 | 0.15 | -7.9 | 5.6 | -8.4 | 4.4 |
| 45 | Eugenia | 01-Mar-04 | 03:55:54 | -4.8 | -6.7 | 0.94 | 96 | 1.64 | 0.18 | -5.9 | 14.3 | -6.7 | 9.7 |
| 45 | Eugenia | 01-Mar-04 | 04:03:50 | -5.1 | -7.4 | 1.02 | 240 | 1.68 | 0.25 | -5.6 | 16.2 | -7.4 | 7.1 |
| 45 | Eugenia | 03-Mar-04 | 04:10:27 | -5.0 | -7.8 | 1.10 | 40 | 1.80 | 0.19 | -6.6 | 10.4 | -7.8 | 5.9 |
| 45 | Eugenia | 04-Mar-04 | 03:32:56 | -4.4 | -8.0 | 0.99 | 300 | 1.59 | 0.15 | -6.9 | 8.8 | -8.1 | 5.3 |
| 45 | Eugenia | 04-Mar-04 | 03:49:08 | -3.8 | -8.3 | 1.30 | 600 | 1.66 | 0.17 | -6.5 | 10.6 | -8.3 | 4.7 |
| 45 | Eugenia | 04-Mar-04 | 04:00:55 | -4.9 | -7.6 | 1.05 | 400 | 1.74 | 0.20 | -6.3 | 12.0 | -7.6 | 6.4 |
| 45 | Eugenia | 07-Mar-04 | 03:17:19 | -3.2 | -7.5 | 1.03 | 300 | 1.58 | 0.15 | -6.7 | 9.8 | -7.5 | 6.7 |
| 45 | Eugenia | 07-Mar-04 | 03:30:42 | -3.4 | -8.1 | 1.18 | 600 | 1.63 | 0.16 | -6.5 | 10.7 | -8.1 | 5.2 |
| 45 | Eugenia | 07-Mar-04 | 03:41:49 | -3.7 | -6.3 | 0.69 | 400 | 1.70 | 0.21 | -5.4 | 17.7 | -6.3 | 11.6 |
| 45 | Eugenia | 08-Mar-04 | 00:56:15 | -5.3 | -7.5 | 0.80 | 300 | 1.38 | 0.15 | -6.9 | 9.1 | -7.5 | 6.7 |
| 45 | Eugenia | 08-Mar-04 | 01:06:29 | -4.2 | -8.0 | 0.94 | 600 | 1.38 | 0.16 | -7.0 | 8.7 | -8.1 | 5.2 |
| 45 | Eugenia | 08-Mar-04 | 01:17:31 | -3.5 | -6.9 | 1.14 | 400 | 1.37 | 0.19 | -6.0 | 13.7 | -7.0 | 8.7 |
| 45 | Eugenia | 09-Mar-04 | 02:09:22 | -4.3 | -8.3 | 0.77 | 300 | 1.41 | 0.13 | -7.1 | 8.3 | -8.3 | 4.8 |
| 45 | Eugenia | 10-Mar-04 | 01:26:36 | -3.5 | -7.4 | 0.94 | 300 | 1.38 | 0.14 | -6.9 | 9.1 | -7.4 | 7.2 |
| 45 | Eugenia | 11-Mar-04 | 00:31:09 | -3.5 | -6.5 | 0.68 | 300 | 1.40 | 0.17 | -5.8 | 14.8 | -6.6 | 10.4 |
| 45 | Eugenia | 11-Mar-04 | 00:45:46 | -3.8 | -7.3 | 0.86 | 300 | 1.39 | 0.16 | -6.6 | 10.4 | -7.3 | 7.4 |
| 45 | Eugenia | 12-Mar-04 | 01:36:44 | -6.3 | -6.4 | 0.70 | 300 | 1.39 | 0.17 | -6.0 | 13.3 | -6.5 | 10.9 |
| 45 | Eugenia | 12-Mar-04 | 01:44:32 | -2.0 | -5.5 | 1.18 | 300 | 1.40 | 0.18 | -4.3 | 29.6 | -5.6 | 16.3 |
| 45 | Eugenia | 12-Mar-04 | 01:53:32 | -3.6 | -7.0 | 1.05 | 300 | 1.41 | 0.16 | -6.2 | 12.4 | -7.0 | 8.4 |
| 45 | Eugenia | 04-Jun-04 | 06:32:55 | -6.4 | -9.4 | 0.81 | 60 | 1.87 | 0.12 | -6.6 | 10.2 | -9.4 | 2.9 |
| 45 | Eugenia | 04-Jun-04 | 06:44:05 | -3.9 | -9.0 | 1.12 | 60 | 2.01 | 0.12 | -6.8 | 9.4 | -9.0 | 3.4 |
| 45 | Eugenia | 15-Jan-05 | 15:02:35 | -5.5 | -7.8 | 0.93 | 360 | 1.61 | 0.12 | -5.6 | 16.6 | -7.9 | 5.7 |
| 45 | Eugenia | 17-Jul-05 | 07:25:44 | -3.9 | -8.7 | 0.92 | 180 | 1.23 | 0.13 | -7.9 | 5.7 | -8.7 | 3.9 |
| 45 | Eugenia | 03-Aug-06 | 14:07:40 | -6.4 | -8.4 | 0.67 | 180 | 1.10 | 0.12 | -7.2 | 7.8 | -8.4 | 4.5 |
| 45 | Eugenia | 09-Sep-07 | 14:31:52 | -5.6 | -9.3 | 0.65 | 720 | 1.17 | 0.10 | -7.6 | 6.5 | -9.5 | 2.7 |
| 45 | Eugenia | 09-Sep-07 | 15:20:50 | -4.4 | -8.9 | 0.88 | 180 | 1.05 | 0.10 | -7.4 | 7.0 | -8.9 | 3.5 |

**Table 3c**: Search for moonlet companions around 107 Camilla.

| ID | Name | Date | UT Time | α | Dm Limit | r Limit arcsec | Int Time s | Airmass | FWHM arcsec | Δm at 2/100xRHill | Diameter (km) at 2/100xRHill | Dm at 1/4xRHill | Diameter (km) at 2/100xRHill |
|---|---|---|---|---|---|---|---|---|---|---|---|---|---|
| 107 | Camilla | 15-Jul-03 | 07:37:26 | -4 | -7.4 | 1.52 | 40 | 1.35 | 0.28 | -5.1 | 21.4 | -7.4 | 7.3 |
| 107 | Camilla | 01-Sep-04 | 05:07:39 | -6.34 | -9.03 | 0.62 | 300 | 1.09 | 0.11 | -8.5 | 4.4 | -9.1 | 3.4 |
| 107 | Camilla | 01-Sep-04 | 05:17:23 | -4.12 | -9.02 | 0.86 | 300 | 1.09 | 0.12 | -7.6 | 6.8 | -9.1 | 3.3 |
| 107 | Camilla | 01-Sep-04 | 08:06:44 | -3.74 | -7.55 | 0.89 | 300 | 1.38 | 0.14 | -6.2 | 13.1 | -7.8 | 6.2 |
| 107 | Camilla | 03-Sep-04 | 06:51:57 | -3.45 | -7.52 | 0.93 | 300 | 1.16 | 0.16 | -5.8 | 15.8 | -7.9 | 5.9 |
| 107 | Camilla | 04-Sep-04 | 09:02:35 | -5.02 | -7.92 | 0.9 | 300 | 1.90 | 0.16 | -6.4 | 11.5 | -8.0 | 5.7 |
| 107 | Camilla | 05-Sep-04 | 04:28:20 | -11.9 | -6.15 | 0.46 | 300 | 1.10 | 0.15 | -5.8 | 15.2 | -7.5 | 6.9 |
| 107 | Camilla | 05-Sep-04 | 08:34:56 | -1.87 | -5.28 | 1.02 | 300 | 1.66 | 0.19 | -3.9 | 37.2 | -5.5 | 17.6 |
| 107 | Camilla | 06-Sep-04 | 03:35:31 | -5.24 | -8.86 | 0.9 | 300 | 1.18 | 0.12 | -7.2 | 8.0 | -8.9 | 3.7 |
| 107 | Camilla | 06-Sep-04 | 07:49:11 | -12.59 | -5.58 | 0.33 | 300 | 1.39 | 0.16 | -4.9 | 22.9 | -6.2 | 12.6 |
| 107 | Camilla | 07-Sep-04 | 02:20:10 | -5.06 | -8.72 | 0.9 | 300 | 1.43 | 0.14 | -6.4 | 11.8 | -8.9 | 3.6 |
| 107 | Camilla | 08-Sep-04 | 06:41:20 | -5.77 | -8.28 | 0.74 | 300 | 1.19 | 0.12 | -7.2 | 8.1 | -8.5 | 4.5 |
| 107 | Camilla | 11-Sep-04 | 04:34:25 | -4.34 | -6.71 | 0.84 | 300 | 1.08 | 0.15 | -5.5 | 17.4 | -7.0 | 9.0 |
| 107 | Camilla | 13-Sep-04 | 03:42:52 | -4.64 | -7.5 | 0.74 | 300 | 1.11 | 0.14 | -6.3 | 12.4 | -7.8 | 6.1 |
| 107 | Camilla | 13-Sep-04 | 05:47:29 | -5.33 | -7.98 | 0.77 | 300 | 1.12 | 0.12 | -6.8 | 9.6 | -8.3 | 4.9 |
| 107 | Camilla | 14-Sep-04 | 04:09:30 | -5.88 | -8.82 | 0.82 | 300 | 1.08 | 0.13 | -7.8 | 6.1 | -8.9 | 3.7 |
| 107 | Camilla | 14-Sep-04 | 07:06:44 | -4.13 | -7.4 | 0.72 | 300 | 1.35 | 0.13 | -6.4 | 11.5 | -7.7 | 6.3 |
| 107 | Camilla | 14-Sep-04 | 07:14:32 | -5.36 | -7.15 | 0.77 | 300 | 1.39 | 0.13 | -6.2 | 12.8 | -7.5 | 7.0 |
| 107 | Camilla | 15-Sep-04 | 04:18:35 | -6.49 | -8.73 | 0.77 | 300 | 1.07 | 0.13 | -7.6 | 6.7 | -8.8 | 3.9 |
| 107 | Camilla | 15-Sep-04 | 04:26:56 | -8.11 | -9.06 | 0.81 | 300 | 1.07 | 0.13 | -8.0 | 5.7 | -9.2 | 3.2 |
| 107 | Camilla | 16-Sep-04 | 04:48:18 | -6.28 | -9.17 | 0.73 | 300 | 1.08 | 0.13 | -7.0 | 9.1 | -9.2 | 3.3 |
| 107 | Camilla | 07-Oct-04 | 01:47:18 | -3.7 | -6.87 | 1.27 | 180 | 1.10 | 0.18 | -5.0 | 22.1 | -6.9 | 9.2 |
| 107 | Camilla | 07-Oct-04 | 02:02:04 | -4.44 | -8.17 | 1.18 | 360 | 1.08 | 0.15 | -6.5 | 11.0 | -8.2 | 5.1 |
| 107 | Camilla | 07-Oct-04 | 02:12:52 | -5.48 | -7.24 | 0.9 | 360 | 1.07 | 0.18 | -5.6 | 16.6 | -7.2 | 8.2 |
| 107 | Camilla | 20-Oct-04 | 00:39:22 | -4.84 | -8.31 | 0.85 | 360 | 1.11 | 0.12 | -7.0 | 9.0 | -8.3 | 4.8 |
| 107 | Camilla | 25-Oct-04 | 06:52:46 | -4.27 | -7.86 | 0.73 | 180 | 1.12 | 0.11 | -6.2 | 12.7 | -7.9 | 6.0 |
| 107 | Camilla | 02-Nov-04 | 07:38:36 | -7.35 | -8.79 | 0.79 | 80 | 1.16 | 0.14 | -7.0 | 8.9 | -8.8 | 3.9 |
| 107 | Camilla | 02-Nov-04 | 15:39:39 | -10.04 | -8.24 | 0.59 | 80 | 1.35 | 0.11 | -7.2 | 8.1 | -8.2 | 5.2 |
| 107 | Camilla | 20-Dec-05 | 10:33:57 | -7.64 | -7.93 | 0.64 | 300 | 1.37 | 0.17 | -6.5 | 11.0 | -8.1 | 5.3 |
| 107 | Camilla | 21-Dec-05 | 09:05:52 | -6.85 | -9.2 | 0.74 | 300 | 1.10 | 0.13 | -7.4 | 7.4 | -9.6 | 2.7 |
| 107 | Camilla | 01-Jan-06 | 10:17:13 | -7.66 | -9.48 | 0.77 | 300 | 1.58 | 0.13 | -7.2 | 8.0 | -9.5 | 2.8 |
| 107 | Camilla | 06-Jan-06 | 09:38:56 | -9.75 | -8.4 | 0.5 | 300 | 1.45 | 0.15 | -6.9 | 9.3 | -8.3 | 4.8 |
| 107 | Camilla | 08-Jan-06 | 05:03:57 | -7.7 | -8.59 | 0.68 | 300 | 1.10 | 0.13 | -7.0 | 8.9 | -8.7 | 4.1 |
| 107 | Camilla | 09-Jan-06 | 05:20:13 | -7.25 | -9.35 | 0.66 | 300 | 1.07 | 0.12 | -6.9 | 9.3 | -8.9 | 3.7 |
| 107 | Camilla | 16-Jan-06 | 05:16:50 | -7.96 | -8.98 | 0.74 | 300 | 1.04 | 0.15 | -6.5 | 10.9 | -9.1 | 3.3 |

**Table 3d**: Search for moonlet companions around 762 Pulcova.

| ID | Name | Date | UT Time | α | Dm Limit | r Limit arcsec | Int Time s | Airmass | FWHM arcsec | Δm at 2/100xRHill | Diameter (km) at 2/100xRHill | Dm at 1/4xRHill | Diameter (km) at 2/100xRHill |
|---|---|---|---|---|---|---|---|---|---|---|---|---|---|
| 762 | Pulcova | 15-Jul-03 | 08:23:30 | -5.7 | -8.1 | 1.00 | 143.1 | 1.47 | 0.10 | -4.7 | 15.5 | -8.2 | 3.2 |
| 762 | Pulcova | 30-Oct-04 | 11:45:25 | -8.0 | -6.6 | 0.55 | 9 | 1.30 | 0.11 | -6.1 | 8.3 | -6.6 | 6.5 |
| 762 | Pulcova | 30-Oct-04 | 11:55:18 | -3.6 | -8.7 | 1.07 | 80 | 1.29 | 0.11 | -5.7 | 10.1 | -8.5 | 2.8 |
| 762 | Pulcova | 30-Oct-04 | 13:45:57 | -4.6 | -8.6 | 0.72 | 40 | 1.05 | 0.11 | -5.5 | 10.9 | -8.6 | 2.6 |
| 762 | Pulcova | 30-Oct-04 | 15:29:56 | -5.9 | -8.1 | 0.77 | 40 | 1.06 | 0.14 | -5.2 | 12.8 | -8.3 | 3.1 |
| 762 | Pulcova | 2-Nov-04 | 12:16:54 | -4.7 | -9.1 | 0.77 | 70 | 1.20 | 0.10 | -6.9 | 5.8 | -9.1 | 2.1 |
| 762 | Pulcova | 2-Nov-04 | 12:22:15 | -7.8 | -8.7 | 0.59 | 120 | 1.18 | 0.13 | -6.5 | 6.8 | -8.1 | 3.4 |
| 762 | Pulcova | 2-Nov-04 | 15:13:27 | -10.0 | -8.9 | 0.59 | 120 | 1.05 | 0.15 | -6.4 | 7.2 | -8.4 | 2.9 |
| 762 | Pulcova | 18-Mar-06 | 12:43:54 | -6.3 | -9.5 | 0.94 | 300 | 1.49 | 0.12 | -6.4 | 7.1 | -9.7 | 1.6 |
| 762 | Pulcova | 16-May-06 | 06:12:35 | -6.9 | -8.3 | 0.79 | 240 | 1.24 | 0.19 | -4.8 | 15.0 | -8.3 | 3.0 |
| 762 | Pulcova | 18-May-06 | 07:17:47 | -13.3 | -8.2 | 0.24 | 300 | 1.27 | 0.17 | -5.1 | 13.2 | -8.0 | 3.4 |
| 762 | Pulcova | 19-May-06 | 08:26:58 | -7.3 | -8.7 | 0.88 | 300 | 1.48 | 0.16 | -5.0 | 13.6 | -8.8 | 2.4 |
| 762 | Pulcova | 23-Jun-06 | 06:18:55 | -7.6 | -8.8 | 0.90 | 300 | 1.43 | 0.14 | -5.0 | 13.8 | -8.9 | 2.3 |

**Table 4a:**
Size, shape and orientation of Kalliope's primary and comparison with a refined model based on Kaasalainen *et al.* (2002) with a pole solution (λ = 197 ± 2° and β = -3 ± 2°) in ECJ2000 and $P_{spin}$ = 4.148299 h. The AO images were fitted by an ellipse function defined by its major axes (2a, 2b) and its orientation (from the celestial east, and counter-clockwise). The average diameter of 22 Kalliope ($D_{AO}$ = 196 km) is 9% larger than STM/IRAS radiometric measurement (Tedesco *et al.* 2002).

| ID | Name | Date | UT | 2a (mas) | 2b (mas) | 2a (km) | 2b (km) | **Observed** Orientation (deg) | a/b | **shape model** Orientation (deg) | a/b | DAO (km) |
|---|---|---|---|---|---|---|---|---|---|---|---|---|
| 22 | Kalliope | 14-Jan-03 | 08:53:59 | 93±6 | NA | 173±12 | NA | -19 | NA | -7 | 1.34 | 173 |
| 22 | Kalliope | 14-Jan-03 | 09:00:19 | 98±6 | 86±7 | 183±11 | 160±12 | -13 | 1.14 | -17 | 1.33 | 171 |
| 22 | Kalliope | 14-Jan-03 | 09:07:39 | 101±6 | 88±7 | 188±11 | 163±12 | -33 | 1.15 | -27 | 1.34 | 176 |
| 22 | Kalliope | 04-Jun-04 | 09:46:02 | 131±11 | 111±12 | 208±18 | 176±20 | -268 | 1.18 | -246 | 1.44 | 192 |
| 22 | Kalliope | 05-Jun-04 | 08:21:22 | 132±11 | 105±13 | 209±18 | 167±20 | -45 | 1.25 | -59 | 1.22 | 188 |
| 22 | Kalliope | 05-Jun-04 | 11:07:04 | 155±10 | 107±13 | 246±16 | 169±20 | -241 | 1.46 | -81 | 1.65 | 208 |
| 22 | Kalliope | 05-Jun-04 | 12:07:29 | 118±12 | 117±12 | 187±19 | 186±19 | -90 | 1.01 | 62 | 1.18 | 187 |
| 22 | Kalliope | 05-Jun-04 | 13:22:58 | 161±10 | 129±11 | 255±16 | 205±18 | -265 | 1.24 | -267 | 1.60 | 230 |
| 22 | Kalliope | 28-Jun-04 | 02:20:44 | 99±6 | NA | 162±10 | NA | -234 | NA | -251 | 1.56 | 162 |
| 22 | Kalliope | 24-Jul-04 | 04:06:17 | 116±5 | 106±6 | 207±10 | 190±10 | -27 | 1.09 | -37 | 1.14 | 198 |
| 22 | Kalliope | 26-Jul-04 | 01:26:47 | 113±6 | 103±6 | 203±10 | 185±11 | 74 | 1.10 | 52 | 1.15 | 194 |
| 22 | Kalliope | 26-Jul-04 | 04:51:14 | 130±5 | 117±5 | 235±9 | 210±10 | -236 | 1.12 | -265 | 1.59 | 223 |
| 22 | Kalliope | 07-Oct-04 | 00:45:10 | 120±5 | 104±6 | 298±13 | 259±15 | -45 | 1.15 | -69 | 1.50 | 278 |
| 22 | Kalliope | 12-Dec-06 | 13:41:59 | 156±1 | 106±3 | 188±1 | 128±4 | 74 | 1.47 | 67 | 1.44 | 158 |

**Table 4b:**
Size , shape and orientation of Eugenia's primary and comparison with Kaasalainen *et al.* (2002) model with a pole solution (λ = 124°± 2°, and β=-30°± 1°) in ECJ2000 and $P_{spin}$ = 5.6991 h . The AO images were fitted by an ellipse function defined by its major axes (2a, 2b) and its orientation (from the celestial east, and counter-clockwise). The a/b ratio and the average diameter ($D_{avg}$) are also labeled. The average diameter of 45 Eugenia ($D_{AO}$ = 193 km) is 9% smaller than STM/IRAS radiometric measurement (Tedesco *et al.* 2002).

| ID | Name | Date | UT | 2a (mas) | 2b (mas) | 2a (km) | 2b (km) | Observed Orientation (deg) | Observed a/b | shape model Orientation (deg) | shape model a/b | DAO (km) |
|---|---|---|---|---|---|---|---|---|---|---|---|---|
| 45 | Eugenia | 06-Dec-03 | 12:42:32 | 136±2 | 120±3 | 212±3 | 186±4 | 78 | 1.13 | 51 | 1.24 | 199 |
| 45 | Eugenia | 06-Dec-03 | 14:20:10 | 150±1 | 128±2 | 232±2 | 198±3 | -22 | 1.16 | -44 | 1.23 | 215 |
| 45 | Eugenia | 04-Jan-04 | 07:03:54 | 137±5 | 115±5 | 186±6 | 156±7 | 36 | 1.20 | 7 | 1.30 | 171 |
| 45 | Eugenia | 05-Jan-04 | 05:56:03 | 164±3 | 132±5 | 221±4 | 179±6 | 17 | 1.23 | 3 | 1.29 | 200 |
| 45 | Eugenia | 06-Jan-04 | 05:10:22 | 154±4 | 135±5 | 207±5 | 182±6 | -13 | 1.14 | -23 | 1.27 | 195 |
| 45 | Eugenia | 07-Jan-04 | 05:43:15 | 165±3 | 135±5 | 221±4 | 181±6 | 61 | 1.22 | 42 | 1.30 | 201 |
| 45 | Eugenia | 07-Jan-04 | 06:41:16 | 194±2 | 165±3 | 260±3 | 222±4 | 17 | 1.17 | -10 | 1.27 | 241 |
| 45 | Eugenia | 11-Feb-04 | 03:11:20 | 139±4 | 118±5 | 186±6 | 158±7 | -85 | 1.18 | -120 | 1.24 | 172 |
| 45 | Eugenia | 11-Feb-04 | 03:20:05 | 158±4 | 135±5 | 212±5 | 180±6 | 79 | 1.17 | 50 | 1.26 | 196 |
| 45 | Eugenia | 11-Feb-04 | 03:28:16 | 165±3 | 134±5 | 221±4 | 179±6 | 73 | 1.24 | 41 | 1.27 | 200 |
| 45 | Eugenia | 11-Feb-04 | 03:43:53 | 168±3 | 139±4 | 225±4 | 187±6 | 58 | 1.21 | 26 | 1.30 | 206 |
| 45 | Eugenia | 12-Feb-04 | 02:52:35 | 150±4 | 107±6 | 202±5 | 144±8 | 24 | 1.40 | 7 | 1.31 | 173 |
| 45 | Eugenia | 12-Feb-04 | 03:00:54 | 163±3 | 121±5 | 218±5 | 162±7 | 15 | 1.35 | 0 | 1.31 | 190 |
| 45 | Eugenia | 12-Feb-04 | 03:08:06 | 161±3 | 129±5 | 217±5 | 173±7 | 10 | 1.26 | -6 | 1.31 | 195 |
| 45 | Eugenia | 12-Feb-04 | 03:33:25 | 159±4 | 130±5 | 213±5 | 174±6 | -5 | 1.22 | -28 | 1.26 | 194 |
| 45 | Eugenia | 12-Feb-04 | 04:09:55 | 155±4 | 128±5 | 208±5 | 172±7 | -31 | 1.22 | 69 | 1.19 | 190 |
| 45 | Eugenia | 14-Feb-04 | 03:41:49 | 158±4 | 115±5 | 213±5 | 155±7 | -2 | 1.37 | -17 | 1.30 | 184 |
| 45 | Eugenia | 15-Feb-04 | 03:30:32 | 167±3 | 121±5 | 225±4 | 164±7 | -45 | 1.38 | 80 | 1.21 | 195 |
| 45 | Eugenia | 16-Feb-04 | 03:42:16 | 150±4 | 118±5 | 204±5 | 160±7 | 22 | 1.27 | 10 | 1.31 | 182 |
| 45 | Eugenia | 04-Mar-04 | 03:32:56 | 135±5 | 109±6 | 197±7 | 159±8 | -1 | 1.23 | -13 | 1.31 | 178 |
| 45 | Eugenia | 04-Mar-04 | 03:49:08 | 160±4 | 135±5 | 232±5 | 196±7 | 8 | 1.18 | -27 | 1.31 | 214 |
| 45 | Eugenia | 07-Mar-04 | 03:17:19 | 140±4 | 113±6 | 207±6 | 166±8 | -25 | 1.25 | -34 | 1.26 | 187 |
| 45 | Eugenia | 07-Mar-04 | 03:30:42 | 143±4 | 124±5 | 210±6 | 183±7 | -20 | 1.14 | 48 | 1.24 | 197 |
| 45 | Eugenia | 08-Mar-04 | 00:56:15 | 139±4 | 115±5 | 206±7 | 170±8 | 40 | 1.21 | 26 | 1.28 | 188 |
| 45 | Eugenia | 08-Mar-04 | 01:06:29 | 136±5 | 116±5 | 202±7 | 172±8 | 45 | 1.17 | 16 | 1.30 | 187 |
| 45 | Eugenia | 09-Mar-04 | 02:09:22 | 119±5 | NA | 177±8 | NA | -77 | NA | -129 | 1.15 | 177 |
| 45 | Eugenia | 10-Mar-04 | 01:26:36 | 125±5 | 107±6 | 187±8 | 160±9 | 67 | 1.16 | 17 | 1.24 | 174 |
| 45 | Eugenia | 11-Mar-04 | 00:31:09 | 144±4 | 123±5 | 216±6 | 185±8 | 27 | 1.16 | 2 | 1.28 | 200 |
| 45 | Eugenia | 11-Mar-04 | 00:45:46 | 151±4 | 123±5 | 227±6 | 185±8 | 7 | 1.23 | -10 | 1.31 | 206 |
| 45 | Eugenia | 12-Mar-04 | 01:53:32 | 148±4 | 131±5 | 223±6 | 198±7 | 26 | 1.12 | 22 | 1.28 | 210 |
| 45 | Eugenia | 15-Jan-05 | 15:02:35 | 115±3 | 94±4 | 224±6 | 184±7 | 37 | 1.21 | 82 | 1.42 | 204 |
| 45 | Eugenia | 17-Jul-05 | 07:25:44 | 172±1 | 94±4 | 236±0 | 129±5 | 74 | 1.82 | 82 | 1.87 | 183 |
| 45 | Eugenia | 03-Aug-06 | 14:07:40 | 120±3 | 106±3 | 173±4 | 152±5 | 94 | 1.13 | -95 | 1.46 | 163 |
| 45 | Eugenia | 09-Sep-07 | 14:31:52 | 100±3 | 81±4 | 206±7 | 167±9 | -10 | 1.23 | 47 | 1.32 | 187 |
| 45 | Eugenia | 09-Sep-07 | 15:20:50 | 115±3 | 77±4 | 237±6 | 158±9 | 133 | 1.51 | 67 | 1.49 | 197 |

**Table 4c:**

Size, shape and orientation of Camilla's primary and comparison with Torppa *et al.* (2003) model with a pole solution (λ = 51° and β = 72°) in ECJ2000 and $P_{spin}$ = 4.84393 h. The AO images were fitted by an ellipse function defined by its major axes (2a, 2b) and its orientation (from the celestial east, and counter-clockwise). The a/b ratio and the average diameter ($D_{avg}$) are also labeled. The average diameter of 107 Camillla ($D_{AO}$ = 246 km) is 10% larger than STM/IRAS radiometric measurement (Tedesco *et al.* 2002), but similar to NEATM/IRAS diameter (See Table 1).

| ID | Name | Date | UT | 2a (mas) | 2b (mas) | 2a (km) | 2b (km) | Observed Orientation (deg) | a/b | shape model Orientation (deg) | a/b | DAO (km) |
|---|---|---|---|---|---|---|---|---|---|---|---|---|
| 107 | Camilla | 01-Sep-04 | 05:07:39 | 123±5 | 94±6 | 238±10 | 182±12 | 11 | 1.46 | 18 | 1.16 | 210 |
| 107 | Camilla | 01-Sep-04 | 05:17:23 | 127±5 | 96±6 | 246±10 | 186±12 | 15 | 1.32 | 19 | 1.23 | 216 |
| 107 | Camilla | 01-Sep-04 | 08:06:44 | 155±4 | 100±6 | 300±7 | 193±12 | 10 | 1.56 | 18 | 1.45 | 246 |
| 107 | Camilla | 03-Sep-04 | 06:51:57 | 173±3 | 120±5 | 334±6 | 232±10 | 11 | 1.44 | 16 | 1.57 | 283 |
| 107 | Camilla | 04-Sep-04 | 09:02:35 | 177±3 | 122±5 | 342±5 | 235±10 | 8 | 1.45 | 15 | 1.57 | 289 |
| 107 | Camilla | 05-Sep-04 | 04:28:20 | 176±3 | 113±6 | 340±5 | 218±11 | 5 | 1.56 | 15 | 1.60 | 279 |
| 107 | Camilla | 06-Sep-04 | 03:35:31 | 127±5 | 96±6 | 244±10 | 185±12 | 15 | 1.32 | 13 | 1.15 | 215 |
| 107 | Camilla | 07-Sep-04 | 02:20:10 | 165±3 | 105±6 | 317±6 | 202±11 | 15 | 1.57 | 20 | 1.54 | 259 |
| 107 | Camilla | 08-Sep-04 | 06:41:20 | 126±5 | 93±6 | 243±10 | 179±12 | 13 | 1.35 | 17 | 1.18 | 211 |
| 107 | Camilla | 11-Sep-04 | 04:34:25 | 155±4 | 115±5 | 298±7 | 221±10 | 14 | 1.35 | 2 | 1.16 | 260 |
| 107 | Camilla | 13-Sep-04 | 03:42:52 | 165±3 | 102±6 | 316±6 | 195±11 | 17 | 1.62 | 19 | 1.57 | 255 |
| 107 | Camilla | 13-Sep-04 | 05:47:29 | 136±5 | 103±6 | 262±9 | 198±11 | 19 | 1.32 | 19 | 1.40 | 230 |
| 107 | Camilla | 14-Sep-04 | 04:09:30 | 172±3 | 89±6 | 330±6 | 171±12 | 16 | 1.92 | 18 | 1.60 | 251 |
| 107 | Camilla | 14-Sep-04 | 07:06:44 | 145±4 | 92±6 | 279±8 | 177±12 | 4 | 1.58 | 12 | 1.48 | 228 |
| 107 | Camilla | 14-Sep-04 | 07:14:32 | 149±4 | 106±6 | 285±8 | 203±11 | 5 | 1.41 | 13 | 1.37 | 244 |
| 107 | Camilla | 15-Sep-04 | 04:18:35 | 167±3 | 85±7 | 321±6 | 164±13 | 14 | 1.96 | 18 | 1.60 | 242 |
| 107 | Camilla | 15-Sep-04 | 04:26:56 | 164±3 | 95±6 | 316±6 | 183±12 | 16 | 1.72 | 18 | 1.60 | 249 |
| 107 | Camilla | 16-Sep-04 | 04:48:18 | 168±3 | 97±6 | 322±6 | 187±12 | 13 | 1.73 | 17 | 1.59 | 254 |
| 107 | Camilla | 07-Oct-04 | 02:02:04 | 177±3 | 117±5 | 349±5 | 232±11 | 12 | 1.50 | 14 | 1.62 | 291 |
| 107 | Camilla | 20-Oct-04 | 00:39:22 | 133±5 | 102±6 | 273±10 | 210±12 | 14 | 1.30 | 19 | 1.32 | 242 |
| 107 | Camilla | 25-Oct-04 | 06:52:46 | 119±3 | 91±4 | 249±6 | 189±8 | 19 | 1.31 | 28 | 1.15 | 219 |
| 107 | Camilla | 02-Nov-04 | 15:39:39 | 126±12 | 122±12 | 272±25 | 264±25 | -47 | 1.03 | 11 | 1.31 | 268 |
| 107 | Camilla | 21-Dec-05 | 09:05:52 | 142±11 | 107±13 | 262±20 | 197±23 | 13 | 1.33 | 15 | 1.34 | 229 |
| 107 | Camilla | 01-Jan-06 | 10:17:13 | 150±11 | 105±13 | 289±20 | 202±24 | -15 | 1.43 | -12 | 1.48 | 246 |
| 107 | Camilla | 08-Jan-06 | 05:03:57 | 127±12 | 115±12 | 252±23 | 228±24 | -3 | 1.11 | -29 | 1.11 | 240 |
| 107 | Camilla | 09-Jan-06 | 05:20:13 | 132±11 | 106±13 | 262±23 | 210±25 | -10 | 1.25 | -27 | 1.09 | 236 |

**Table 4d:** Size, shape and orientation of Pulcova's primary from data collected with Gemini North AO system. No shape models are available for this asteroid. The average diameter of the primary is 216 ± 40 km.

| ID | Name | Date | UT | 2a (mas) | 2b (mas) | 2a (km) | 2b (km) | Observed Orientation (deg) | a/b | DAO (km) |
|---|---|---|---|---|---|---|---|---|---|---|
| 762 | Pulcova | 30-Oct-04 | 11:55:18 | 133±11 | 108±13 | 229±19 | 186±22 | 20 | 1.23 | 208 |
| 762 | Pulcova | 30-Oct-04 | 13:45:57 | 175±10 | 146±11 | 302±16 | 251±19 | -29 | 1.20 | 276 |
| 762 | Pulcova | 2-Nov-04 | 12:16:54 | 120±12 | NA | 204±20 | NA | -38 | NA | 204 |
| 762 | Pulcova | 18-Mar-06 | 12:43:54 | 133±11 | 108±13 | 194±16 | 157±18 | 20 | 1.23 | 176 |

**Table 5a:**

Characteristics of Linus, moonlet of 22 Kalliope measured on the AO images collected with VLT-UT4 and Gemini North in 2003-2004. Additional observations collected in 2001-2003 using Lick-3m and Palomar-5m telescopes (Marchis *et al.* 2003, Margot and Brown, 2003) were added to the orbit analysis. The X and Y relative positions with respect to the primary of the system are measured by fitting their centroid profile with a Moffat-Gauss function. The 1-σ error on the fitted astrometric positions is added. The size of satellite is estimated by calculating the integrated flux ratio of the primary and the secondary and also by measuring directly the size of the primary on the resolved AO images (see Table 4a).

| ID | Primary Name | Date | UT | X | Y | X Error | Y Error | separation | Δm (integrated) | Satellite Size |
|----|---|---|---|---|---|---|---|---|---|---|
|  |  |  |  | arcsec | arcsec | arcsec | arcsec | arcsec |  | km |
| 22 | Kalliope | 31-Aug-01 | 14:32:38 | -0.111 | 0.555 |  |  | 0.566 | NA | NA |
| 22 | Kalliope | 01-Sep-01 | 13:40:48 | 0.234 | 0.072 |  |  | 0.245 | NA | NA |
| 22 | Kalliope | 07-Sep-01 | 12:31:41 | -0.225 | 0.449 |  |  | 0.502 | NA | NA |
| 22 | Kalliope | 08-Sep-01 | 12:11:31 | 0.176 | 0.285 |  |  | 0.335 | NA | NA |
| 22 | Kalliope | 09-Sep-01 | 10:58:04 | 0.149 | -0.525 |  |  | 0.546 | NA | NA |
| 22 | Kalliope | 03-Oct-01 | 13:11:01 | 0.107 | 0.461 | 0.023 | 0.025 | 0.474 | NA | NA |
| 22 | Kalliope | 03-Oct-01 | 13:23:52 | 0.123 | 0.450 | 0.012 | 0.013 | 0.466 | NA | NA |
| 22 | Kalliope | 04-Oct-01 | 09:47:16 | 0.147 | -0.489 | 0.009 | 0.011 | 0.511 | NA | NA |
| 22 | Kalliope | 04-Oct-01 | 10:56:08 | 0.148 | -0.530 | 0.011 | 0.010 | 0.551 | NA | NA |
| 22 | Kalliope | 04-Oct-01 | 13:10:21 | 0.118 | -0.593 | 0.010 | 0.015 | 0.604 | NA | NA |
| 22 | Kalliope | 06-Oct-01 | 10:09:23 | -0.109 | 0.631 | 0.013 | 0.022 | 0.640 | NA | NA |
| 22 | Kalliope | 06-Oct-01 | 10:38:26 | -0.093 | 0.638 | 0.009 | 0.012 | 0.645 | NA | NA |
| 22 | Kalliope | 06-Oct-01 | 12:50:44 | -0.078 | 0.686 | 0.014 | 0.015 | 0.691 | NA | NA |
| 22 | Kalliope | 02-Nov-01 | 07:23:22 | 0.145 | -0.685 | 0.002 | 0.002 | 0.700 | NA | NA |
| 22 | Kalliope | 02-Nov-01 | 07:30:25 | 0.141 | -0.675 | 0.004 | 0.004 | 0.690 | NA | NA |
| 22 | Kalliope | 02-Nov-01 | 11:33:26 | 0.075 | -0.768 | 0.003 | 0.003 | 0.771 | NA | NA |
| 22 | Kalliope | 02-Nov-01 | 11:40:52 | 0.074 | -0.770 | 0.003 | 0.003 | 0.773 | NA | NA |
| 22 | Kalliope | 03-Nov-01 | 07:15:36 | -0.197 | -0.269 | 0.003 | 0.005 | 0.333 | NA | NA |
| 22 | Kalliope | 03-Nov-01 | 07:19:01 | -0.208 | -0.261 | 0.004 | 0.005 | 0.334 | NA | NA |
| 22 | Kalliope | 10-Dec-01 | 06:15:50 | -0.151 | 0.874 |  |  | 0.887 | NA | NA |
| 22 | Kalliope | 02-Apr-02 | 04:09:15 | 0.106 | -0.421 |  |  | 0.434 | NA | NA |
| 22 | Kalliope | 29-Dec-02 | 14:12:29 | 0.505 | 0.131 |  |  | 0.522 | NA | NA |
| 22 | Kalliope | 29-Dec-02 | 16:16:19 | 0.493 | 0.215 |  |  | 0.538 | NA | NA |
| 22 | Kalliope | 14-Jan-03 | 08:53:59 | -0.530 | 0.241 | 0.0047 | 0.0043 | 0.582 | -3.3 | 41 |
| 22 | Kalliope | 14-Jan-03 | 09:00:19 | -0.526 | 0.239 | 0.0050 | 0.0047 | 0.578 | -3.8 | 33 |
| 22 | Kalliope | 14-Jan-03 | 09:07:39 | -0.530 | 0.232 | 0.0056 | 0.0052 | 0.578 | -3.9 | 32 |
| 22 | Kalliope | 14-Feb-03 | 08:21:57 | 0.625 | 0.161 | 0.0194 | 0.0138 | 0.646 | -5.0 | 26 |
| 22 | Kalliope | 15-Jul-03 | 00:58:10 | 0.425 | -0.126 | 0.0224 | 0.0115 | 0.443 | NA | NA |
| 22 | Kalliope | 16-Jul-03 | 00:01:32 | 0.105 | 0.438 | 0.0110 | 0.0237 | 0.450 | -3.5 | 41 |
| 22 | Kalliope | 03-Mar-04 | 10:08:26 | -0.188 | -0.355 | 0.0074 | 0.0073 | 0.402 | -5.5 | 20 |
| 22 | Kalliope | 04-Mar-04 | 10:13:54 | 0.158 | -0.245 | 0.0089 | 0.0097 | 0.291 | -4.9 | 26 |
| 22 | Kalliope | 04-Jun-04 | 09:46:02 | -0.340 | -0.040 | 0.0108 | 0.0100 | 0.343 | -3.2 | 43 |
| 22 | Kalliope | 05-Jun-04 | 08:21:22 | -0.051 | -0.703 | 0.0086 | 0.0098 | 0.705 | -3.1 | 43 |
| 22 | Kalliope | 05-Jun-04 | 11:07:04 | 0.040 | -0.643 | 0.0084 | 0.0096 | 0.644 | -3.7 | 36 |
| 22 | Kalliope | 05-Jun-04 | 12:07:29 | 0.031 | -0.697 | 0.0039 | 0.0044 | 0.698 | -3.2 | 41 |
| 22 | Kalliope | 05-Jun-04 | 13:22:58 | 0.069 | -0.624 | 0.0091 | 0.0106 | 0.628 | -4.6 | 26 |
| 22 | Kalliope | 28-Jun-04 | 02:20:44 | 0.389 | 0.346 | 0.0042 | 0.0043 | 0.520 | -3.3 | 39 |
| 22 | Kalliope | 24-Jul-04 | 04:06:17 | -0.103 | 0.492 | 0.0037 | 0.0042 | 0.502 | -4.2 | 29 |
| 22 | Kalliope | 25-Jul-04 | 05:40:48 | -0.307 | -0.444 | 0.0052 | 0.0054 | 0.540 | -6.1 | 16 |
| 22 | Kalliope | 25-Jul-04 | 05:47:42 | -0.341 | -0.431 | 0.0037 | 0.0037 | 0.549 | -6.7 | 15 |
| 22 | Kalliope | 26-Jul-04 | 01:26:47 | 0.154 | -0.440 | 0.0049 | 0.0052 | 0.467 | -4.1 | 28 |
| 22 | Kalliope | 26-Jul-04 | 03:23:10 | 0.214 | -0.378 | 0.0051 | 0.0052 | 0.434 | -6.7 | 12 |
| 22 | Kalliope | 26-Jul-04 | 04:51:14 | 0.239 | -0.329 | 0.0072 | 0.0078 | 0.407 | -6.6 | 10 |
| 22 | Kalliope | 07-Oct-04 | 00:31:15 | 0.150 | 0.347 | 0.0053 | 0.0063 | 0.378 | -6.0 | 24 |
| 22 | Kalliope | 07-Oct-04 | 00:45:10 | 0.159 | 0.366 | 0.0031 | 0.0035 | 0.399 | -4.8 | 29 |
| 22 | Kalliope | 12-Dec-06 | 13:41:59 | 0.153 | -0.854 | 0.0021 | 0.0019 | 0.868 | -4.2 | 26 |

**Table 5b:**

Characteristic of the moonlet orbiting around 45 Eugenia (Petit-Prince) measured on the AO images collected with VLT-UT4 in 2003-2004, then Keck in 2005-2007. Three additional points extracted from CFHT Science Archive taken with PUEO instrument in 1998 were added for the orbit analysis. The size of satellite is estimated by calculating the integrated flux ratio of the primary and the secondary and also by measuring directly the size of the primary on the resolved AO images (see Table 4b).

| ID | Primary Name | Date | UT | X | Y | X Error | Y Error | separation | Δm (integrated) | Satellite Size |
|----|----|----|----|----|----|----|----|----|----|----|
| | | | | arcsec | arcsec | arcsec | arcsec | arcsec | | km |
| 45 | Eugenia | 01-Nov-98 | 12:58:39 | 0.461 | 0.609 | NA | NA | 0.764 | NA | NA |
| 45 | Eugenia | 06-Nov-98 | 14:55:13 | 0.207 | 0.740 | NA | NA | 0.768 | NA | NA |
| 45 | Eugenia | 06-Nov-98 | 15:23:43 | 0.176 | 0.714 | NA | NA | 0.735 | NA | NA |
| 45 | Eugenia | 06-Dec-03 | 12:42:32 | -0.638 | 0.435 | 0.002 | 0.002 | 0.772 | -7.4 | 7 |
| 45 | Eugenia | 06-Dec-03 | 14:20:10 | -0.673 | 0.365 | 0.004 | 0.004 | 0.766 | -8.4 | 5 |
| 45 | Eugenia | 04-Jan-04 | 7:03:54 | -0.827 | -0.012 | 0.007 | 0.005 | 0.827 | -7.0 | 9 |
| 45 | Eugenia | 05-Jan-04 | 5:56:03 | -0.118 | -0.709 | 0.005 | 0.007 | 0.719 | -6.7 | 10 |
| 45 | Eugenia | 06-Jan-04 | 5:10:22 | 0.788 | -0.410 | 0.006 | 0.006 | 0.888 | -7.0 | 9 |
| 45 | Eugenia | 07-Jan-04 | 5:43:15 | 0.522 | 0.522 | 0.005 | 0.005 | 0.738 | -6.8 | 10 |
| 45 | Eugenia | 11-Feb-04 | 3:11:20 | -0.822 | -0.192 | 0.005 | 0.005 | 0.844 | -6.9 | 9 |
| 45 | Eugenia | 11-Feb-04 | 3:20:05 | -0.826 | -0.197 | 0.007 | 0.006 | 0.849 | -8.4 | 5 |
| 45 | Eugenia | 11-Feb-04 | 3:43:53 | -0.834 | -0.210 | 0.005 | 0.004 | 0.860 | -8.4 | 5 |
| 45 | Eugenia | 12-Feb-04 | 2:52:35 | 0.081 | -0.725 | 0.004 | 0.005 | 0.730 | -6.6 | 10 |
| 45 | Eugenia | 12-Feb-04 | 3:00:54 | 0.083 | -0.727 | 0.006 | 0.007 | 0.732 | -7.9 | 6 |
| 45 | Eugenia | 12-Feb-04 | 3:33:25 | 0.115 | -0.723 | 0.006 | 0.009 | 0.732 | -7.3 | 8 |
| 45 | Eugenia | 12-Feb-04 | 4:09:55 | 0.131 | -0.720 | 0.005 | 0.005 | 0.732 | -8.1 | 5 |
| 45 | Eugenia | 14-Feb-04 | 3:41:49 | 0.317 | 0.658 | 0.006 | 0.007 | 0.730 | -6.4 | 11 |
| 45 | Eugenia | 15-Feb-04 | 3:30:32 | -0.714 | 0.426 | 0.008 | 0.007 | 0.832 | -6.6 | 10 |
| 45 | Eugenia | 16-Feb-04 | 3:42:16 | -0.622 | -0.456 | 0.005 | 0.004 | 0.771 | -6.6 | 10 |
| 45 | Eugenia | 01-Mar-04 | 4:03:50 | -0.672 | -0.342 | 0.006 | 0.006 | 0.754 | -10.5 | 3 |
| 45 | Eugenia | 03-Mar-04 | 4:10:27 | 0.826 | 0.052 | 0.008 | 0.007 | 0.828 | -9.3 | 4 |
| 45 | Eugenia | 04-Mar-04 | 3:32:56 | 0.139 | 0.626 | 0.006 | 0.008 | 0.641 | -7.3 | 8 |
| 45 | Eugenia | 04-Mar-04 | 3:49:08 | 0.140 | 0.627 | 0.005 | 0.006 | 0.642 | -9.2 | 3 |
| 45 | Eugenia | 07-Mar-04 | 3:17:19 | 0.517 | -0.508 | 0.004 | 0.004 | 0.725 | -7.8 | 6 |
| 45 | Eugenia | 07-Mar-04 | 3:30:42 | 0.533 | -0.498 | 0.005 | 0.004 | 0.730 | -9.0 | 4 |
| 45 | Eugenia | 08-Mar-04 | 0:56:15 | 0.770 | 0.172 | 0.004 | 0.004 | 0.789 | -7.4 | 7 |
| 45 | Eugenia | 08-Mar-04 | 1:06:29 | 0.763 | 0.185 | 0.007 | 0.006 | 0.785 | -8.4 | 5 |
| 45 | Eugenia | 09-Mar-04 | 2:09:22 | -0.105 | 0.624 | 0.006 | 0.007 | 0.633 | -6.5 | 10 |
| 45 | Eugenia | 10-Mar-04 | 1:26:36 | -0.771 | 0.089 | 0.009 | 0.009 | 0.776 | -7.2 | 8 |
| 45 | Eugenia | 11-Mar-04 | 0:45:46 | -0.289 | -0.560 | 0.005 | 0.004 | 0.630 | -9.4 | 3 |
| 45 | Eugenia | 17-Jul-05 | 7:25:44 | 0.328 | 0.095 | 0.005 | 0.004 | 0.341 | -6.8 | 8 |
| 45 | Eugenia | 03-Aug-06 | 14:07:40 | 0.149 | -0.636 | 0.006 | 0.007 | 0.653 | -8.6 | 4 |
| 45 | Eugenia | 9-Sep-07 | 14:31:52 | 0.412 | 0.229 | 0.004 | 0.004 | 0.472 | -7.1 | 8 |
| 45 | Eugenia | 9-Sep-07 | 15:20:50 | 0.420 | 0.265 | 0.004 | 0.004 | 0.497 | -7.3 | 7 |

**Table 5c**: Characteristics of the moonlet of 107 Camilla (named S/2002(107)1) measured on the AO images collected mainly with VLT/Yepun in 2004 and Gemini North in 2006. The average diameter of the moonlet companion is 16±6 km.

| ID | Primary Name | Date | UT | X | Y | X Error | Y Error | separation | Δm (integrated) | Satellite Size |
|----|---|---|---|---|---|---|---|---|---|---|
|  |  |  |  | arcsec | arcsec | arcsec | arcsec | arcsec |  | km |
| 107 | Camilla | 15-Jul-03 | 07:37:26 | -0.555 | 0.188 | 0.0216 | 0.0167 | 0.586 | -5.9 | 37 |
| 107 | Camilla | 1-Sep-04 | 05:07:39 | 0.490 | -0.162 | 0.0049 | 0.0045 | 0.516 | -5.3 | 19 |
| 107 | Camilla | 1-Sep-04 | 05:17:23 | 0.488 | -0.160 | 0.0085 | 0.0067 | 0.514 | -6.4 | 13 |
| 107 | Camilla | 1-Sep-04 | 08:06:44 | 0.557 | -0.161 | 0.0079 | 0.0056 | 0.580 | -7.0 | 11 |
| 107 | Camilla | 3-Sep-04 | 06:51:57 | -0.631 | 0.171 | 0.0035 | 0.0034 | 0.653 | -6.9 | 13 |
| 107 | Camilla | 5-Sep-04 | 04:28:20 | 0.611 | -0.142 | 0.0074 | 0.0056 | 0.627 | -7.5 | 9 |
| 107 | Camilla | 7-Sep-04 | 02:20:10 | -0.646 | 0.157 | 0.0071 | 0.0053 | 0.665 | -5.9 | 18 |
| 107 | Camilla | 11-Sep-04 | 04:34:25 | -0.539 | 0.092 | 0.0078 | 0.0055 | 0.547 | -7.1 | 11 |
| 107 | Camilla | 13-Sep-04 | 03:42:52 | 0.461 | -0.086 | 0.0087 | 0.0053 | 0.469 | -6.9 | 12 |
| 107 | Camilla | 13-Sep-04 | 05:47:29 | 0.393 | -0.048 | 0.0135 | 0.0047 | 0.395 | NA | NA |
| 107 | Camilla | 14-Sep-04 | 04:09:30 | -0.496 | 0.145 | 0.0034 | 0.0029 | 0.517 | -5.4 | 22 |
| 107 | Camilla | 14-Sep-04 | 07:06:44 | -0.567 | 0.159 | 0.0034 | 0.0033 | 0.589 | -6.1 | 16 |
| 107 | Camilla | 14-Sep-04 | 07:14:32 | -0.570 | 0.171 | 0.0061 | 0.0052 | 0.595 | -6.7 | 12 |
| 107 | Camilla | 15-Sep-04 | 04:18:35 | -0.304 | 0.036 | 0.0164 | 0.0058 | 0.306 | -5.3 | 22 |
| 107 | Camilla | 15-Sep-04 | 04:26:56 | -0.343 | 0.063 | 0.0124 | 0.0067 | 0.349 | -5.9 | 18 |
| 107 | Camilla | 16-Sep-04 | 04:48:18 | 0.583 | -0.179 | 0.0058 | 0.0049 | 0.610 | -5.3 | 22 |
| 107 | Camilla | 7-Oct-04 | 02:02:04 | -0.536 | 0.105 | 0.0115 | 0.0073 | 0.546 | -7.1 | 11 |
| 107 | Camilla | 20-Oct-04 | 00:39:22 | 0.554 | -0.140 | 0.0098 | 0.0068 | 0.572 | -6.2 | 14 |
| 107 | Camilla | 25-Oct-04 | 06:52:46 | -0.571 | 0.150 | 0.0042 | 0.0038 | 0.590 | -7.1 | 9 |
| 107 | Camilla | 21-Dec-05 | 09:05:52 | 0.705 | 0.011 | 0.0115 | 0.0096 | 0.705 | -6.0 | 16 |
| 107 | Camilla | 1-Jan-06 | 10:17:13 | 0.664 | -0.039 | 0.0114 | 0.0094 | 0.665 | -6.4 | 13 |
| 107 | Camilla | 9-Jan-06 | 05:20:13 | 0.576 | 0.153 | 0.0109 | 0.0091 | 0.596 | -5.7 | 18 |
| 107 | Camilla | 16-Jan-06 | 05:16:50 | 0.606 | -0.013 | 0.0150 | 0.0105 | 0.606 | -6.8 | 13 |

**Table 5d:** Characteristics of the moonlet of 762 Pulcova (named S/2002(762)1) measured on the AO images collected with VLT in 2003 & Gemini North telescope in 2004 and 2006. The X and Y relative positions with respect to the primary of the system are measured by fitting their centroid profile with a Moffat-Gauss function. The size of the moonlet is 19 ± 7 km.

| ID | Primary Name | Date | UT | X | Y | X Error | Y Error | separation | Δm (integrated) | Satellite Size |
|---|---|---|---|---|---|---|---|---|---|---|
| | | | | arcsec | arcsec | arcsec | arcsec | arcsec | | km |
| 762 | Pulcova | 15-Jul-03 | 8:23:30 | -0.271 | 0.027 | 0.0110 | 0.0099 | 0.272 | -4.2 | 34.4 |
| 762 | Pulcova | 30-Oct-04 | 11:55:18 | -0.329 | -0.198 | 0.0119 | 0.0105 | 0.383 | -6.5 | 9.31 |
| 762 | Pulcova | 30-Oct-04 | 13:45:57 | -0.358 | -0.225 | 0.0108 | 0.0103 | 0.423 | -5.2 | 18.3 |
| 762 | Pulcova | 30-Oct-04 | 15:29:56 | -0.356 | -0.232 | 0.0082 | 0.0071 | 0.425 | -6.0 | 15.3 |
| 762 | Pulcova | 18-Mar-06 | 12:43:54 | -0.182 | -0.351 | 0.0098 | 0.0102 | 0.395 | -4.6 | 20.7 |
| 762 | Pulcova | 28-Apr-06 | 08:32:21 | -0.428 | -0.078 | NA | NA | 0.435 | NA | NA |
| 762 | Pulcova | 16-May-06 | 06:12:35 | 0.332 | -0.215 | 0.0155 | 0.0130 | 0.396 | -6.3 | 17.4 |
| 762 | Pulcova | 18-May-06 | 07:17:47 | -0.321 | 0.281 | 0.0139 | 0.0128 | 0.427 | -6.7 | 12.8 |
| 762 | Pulcova | 19-May-06 | 08:26:58 | -0.258 | -0.215 | 0.0143 | 0.0132 | 0.336 | -5.7 | 19.1 |
| 762 | Pulcova | 23-Jun-06 | 06:18:55 | 0.269 | 0.133 | 0.0166 | 0.0123 | 0.300 | -5.8 | 19.2 |

**Table 6:** Best-fitted orbital elements of the asteroidal companions of 22 Kalliope, 45 Eugenia, 107 Camilla, and 762 Pulcova. The orbits of the satellite and its relative location with respect to the primary are displayed in Fig. 3a-d.

|  | Kalliope I Linus | Eugenia I Petit-Prince | S/2002(107)1 | S/2001(762)1 |
|---|---|---|---|---|
| Period (days) | 3.596± 0.001 | 4.765 ± 0.001 | 3.722 ± 0.003 | 4.438 ± 0.001 |
| Semi-major axis (km) | 1095 ± 11 | 1180 ± 8 | 1250 ± 10 | 703 ± 14 |
| Eccentricity | <0.002 | <0.002 | <0.002 | 0.03 ± 0.01 |
| Inclination in J2000 (degree) | 99.6 ± 0.5 | 109 ±2 | 17 ± 5 | 132 ± 2 |
| Pericenter argument (degrees) | 268 ± 60 | 112 ± 20 | 32 ± 30 | 170 ± 20 |
| Time of pericenter (Julian days) | 2452183.997± 0.23 | 2453064.806± 0.009 | 2453246.554 ± 0.126 | 2453813.5 ± 0.012 |
| Ascending Node (degrees) | 284.5 ± 0.5 | 203 ± 2 | 141 ± 2 | 235 ± 2 |
| Pericenter longitude (degrees) | 192 | 315 | 109 | 46 |
| Mean longitude (degrees) | 242 | 196 | 162 | 122 |
| Reference epoch | 2452186.5 | 2453068.5 | 2453249.5 | 2453813.5 |

**Table 7:** Physical properties of the binary asteroidal systems

|  | Kalliope I Linus | Eugenia I Petit-Prince | S/2002(107)1 | S/2001(762)1 |
|---|---|---|---|---|
| Mass System (kg) | $8.1\pm0.2 \times 10^{18}$ | $5.69 \pm 0.12 \times 10^{18}$ | $11.2\pm0.3 \times 10^{18}$ | $1.40 \pm 0.1 \times 10^{18}$ |
| $R_{Hill}$ (km) | 48000 | 40 000 | 64 000 | 26000 |
| a in $R_{Hill}$ | $1/40 \times R_{Hill}$ | $3/100 \times R_{Hill}$ | $2/100 \times R_{Hill}$ | $1/40 \times R_{Hill}$ |
| a in $R_{avg}$ | $12.5 \times R_p$ | $11 \times R_p$ | $10 \times R_p$ | $10 \times R_p$ |
| $R_{satellite}/R_{primary}$ | 0.15 | 0.03 | 0.06 | 0.14 |
| Density of Primary with $D_{STM/NEATM}$ and mass error (g.cm$^{-3}$) | $2.6/2.8 \pm 0.08$ | $1.09/1.06 \pm 0.02$ | $1.9/1.4 \pm 0.01$ | $1.0/0.9 \pm 0.01$ |
| Bulk Density[1] of Primary (g.cm$^{-3}$) | $2.8 \pm 0.2$ | $1.1 \pm 0.1$ | $1.4 \pm 0.3$ | $0.9 \pm 0.1$ |
| Spin Pole Solution in ECJ2000 and degrees | $195° \pm 7°$<br>$-4° \pm 5°$ | $102° \pm 2°$<br>$-30° \pm 1$ | $75° \pm 5°$<br>$55° \pm 10$ | $170° \pm 3°$<br>$-52° \pm 4$ |
| Inclination w.r.t. equator | ~0 | 12±1 | ~0 | ~0 |
| theoretical J2 | 0.192 | 0.192 | 0.130 | No shape model |
| measured J2 | Not detectable | 0.15 | Not detectable | Not detectable |

1. The errors on these bulk densities encompass the uncertainty on the size of the primary from the STM and NEATM models plus the error on the mass.

**Table 8:** Bulk-density of various binary "C-group" asteroids including C-type and the sub-class (X,Cb, G,Cb)

| Asteroid | Taxonomic class | Bulk-Density g/cm$^3$ | reference |
|---|---|---|---|
| 45 Eugenia | C-type | 1.1±0.1 | This work |
| 87 Sylvia | X-type | 1.2±0.1 | Marchis *et al.* 2005b |
| 90 Antiope | C-type | 1.25±0.05 | Descamps *et al.* 2007 |
| 107 Camilla | X-type | 1.4±0.3 | This work |
| 121 Hermione | C-type | 1.1±0.3 | Marchis *et al.* 2005a |
| 130 Elektra | G-type | 1.2±0.6 | Marchis *et al.* 2008 |
| 283 Emma | X-type | 0.85±0.05 | Marchis *et al.* 2008 |
| 379 Huenna | C-type | 0.85±0.05 | Marchis *et al.* 2008 |
| 762 Pulcova | Cb-Type | 0.9±0.1 | This work |

**Table 9:** List of main-belt binary asteroidal systems for which we have a good estimate of their mutual orbits (see Section 5 for references). Using Weidenschilling *et al.* (1989) describing the evolution of a binary asteroidal system due to tidal dissipations, we derive the evolution time scale for the semi-major axis in function of the of the tidal parameter µQ. µQ=$10^{10}$ N/m$^2$ gives realistic evolution time scale for these systems. The coefficient $\alpha_{evol}$ ($de/e = \alpha_{evol}$ x $da/a$ see Eq. 4) is always >0 indicating that the eccentricity of these systems should increase due to the tides assuming the same density, rigidity, and Q between the satellite and the primary.

| Binary system | Stability | Evolution time scale for *a* | | $\alpha_{evol}$ |
|---|---|---|---|---|
| | | µQ=$10^{10}$ | µQ=$10^{12}$ | |
| 22 Kalliope | No | 10 Myrs | 1 Byrs | 1.9 |
| 45 Eugenia | No | 2 Byrs | 260 Byrs | 2.2 |
| 87 Sylvia Romus I | No | 10 Myrs | 1Byrs | 1.9 |
| 87 Sylvia Remulus II | No | 2 Myrs | 280 Myrs | 2.2 |
| 90 Antiope | Yes | N/A | N/A | N/A |
| 107 Camilla | No | 90 Myrs | 9 Byrs | 2.1 |
| 121 Hermione | No | 40 Myrs | 4 Byrs | 2.2 |
| 130 Elektra | No | 4.3 Byrs | 400 Byrs | 2.2 |
| 283 Emma | No | 30 Myrs | 3 Byrs | 1.9 |
| 379 Huenna | Yes | N/A | N/A | 2.2 |
| 762 Pulcova | No | 80 Myrs | 8 Byrs | 1.1 |
| 3749 Balam | Yes | N/A | N/A | N/A |

**Figure 1a:** Search for moonlets around 22 Kalliope. On the left-top figure a VLT/NACO observation of Kalliope taken on Jul. 26, 2004 is displayed. The right top figure corresponds to the same observations after subtracting its azimuthally average. The detection of the moonlet companion is easier. The plot below is the azimuthally averaged 2-σ detection function for this observation. It is approximated using two linear functions which depends of three parameters: α, the coefficient of the slope of the linear regimes, $r_{lim}$ the separation between 2 regimes, Δm, the difference in magnitude in the stable regime. The minimum size of a moonlet to be detected can be derived from this profile. The characteristics of the 2- σ detection curve profile for all Kalliope observations can be found in Table 3a. Kalliope Linus I is visible on the images at 7 o'clock and 0.5 arcsec from the primary.

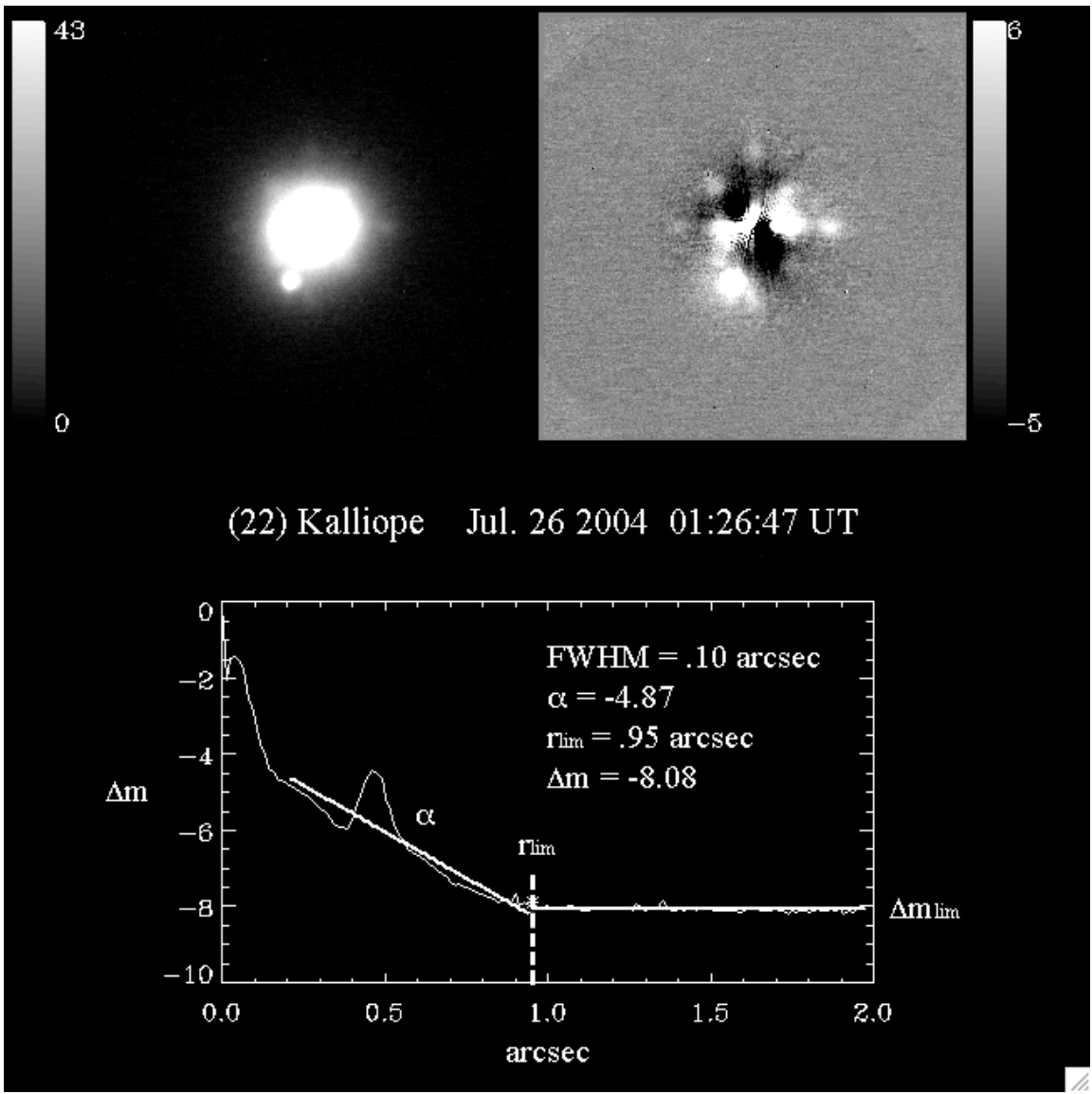

(22) Kalliope   Jul. 26 2004   01:26:47 UT

FWHM = .10 arcsec
$\alpha$ = -4.87
$r_{lim}$ = .95 arcsec
$\Delta m$ = -8.08

**Figure 1b:** Search for moonlets around 45 Eugenia for Feb. 15, 2004 VLT/NACO observations. Eugenia Petit-Prince I is located at 0.8 arcsec from the primary and is visible at 2 o'clock on the images. The second moonlet (temporary name S/2004(45)1) discovered by Marchis *et al.* (2007b) can be seen on the right top image at 1 o'clock.

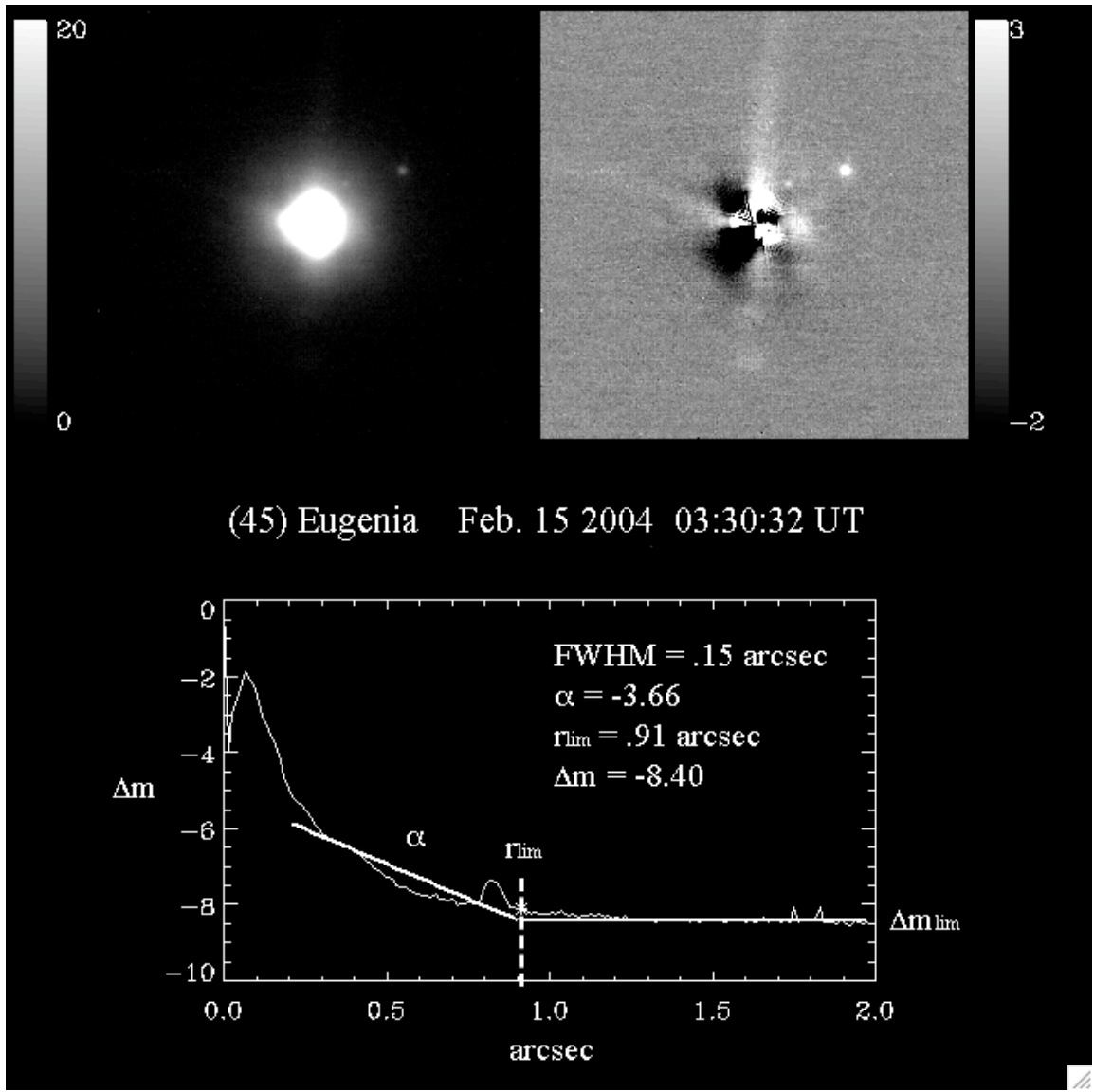

**Figure 1c:** Search for moonlets around 107 Camilla for Oct. 20, 2004 VLT/NACO observation. S/2002 (107) 1, known satellite of Camilla, is visible at 8 o'clock position with an angular separation of 0.6 arcsec.

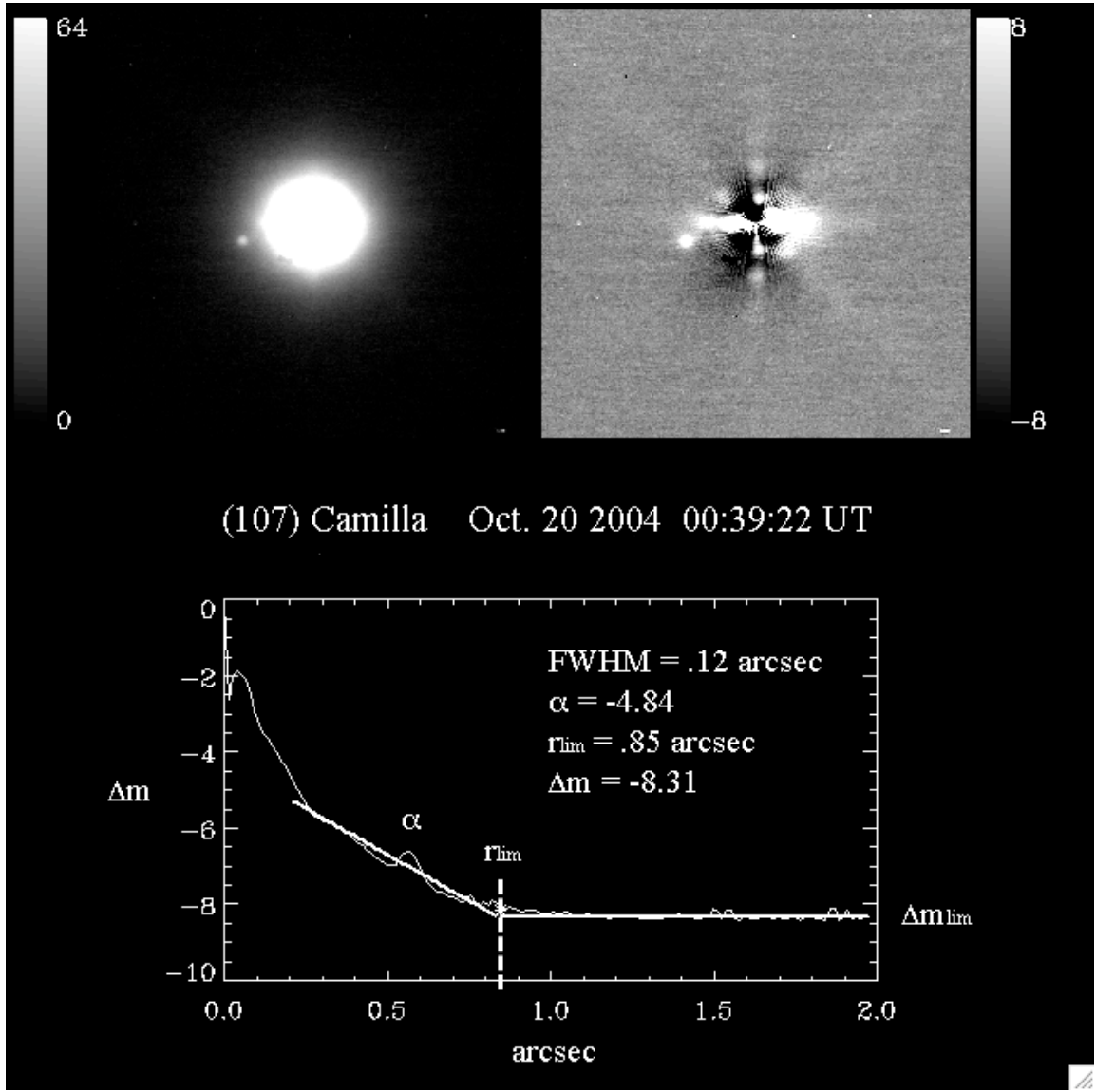

**Figure 1d:** Search for moonlets around 762 Pulcova on Jul. 15, 2003, taken with Gemini North telescope and Altair AO systems (NIRI camera). Its known satellite called S/2001(762)1 is located at 3 o'clock position with an angular separation of 0.3 arcsec.

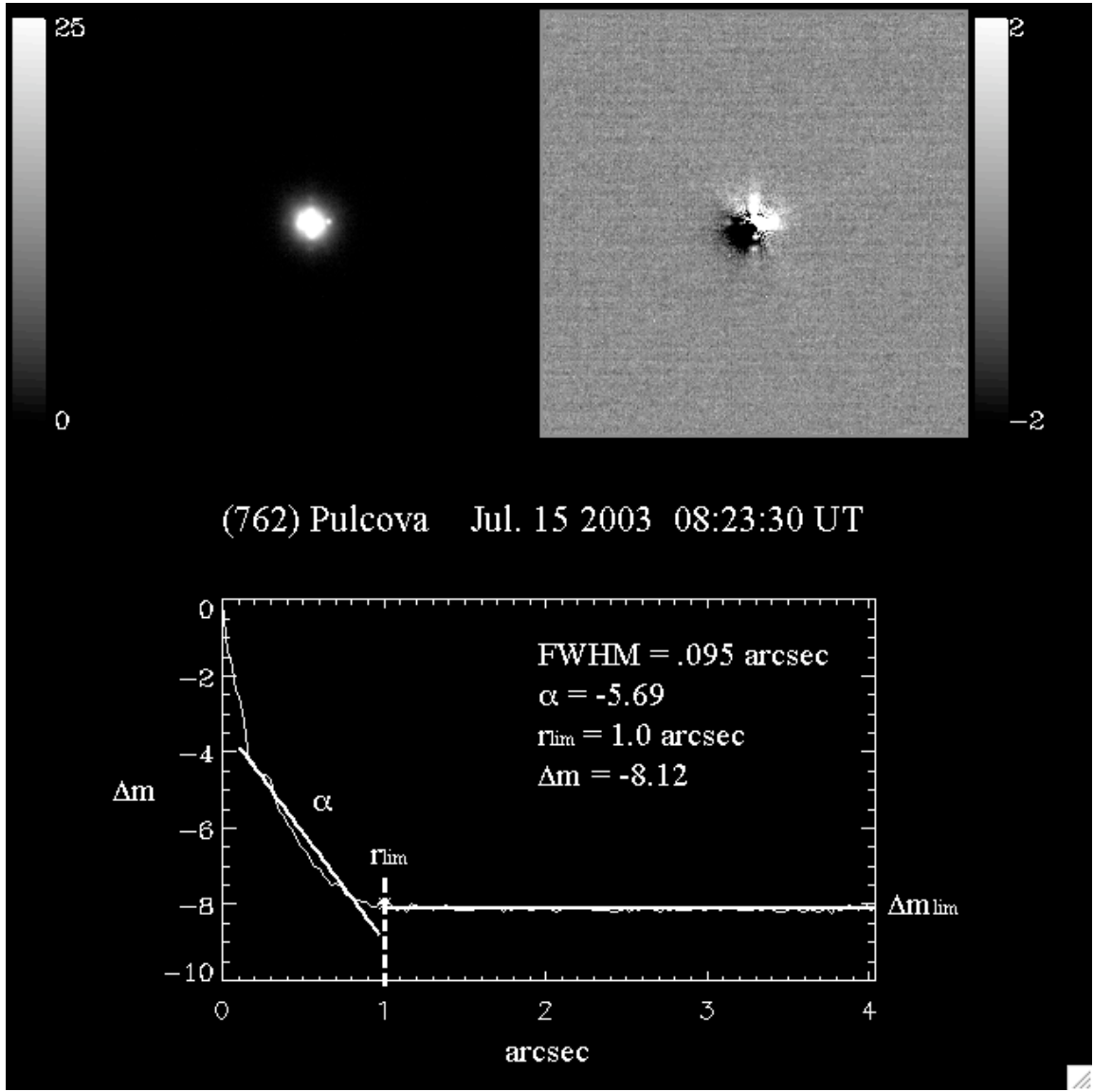

**Figure 2a:** Shape and orientation of 22 Kalliope and comparison with a refined Kaasalainen *et al.* (2002) 3D-shape model rendered with a Minnaert law (k = 0.65) . The global orientation of the model and observation images is excellent with an average shift of 10 degrees. The a/b ratio described in Table 4a is however significantly different suggesting that some corrections should be introduced in the shape model.

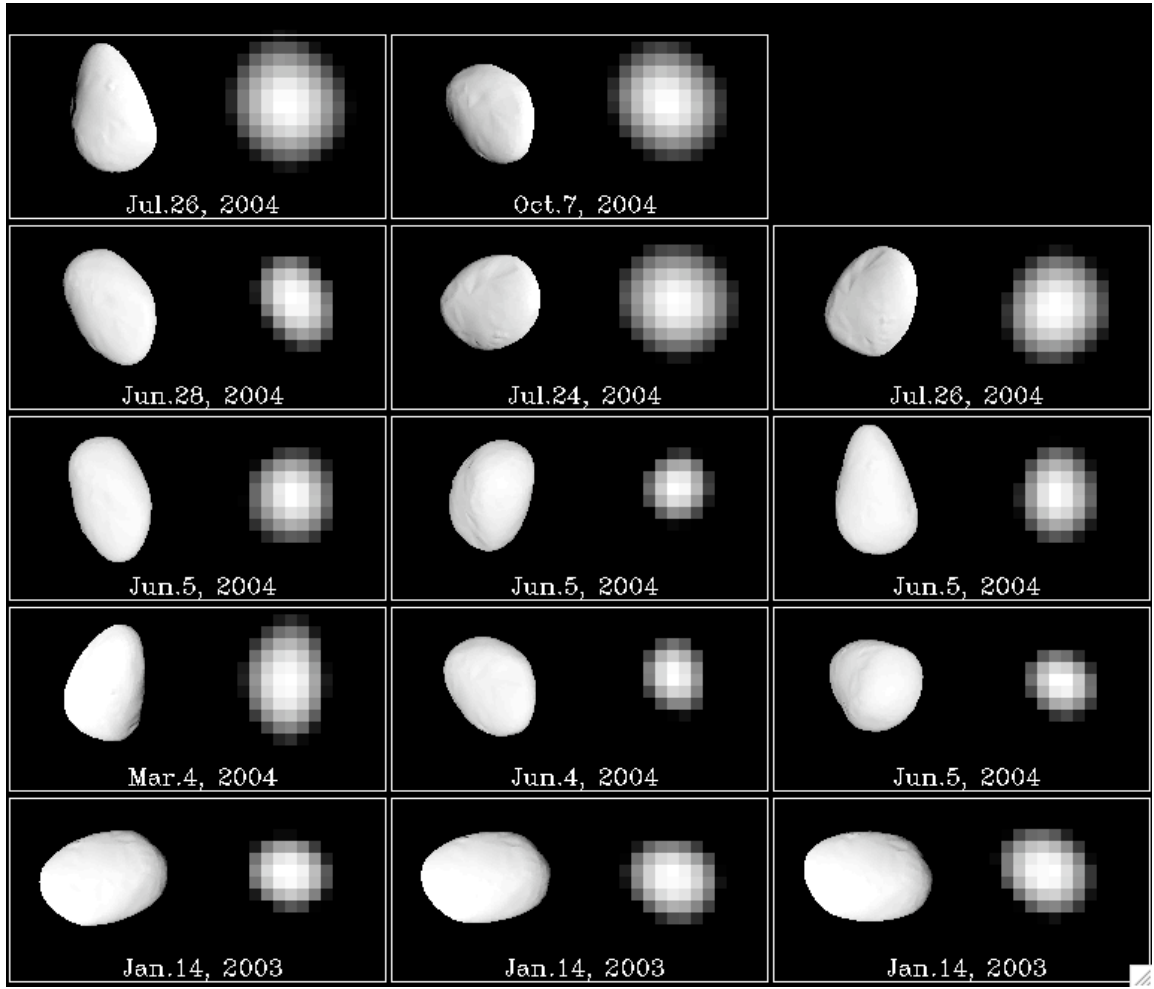

**Figure 2b**: Shape and orientation of 45 Eugenia and comparison with Kaasalainen *et al.* (2002) 3D-shape model rendered with a Minnaert law (k=0.65). The apparent diameter of the primary varies because of the distance to Earth. The model size was not adjusted to reproduce this effect; only its shape and orientation are comparable.

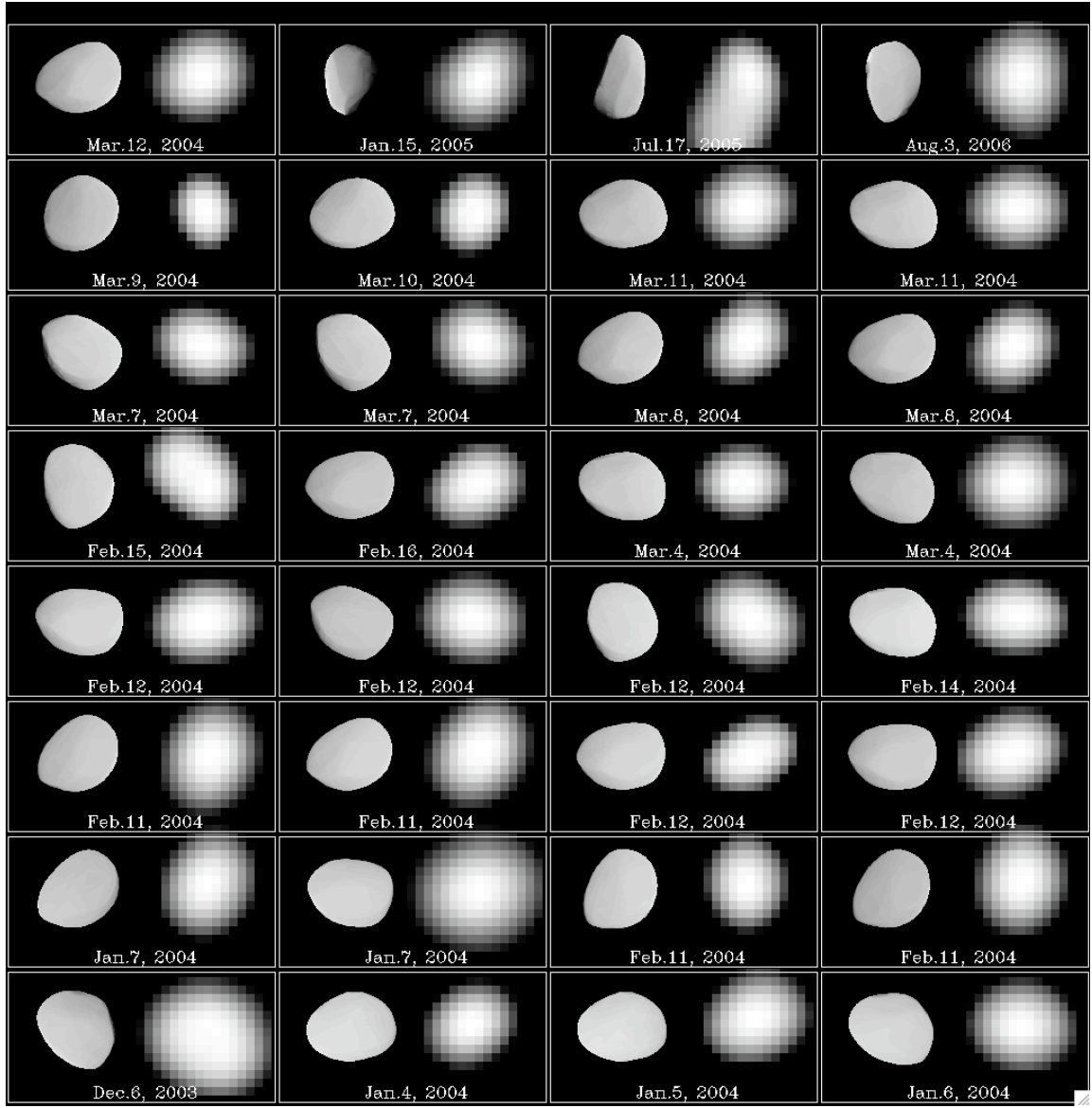

**Figure 2c**: Shape and orientation of 107 Camilla and comparison with Torppa *et al.* (2003) 3D-shape model (λ = 72º, β=51º ECJ2000) rendered with a Minnaert law (k=0.65). The apparent diameter of the primary varies because of the distance to Earth and the pixel scale of the detectors. The model size was not adjusted to reproduce this effect; only its shape and orientation are comparable.

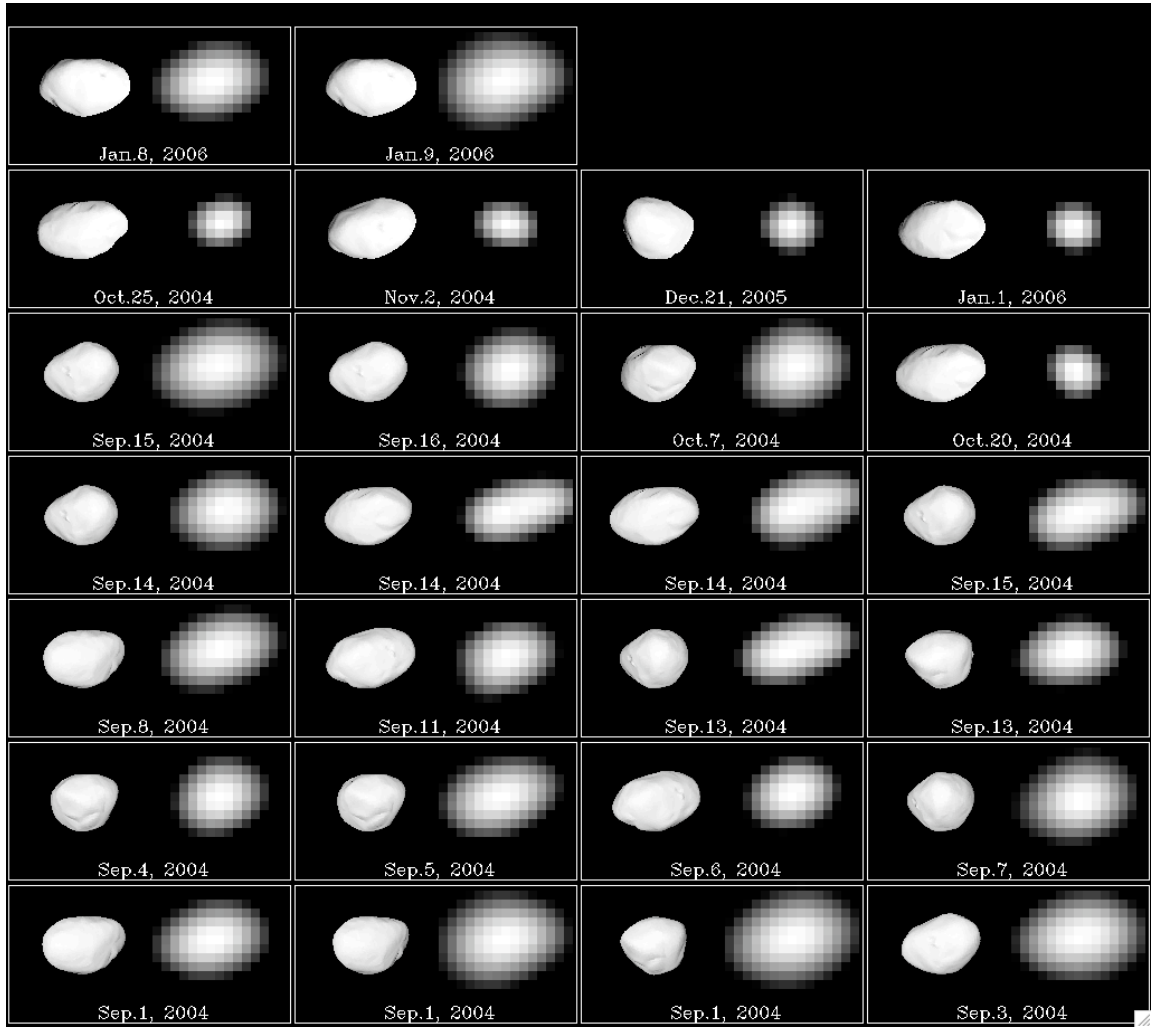

**Figure 3a**: Measured astrometric positions (crosses) from Table 5a of Linus, companion of Kalliope, and positions from our model (dots) are displayed.

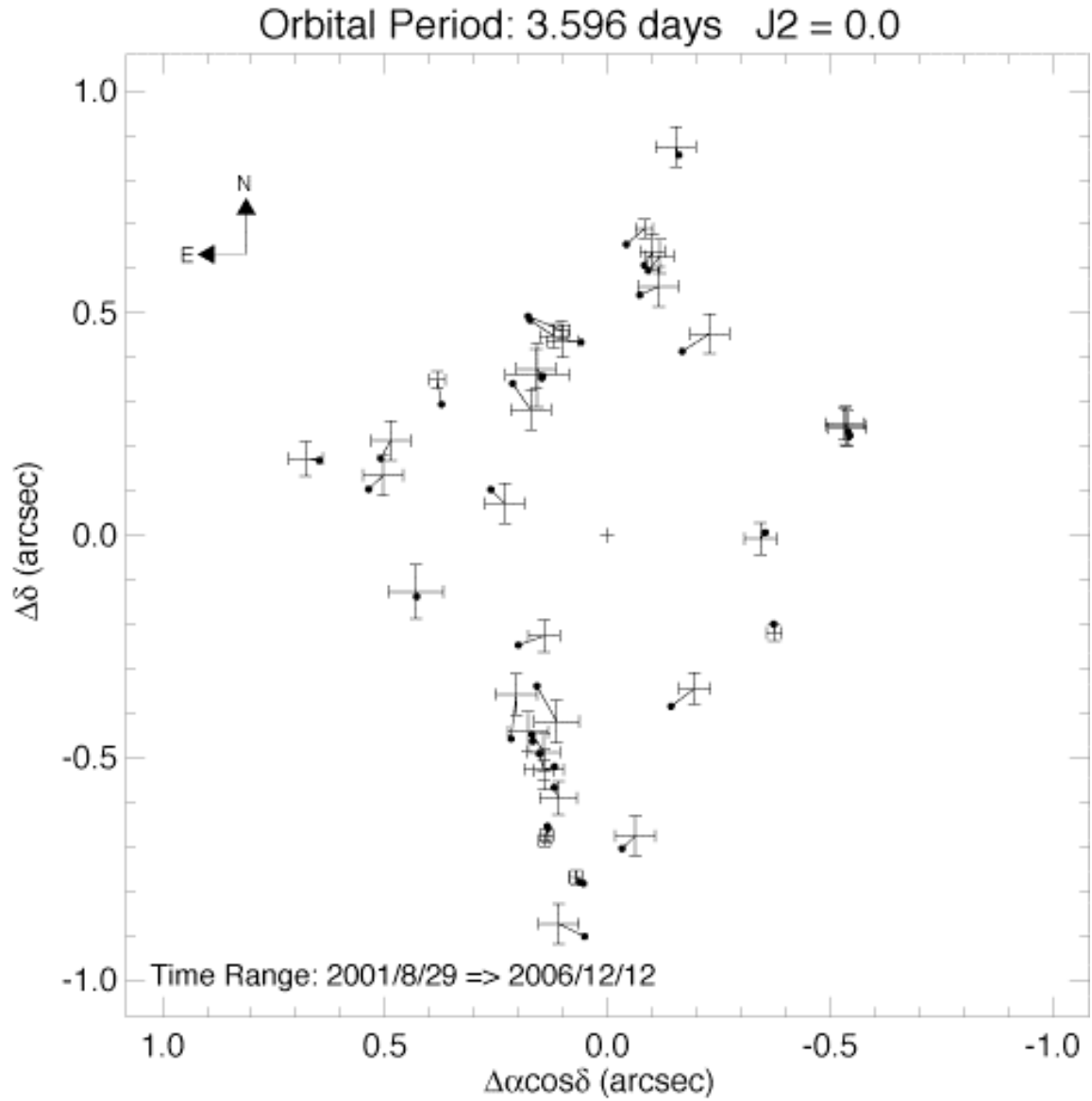

**Figure 3b:** Measured astrometric positions (crosses) from Table 5b of Petit-Prince, companion of Eugenia.

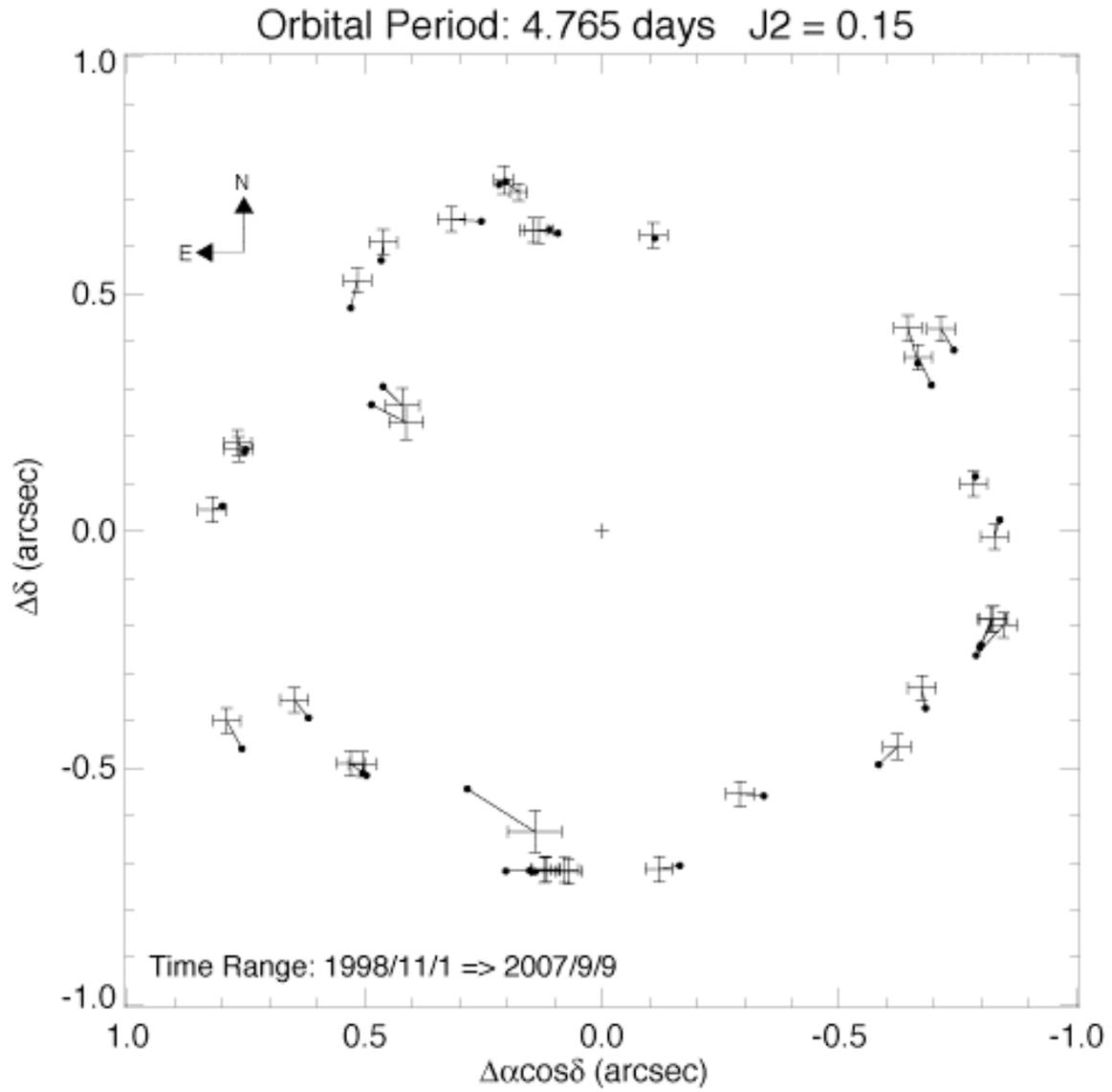

**Figure 3c:** Measured astrometric positions (crosses) from Table 5c of S/2002(107)1, companion of Camilla. The orbit is roughly seen in the edge on configuration.

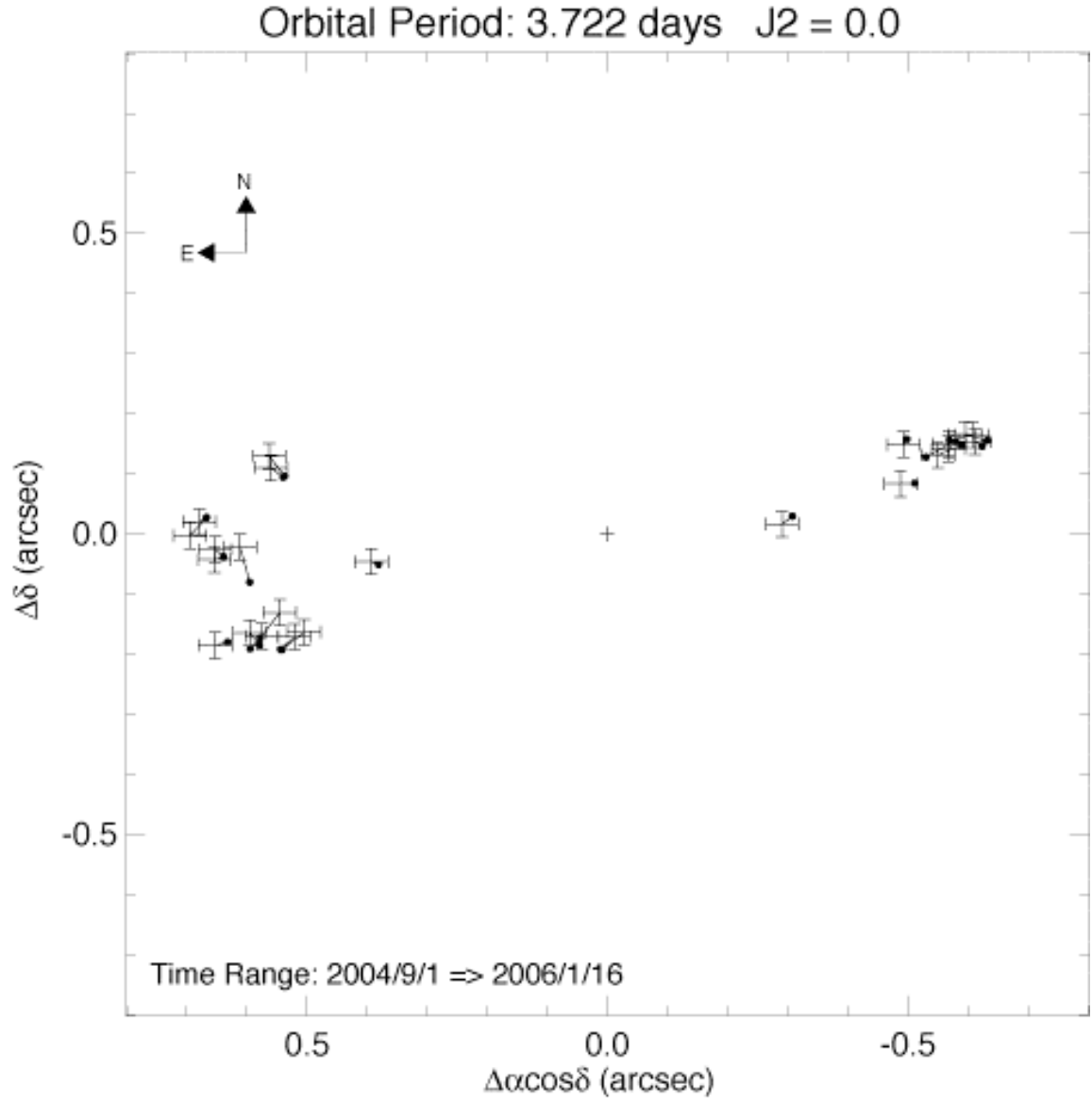

**Figure 3d:** Measured astrometric positions (crosses) from Table 5d of S/2001(762)1, companion of Pulcova. The distribution of points is not uniform along the orbit, so the orbital solution remains an approximation of the real orbit.

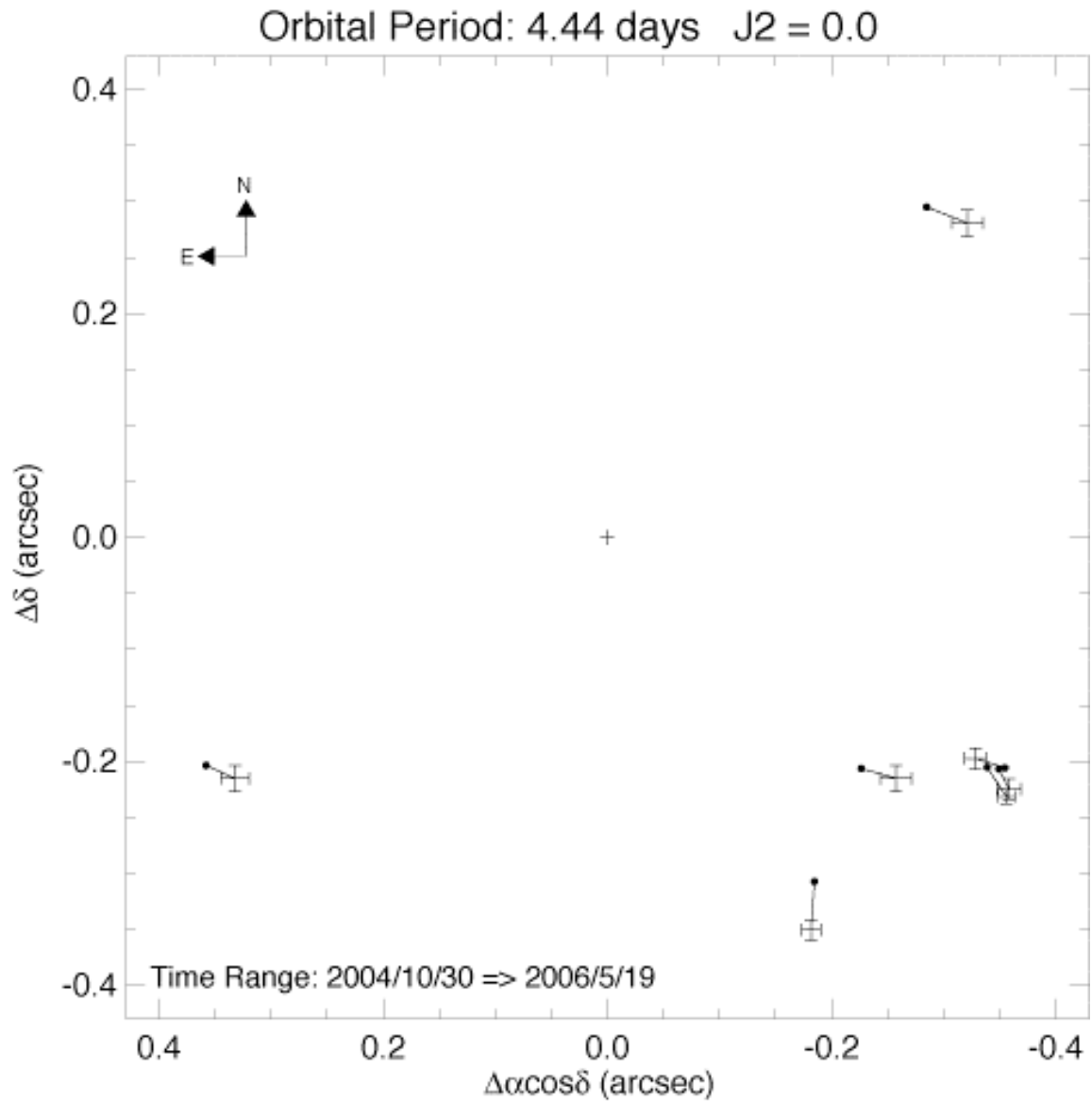

**Figure 4:** Evolution of binary asteroid mutual orbits due to tidal dissipation. This diagram shows the three domains of evolution for a binary asteroid system. A system with characteristics making it located above the synchronous stability (e.g. 90 Antiope) will be stable and do not evolve due to the tides. A system located underneath the eccentricity excitation limit, drawn here assuming $Q_s = Q_p$, $k_s=k_p$, $\rho_s=\rho_p$, will have a satellite orbiting describing an eccentric orbit due to excitation by the tides (e.g. the binary systems labeled in blue such as 130 Elektra and 283 Emma). Finally a binary system placed between these two limits will have its eccentricity damped due to the tides (e~0). 22 Kalliope and 762 Pulcova (red dots) orbits presented in this work are in agreement with this theoretical work (based on Weidenschilling *et al.* 1989). However, other binary systems such as 45 Eugenia, 107 Camilla, and 121 Hermione should not have a circular mutual orbit.

The almost-vertical dash lines define the timescale for the tides to act on the binary system. They were drawn assuming a density of 1.1 g/cm$^3$ for the primary and secondary and a product of rigidity and specific dissipation parameter $\mu Q \sim 10^{10}$, a realistic value for these rubble-pile asteroids (see Table 9). All the systems, but 130 Elektra, appear quite young (less than 1 Byrs old).

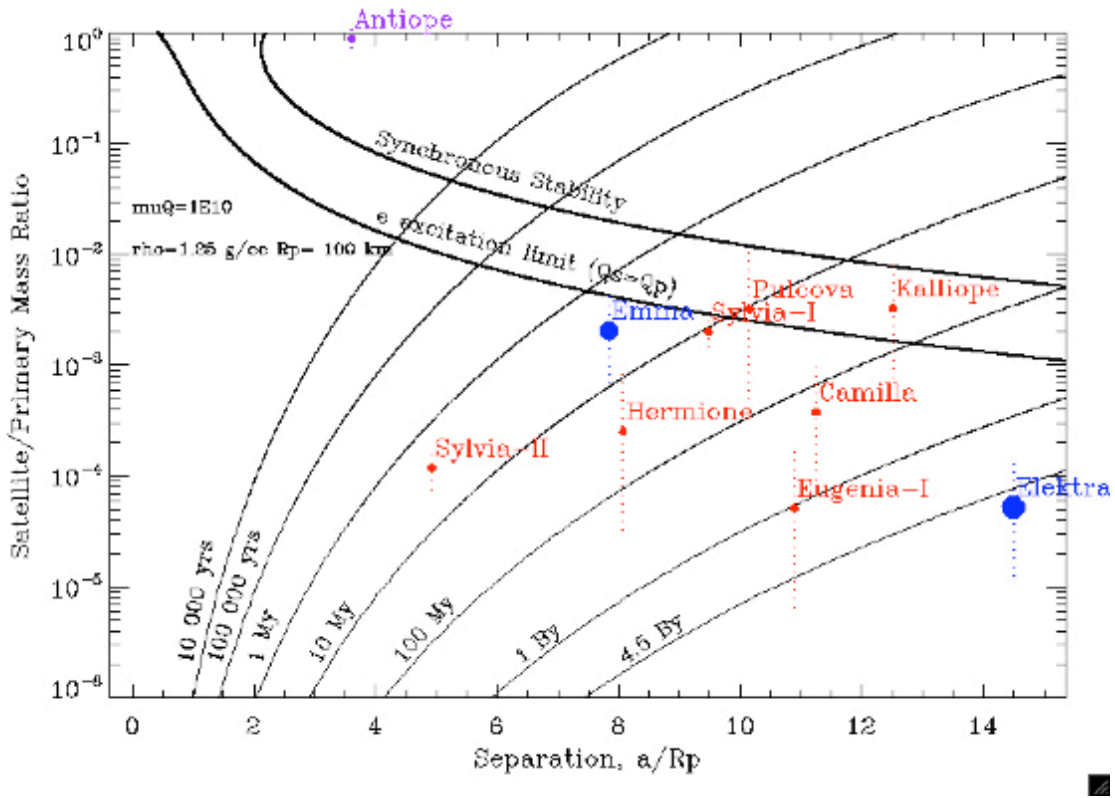